%% file: SDarXiv.tex
\documentclass[a4paper,12pt]{amsart}

\usepackage{geometry}
\usepackage{xr-hyper}
\makeatletter
\newcommand*{\addFileDependency}[1]{
   \typeout{(#1)}
   \@addtofilelist{#1}
   \IfFileExists{#1}{}{\typeout{No file #1.}}
}
\makeatother

\newcommand*{\myexternaldocument}[1]{%
     \externaldocument[][nocite]{#1}%
     \addFileDependency{#1.tex}%
     \addFileDependency{#1.aux}%
}

\myexternaldocument{SDsupplement}

\usepackage[comma]{natbib}
\usepackage[inline]{enumitem}
\usepackage{graphicx}
\usepackage[textfont=it,tableposition=above,font=small]{caption}		

\usepackage{placeins}
\usepackage{titling}        

\usepackage{framed}			
\usepackage[foot]{amsaddr}	
\usepackage{url}			
\usepackage{xargs}			
\usepackage{amssymb}		

\usepackage{silence}
\usepackage{todonotes}
\usepackage{pgffor}

\RequirePackage[colorlinks,citecolor=blue,urlcolor=blue]{hyperref}
\usepackage[norefs,nocites]{refcheck}
\usepackage{multirow}				
\usepackage{arydshln}				
\usepackage{psfrag}
\usepackage[symbol]{footmisc}
\usepackage{mathtools}				
\usepackage{booktabs}               
\usepackage{rotating}

\usepackage{color}
\usepackage{xcolor}
\usepackage{colortbl}

\newtheorem{theorem}{Theorem}

\newtheorem{corollary}[theorem]{Corollary}

\theoremstyle{definition}

\newcommand{\PP}{\mathsf P}								
\newcommand{\R}{\mathbb R}								
\newcommand{\I}[1]{\mathbb{I}\left[{#1}\right]}			
\newcommand{\dd}{\,\mathrm{d}\,}						
\DeclareMathOperator{\E}{E}								
\DeclareMathOperator{\Var}{Var}							
\newcommand{\norm}[1]{\left\Vert #1 \right\Vert}        
\newcommand{\eqd}{\stackrel{\mathclap{\mbox{\tiny{\emph{d}}}}}{=}}

\newcommand{\Sph}[1][d-1]{\mathbb{S}^{#1}}

\newcommand{\Simplex}[1]{\mathsf{S}\left( #1 \right)}

\newcommand{\calA}{\mathcal A}

\newcommand{\Prob}[1][\R^d]{\mathcal P\left({#1}\right)}  

\definecolor{yllw}{rgb}{0.9,0.9,0}

\title{Resampling simplicial depth}

\author{Stanislav Nagy$^{1}$}
\author{Martin Wendler$^2$}
\author{Carsten Jentsch$^3$}

\email{nagy@karlin.mff.cuni.cz}

\address{\hspace{-1em}$^1$
	Faculty of Mathematics and Physics,
	Charles University, Prague,
	Czech Republic
}

\address{$^2$
    Institute for Mathematical Stochastics,
    Otto-von-Guericke-Universit\"at Magdeburg, Magdeburg, Germany}

\address{$^3$
    Department of Statistics, 
    TU~Dortmund University, 
    Dortmund, Germany}    

\date{\today}

\begin{document}

\begin{abstract}
The simplicial depth (SD) is a commonly used indicator of the centrality of points $x\in\R^d$ with respect to distributions $P$ on $\R^d$. Asymptotic theory for the sample SD is based on its representation as a $U$-statistic, which can be either non-degenerate or degenerate. For $d=2$, we prove under mild conditions that this $U$-statistic is degenerate with rate $n$ if and only if $x$ is a center of symmetry of $P$. Otherwise, the asymptotic distribution of SD is non-degenerate with rate $\sqrt{n}$. Because the location of the center of symmetry of $P$ is usually unknown, these two modes of behavior complicate the estimation of the sample distribution of SD at $x$. 

We propose a two-step adaptive subsampling procedure for estimating that distribution. First, an estimator $\widehat \gamma$ of a parameter $\gamma\in\{1/2, 1\}$ characterizing the correct rate of convergence $n^\gamma$ of SD is constructed based on subsampling. Our estimator uses a bias correction suitable for $U$-statistics. Second, $\widehat \gamma$ is employed for approximating the distribution of the sample SD. We prove the consistency of this subsampling approach and illustrate its usefulness (i) in the construction of confidence intervals for SD, and (ii) in an SD-based supervised classification task. 

\vspace{2mm}

\vspace{2mm}

\end{abstract}

\maketitle

\section{Introduction: Simplicial depth and its asymptotics}

Nonparametric statistical methods are of great importance in the analysis of univariate data. In contrast, nonparametric statistics for multivariate data is considerably less developed, the main difficulty stemming from the fact that no canonical ordering of points in multivariate spaces $\R^d$ exists. No universally accepted notions of quantiles, ranks, or orderings exist for data living in $\R^d$. One way of dealing with this problem offer statistical depth functions \citep{Tukey1975, Liu1988, Liu1990, Liu_etal1999, Zuo_Serfling2000}. Instead of defining a universally valid ordering in $\R^d$, depth functions order points in $\R^d$ with respect to (w.r.t.) a reference distribution $P$ on $\R^d$. The depth of $x \in \R^d$ w.r.t. $P$ is a number $D(x; P) \in [0,1]$ describing how much ``centrally positioned'' is $x$ within the mass of $P$. Data points are then ranked from the most central ones (with high depth) towards the peripheral ones, according to their decreasing depth. The (data) point $x$ with the highest depth value $D(x; P)$ in $\R^d$ is a multivariate generalization of the median of $P$, and loci of points whose depth exceeds given thresholds $\alpha > 0$ may serve as regions in $\R^d$ containing the ``$\alpha$-central part'' of the distribution $P$. 

Many different depth functions have been proposed in the literature. We mention the halfspace depth \citep{Tukey1975}, the spatial depth \citep{Chaudhuri1996}, and the zonoid depth \citep{Koshevoy_Mosler1998}. In this paper, we are interested in another classical depth function, the simplicial depth \citep{Liu1988, Liu1990}. Let $\Prob[\R^d]$ denote the set of all (Borel) probability distributions on $\R^d$. For $x \in \R^d$ and $P \in \Prob$, the \emph{simplicial depth} (SD) is given by
    \begin{equation}    \label{eq: SD}
    SD(x; P) = \PP\left( x \in \Simplex{X_1,\dots, X_{d+1}} \right),
    \end{equation}
where $X_1, \dots, X_{d+1}$ are independent and identically distributed (i.i.d.) random variables with distribution $P$, and $\Simplex{x_1,\dots, x_{d+1}} \subset \R^d$ is the closed convex hull (the simplex) determined by its vertices $x_1, \dots, x_{d+1} \in \R^d$. The sample version of the simplicial depth~\eqref{eq: SD} is a U-statistic of order $d+1$ given by
    \begin{equation}   \label{eq: SDn} 
    SD_n(x; \mathcal X_n) = \binom{n}{d+1}^{-1} \sum_{1 \leq i_1 < i_2 < \dots < i_{d+1} \leq n} \I{x \in \Simplex{X_{i_1}, X_{i_2}, \dots, X_{i_{d+1}}}},
    \end{equation}
where $\mathcal X_n = \{ X_1, \dots, X_n\}$ is a random sample from $P$, and $\I{\cdot}$ is the indicator function. As the sample simplicial depth~\eqref{eq: SDn} is a U-statistic, many statistical properties of SD and its sample version are already well explored in the literature \citep{Serfling1980, Liu1990, Arcones_Gine1992, delaPena_Gine1999}. 

In this paper, we are concerned with the estimation of both the finite sample, and the asymptotic distribution of the sample SD from~\eqref{eq: SDn}. We combine the surprisingly fast computation of $SD_n$ in the plane with a suitable resampling principle. For this purpose, we design a bootstrap method to approximate the distribution of $SD_n(x; \mathcal X_n)$. This method is adaptive in the sense that the resampling procedure automatically adapts to the (unknown) degree of degeneracy of the U-statistic from~\eqref{eq: SDn}. We suppose that a point $x \in \R^d$ (in what follows, we focus mainly on $d=2$) and a distribution $P \in \Prob[\R^d]$ are given, and $\mathcal X_n \subset \R^d$ is a random sample of size $n$ from $P$. We consider the quantity
    \begin{equation}  \label{eq: Rn}
    R_n = \tau(n)\left( SD_n(x; \mathcal X_n) - SD(x;P) \right), 
    \end{equation}
for an appropriate sequence of constants $\left\{ \tau(n) \right\}_{n=d+1}^\infty$ such that the limiting distribution of $R_n$ is non-degenerate. According to the general theory of U-statistics \citep[Section~5.5]{Serfling1980}, we know that the asymptotic distribution of $R_n$ heavily depends on whether the associated U-statistic~\eqref{eq: SDn} degenerates or not: 
   \begin{itemize}
        \item In case of non-degeneracy of $SD_n(x; \mathcal X_n)$, the appropriate rate of $R_n$ is $\tau(n) = \sqrt{n}$, and the asymptotic distribution of $R_n$ is normal \citep[Section~5.5.1]{Serfling1980}
            \[  
            R_n = \sqrt{n}\left( SD_n(x; \mathcal X_n) - SD(x; P)\right) \xrightarrow[n\to\infty]{\mathcal D} N(0, (d+1)^2 \zeta_1),
            \]
        for $\zeta_1 = \Var{h_1(X_1)} > 0$ with $h_1(y) = \E \big[\I{x \in \Simplex{y, X_{2}, \dots, X_{{d+1}}}}\big]$.
        \item If the U-statistic $SD_n(x; \mathcal X_n)$ is degenerate (of order $1$), the rate is $\tau(n) = n$, and the asymptotic distribution of $R_n$ is a weighted sum of $\chi^2_1$-distributed random variables \citep[Section~5.5.2]{Serfling1980}
        \[  
        R_n = n \left( SD_n(x; \mathcal X_n) - SD(x; P)\right) \xrightarrow[n\to\infty]{\mathcal D} \frac{d(d+1)}{2} Y,
        \]
    where the random variable $Y$ can be written as $Y = \sum_{j=1}^\infty \lambda_j \left( \chi^2_{1,j} - 1 \right) $ for $\chi^2_{1,1}, \chi^2_{1,2}, \dots$ a sequence of i.i.d. $\chi^2_1$-distributed random variables, and an appropriate sequence of real constants $\lambda_1, \lambda_2, \dots$. 
    \end{itemize}
A well-known fact \citep[Remark~B]{Liu1990} is that for $P$ having a density and $x \in \R^d$ the center of (angular or halfspace) symmetry of $P$,\footnote{Formal definitions of these notions of symmetry are given in Section~\ref{section: degeneracy} below.} the U-statistic~\eqref{eq: SDn} is degenerate. In the plane, it then must be degenerate of order $1$, see also \citet[Section~4]{Killeen_Hettmansperger2021}. Besides these results, it is, however, interesting to note that not much more is known about the degeneracy of the SD. For example, it appears to be unknown whether the U-statistic $SD_n(x; \mathcal X_n)$ is degenerate (of order $1$) only if $x$ is the center of symmetry of $P$. This incomplete picture of the asymptotic properties of SD substantially complicates statistical inference based on this depth \citep{Dyckerhoff_etal2015, Malcherczyk_etal2021, Muller_etal2025}.

As our first result, we show in Section~\ref{section: degeneracy} that under mild conditions, the U-statistic~\eqref{eq: SDn} in the plane is degenerate if and only if $x$ is the center of angular (or halfspace) symmetry of $P$, thus resolving the characterization problem of degeneracy of SD in the plane. The proof of our result relies heavily on the geometry of the plane, and it is not clear whether an analogous statement holds true in higher dimensions. Such a substantial disparity between the theory of SD in the plane and in dimension $d>2$ is well known in the literature, and for exactly the same reasons, much of what can be found about the theory of SD is limited to dimension $d = 2$ \citep{Oja_Nyblom1989, Cascos2009, Dyckerhoff_etal2015, Killeen_Hettmansperger2021, Malcherczyk_etal2021, Mendros_Nagy2025}.

Our result on characterizing the degree of degeneracy of $R_n$ in dimension $d = 2$ relates to several well-known tests for bivariate symmetry that can be interpreted as SD-based tests. Testing procedures based on SD for deciding whether $x \in \R^2$ is a center of symmetry of $P \in \Prob[\R^2]$ are constructed in \citet[Section~3]{Oja_Nyblom1989} and \citet{Killeen_Hettmansperger2021}. In both these references, we find explicit expressions for the limiting distribution of the test statistic $R_n$ from~\eqref{eq: Rn} with $\tau(n) = n$ and $SD(x; P) = 1/4$ under the null hypothesis of symmetry around $x$, that is, in the degenerate case.\footnote{In \citet[Example~5.1]{Oja_Nyblom1989}, there appears to be a typo. Comparing their formula to the expression from \citet[Theorem~2]{Killeen_Hettmansperger2021}, we observe that the $\chi^2_1$-distributions given in \citet[Example~5.1]{Oja_Nyblom1989} should be centered around their expectation $2$.} In \citet{Killeen_Hettmansperger2021}, also the exact finite-sample distribution of $R_n$ under the null hypothesis (that is, in the degenerate case again) is studied. That distribution is found to relate to a relative number of sub-sequences of form 010 or 101 in a random sample of $n$ fair Bernoulli trials \citep[Theorem~1]{Killeen_Hettmansperger2021}. The last quantity (the number of alternating sub-sequences of length $3$) is, in turn, equivalent to the so-called $3$-sign depth of time series considered extensively in \citet{Malcherczyk_etal2021} and the references therein. In \citet[Theorem~2.2]{Malcherczyk_etal2021}, the asymptotic distribution of the $3$-sign depth is derived in the degenerate setting. The last result can be found to be equivalent to those from \citet{Oja_Nyblom1989} and \citet{Killeen_Hettmansperger2021}, for a proof of equivalence see~\citet{Mendros2025}. Note that all these important results on the behavior of $R_n$ cover only the situation under the assumption of degeneracy of $SD_n(x; \mathcal X_n)$. 

Compared to the references above, the scope of this paper is wider. We consider a scenario when only $x \in \R^2$ and a random sample $\mathcal X_n = \left\{ X_1, \dots, X_n \right\}$ from an unknown distribution $P \in \Prob[\R^2]$ are given, and our task is to approximate the (finite sample, or asymptotic) distribution of the quantity $R_n$ from~\eqref{eq: Rn}. Without making prior assumptions on the degree of degeneracy of $SD_n(x; \mathcal X_n)$, we design fast bootstrap procedures for approximating the distribution of $R_n$. These methods are adaptive in the sense that the resampling automatically adapts to the degree of degeneracy of $SD_n(x; \mathcal X_n)$. One novelty of our approach is the use of bootstrap procedures in conjunction with SD. This has not been considered in the literature before, mainly because of the prohibitive computational cost of $SD_n(x; \mathcal X_n)$ in $\R^d$. For $d=2$, however, quite efficient computational methods are available \citep{Rousseeuw_Ruts1996, Aloupis_etal2002}, which make bootstrapping SD not only possible but also extremely fast.

Building upon the discussion on the degeneracy of SD in the plane in Section~\ref{section: degeneracy}, in Section~\ref{section: gamma estimators}, we use subsampling to estimate the rate of convergence (and thus the degree of degeneracy) of the sample SD. Several versions of the estimators are considered, which are all proven to be consistent. These rate estimators are, nevertheless, found to suffer from systematic biases. To address this problem, in Section~\ref{section: bias correction}, we propose two heuristic methods for correcting the bias in these estimators. The performance of our methods is investigated in a numerical study in Section~\ref{section: simulations}, where we tune the parameters of the estimators, and demonstrate good finite sample performance of our novel methods. The paper is concluded in Section~\ref{section: applications}, where our method is applied in the practical tasks of (i) constructing confidence intervals for the true simplicial depth $SD(x; P)$ in Section~\ref{section: application 1}, and (ii) assessing the credibility of an SD-based supervised classification procedure in the plane in Section~\ref{section: application 2}. Technical proofs and additional detailed results of our simulations can be found in the online Supplementary Material. The online Supplementary Material also contains the complete \textsf{R} codes of our simulation studies and a new \textsf{R} package \textsf{SD} implementing both the computation of the bivariate SD and our resampling estimators in \textsf{C++} and \textsf{R} via the \textsf{RcppArmadillo} framework \citep{RcppArmadillo}.


\subsection*{Notations} We write $(\Omega, \calA, \PP)$ for the probability space on which all random elements are defined. The set of Borel probability measures on a topological space $\R^d$ is denoted by $\Prob[\R^d]$. By $X \sim P$ with $P \in \Prob[\R^d]$ we mean (an $\R^d$-valued) random variable $X$ with distribution $P$. By $X \eqd Y$ we mean that the random vectors $X$ and $Y$ have the same distribution. The convergence of a sequence of variables $\left\{X_n\right\}_{n=1}^\infty$ to $X$ in probability as $n \to \infty$ is denoted by $X_n \xrightarrow{\mathcal{P}} X$.

\section{Characterizing degeneracy of SD in the plane}   \label{section: degeneracy}

Recall that a random vector $X \sim P \in \Prob$ is called \emph{angularly symmetric} around a point $x \in \R^d$ \citep{Liu1990, Zuo_Serfling2000c} if 
    \[  \frac{X - x}{\norm{X - x}} \eqd -\frac{X - x}{\norm{X - x}},  \]
with the convention that $0/0 = 0 \in \R^d$ if $\PP(X = x)>0$. According to \citet[Theorems~1 and~2]{Rousseeuw_Struyf2004}, for any $P$ without atoms, angular symmetry of $P$ around $x$ is equivalent with the \emph{halfspace symmetry} of $P$ around $x$, which means that $P(H) \geq 1/2$ for each closed halfspace $H$ whose boundary hyperplane passes through $x$. Both these notions of symmetry are weaker than the standard concept of \emph{central symmetry} of $X \sim P$ around $x$, meaning that $X-x$ has the same distribution as $x-X$; for a thorough discussion, see \citet{Zuo_Serfling2000c} and \citet{Rousseeuw_Struyf2004}.

Suppose for simplicity that we are given $P \in \Prob[\R^2]$ that is absolutely continuous, and $\mathcal X_n = \left\{ X_1, \dots, X_n \right\}$ is a random sample from $P$. We are interested in the degeneracy of the U-statistic $SD_n(x; \mathcal X_n)$ involved in $R_n$ in~\eqref{eq: Rn}. An important fact is that under the assumption of angular (or spherical) symmetry of $P$ around $x \in \R^2$, the distribution of $R_n$ is always the same (that is, it does not depend on the exact form of $P$), and degenerate of order $1$ \citep{Oja_Nyblom1989, Liu1990, Killeen_Hettmansperger2021, Malcherczyk_etal2021}. The main result of the present section is Theorem~\ref{thm: Thm1}, where we provide a converse to this claim, thus resolving the problem of degeneracy of SD in the plane. We prove that for $d=2$, the U-statistic $SD_n(x; \mathcal X_n)$ is degenerate of order $1$ if and only if $x \in \R^2$ is the center of (angular or halfspace) symmetry of $P$. Otherwise, the U-statistic $SD_n(x; \mathcal X_n)$ is non-degenerate.



\begin{theorem}[Characterization of degeneracy of $SD_n$] \label{thm: Thm1}
Let $P \in \Prob$ be a distribution with a Lebesgue density. Then the following holds true.
    \begin{enumerate}[label=(\roman*)] 
    \item If $P$ is angularly (or halfspace) symmetric around $x \in \R^d$, then the U-statistic~\eqref{eq: SDn} is degenerate of order 1.
    \item For $d=2$, the U-statistic~\eqref{eq: SDn} is degenerate if and only if either $SD(x; P) = 0$, or if $x$ is the center of angular (or halfspace) symmetry of $P$.
    \end{enumerate}
\end{theorem}

The proof of Theorem~\ref{thm: Thm1} can be found in the Supplementary Material, Section~\ref{section: proofs}.
In dimension $d>2$, Theorem~\ref{thm: Thm1} states only one implication: for $x \in \R^d$ being the center of halfspace symmetry of $P$, the U-statistic $SD_n(x; \mathcal X_n)$ is degenerate. It appears to be an open problem whether the other implication is valid in higher dimensions. Our proof of Theorem~\ref{thm: Thm1} relies on the geometry of the plane and does not directly extend to dimension $d>2$.

\section{Subsampling simplicial depth}    \label{section: rate}

\subsection{A subsampling estimator of the convergence rate of \texorpdfstring{$SD_n$}{SD}} \label{section: gamma estimators}

The convergence rate of the sample SD depends on the (non-)degeneracy of the corresponding U-statistic $SD_n(x; \mathcal X_n)$. Thus, in view of Theorem~\ref{thm: Thm1}, this rate is unknown, and it has to be estimated. We propose a subsampling procedure for estimating the rate of convergence in a more general situation: Assume that our quantity of interest is $\theta \in \R$, and we have an estimator $\widehat{\theta}_n$ based on a dataset $\mathcal X_n = \{ X_1, \dots, X_n \}$. We suppose that $n^{\gamma}(\widehat{\theta}_n-\theta)$ converges in distribution to an unknown, but non-degenerate distribution, and the rate parameter $\gamma>0$ is also unknown. Our procedure consists of four steps.

\begin{enumerate}
    \item Choose integers $1 < m_\ell<m_u<n$, such that $m_\ell\rightarrow \infty$, $m_u/m_\ell\rightarrow \infty$ and $n/m_u\rightarrow \infty$ as $n\rightarrow \infty$.
    \item For $s=1,\dots,S$:
    \begin{enumerate}
    \item Choose a subsample of size $m_u$ from $\mathcal X_n$ (drawing $m_u$ observations uniformly from $X_1, \dots, X_n$ without replacement).
    \item Calculate the estimator $\widetilde\theta_{m_u,s}$ from the $m_u$ observations from step (a).
    \item  From the $m_u$ observations of this subsample from step (a), draw another subsubsample of size $m_\ell$.
    \item Calculate the estimator $\widetilde\theta_{m_\ell,s}$ from the $m_\ell$ observations from step (c).
    \end{enumerate}
    \item Calculate $T_u$ and $T_\ell$ defined by
        \[
        T_u=\operatorname{median}\{|\widetilde\theta_{m_u,s}-\widehat{\theta}_n| \colon s=1,\dots, S \} \quad \mbox{and} \quad T_\ell=\operatorname{median}\{|\widetilde\theta_{m_\ell,s}-\widehat{\theta}_n |\colon s=1,\dots,S\}.
        \]
    \item The (raw) estimator $\widehat{\gamma}$ is the solution of $  m_u^{\widehat{\gamma}}T_u=m_\ell^{\widehat{\gamma}}T_\ell$, so
    \begin{equation} \label{eq: gamma hat}
    \widehat{\gamma}=\frac{\log T_\ell - \log T_u}{\log m_u-\log m_\ell}.
    \end{equation}
   
\end{enumerate}
In the situation with the sample SD, we know by Theorem~\ref{thm: Thm1} that $\gamma\in\{1/2,1\}$. Hence, instead of the unrestricted estimator $\widehat \gamma$ we can also use the trimmed version of $\widehat{\gamma}$ from~\eqref{eq: gamma hat} in the form
    \begin{equation*}
    \widehat{\gamma}_\mathrm{t}=\max\left\{\min\left\{\frac{\log T_\ell - \log T_u}{\log m_u-\log m_\ell},1\right\},1/2\right\},
    \end{equation*}
which, by construction, takes values in $[1/2,1]$, or the rounded estimator
         \[  \widehat{\gamma}_\mathrm{r} = 
            \begin{cases}
            1 & \mbox{if } \frac{\log T_\ell - \log T_u}{\log m_u-\log m_\ell} > 3/4, \\
            1/2 & \mbox{otherwise},
            \end{cases}  
            \]
guaranteed to lie in $\{1/2,1\}$. Using the consistency of subsampling (see \citealp{Politis_Romano1999}), it is possible to show also the consistency of the three estimators $\widehat{\gamma}$, $\widehat{\gamma}_\mathrm{t}$ and $\widehat{\gamma}_\mathrm{r}$. The proof of the following theorem is in the Supplementary Material, Section~\ref{section: proofs}.

\begin{theorem}[Consistent estimation of $\gamma$]\label{lem:gammaest} Let $\left\{X_n\right\}_{n=1}^\infty$ be an i.i.d. sequence of random variables and $\widehat{\theta}_n=\widehat{\theta}_n(X_1,...,X_n)$ be a sequence of estimators of $\theta \in \R$, such that for some $\gamma>0$, $n^{\gamma}|\widehat{\theta}_n-\theta|$ converges to a non-degenerate limit distribution, which has a uniquely defined median $M \in \R$. 
    \begin{enumerate}[label=(\roman*)]
        \item If  $m_\ell<m_u<n$,  $m_\ell\rightarrow \infty$,  $n/m_u\rightarrow \infty$, $m_u/m_\ell \geq c_1n^{c_2}$ for some $c_1,c_2>0$ and $S\rightarrow \infty$, then $\frac{n^{\widehat{\gamma}}}{n^\gamma}\xrightarrow{\mathcal{P}}1$.
        \item If additionally $\gamma\in[1/2,1]$, then  $\frac{n^{\widehat{\gamma}_\mathrm{t}}}{n^\gamma}\xrightarrow{\mathcal{P}}1$, and if $\gamma\in\{1/2,1\}$, then $\frac{n^{\widehat{\gamma}_\mathrm{r}}}{n^\gamma}\xrightarrow{\mathcal{P}}1$.
    \end{enumerate}  
\end{theorem}

\subsection{Bias correction for the rate estimator}    \label{section: bias correction} 

As will be seen in the simulations in Section \ref{section: simulations}, all the estimators of $\gamma$ from Section~\ref{section: gamma estimators} suffer from a positive bias. The reason is that for $m$ close to $n$, the variance of $SD_{m}(x; \mathcal X_m)-SD_n(x; \mathcal X_n)$ is smaller than the variance of $SD_m(x; \mathcal X_m)$ (in the extreme case of $m=n$ we get zero variance). Because $m_u$ is closer to $n$ than $m_\ell$, this effect is stronger for $m_u$, and the variance seems to go to 0 faster. This leads to  $\gamma$ being overestimated in~\eqref{eq: gamma hat}. To derive a correction for this bias, note that in the non-degenerate case, we can use the Hoeffding decomposition of a $U$-statistic: We write
\begin{equation} \label{eq: correction}
SD_n(x; \mathcal X_n)=\frac{d+1}{n}\sum_{i=1}^n h_1(X_i)+U_{2,n}
\end{equation}
with $h_1(y)=\E\big[\I{x \in \Simplex{y, X_{i_2}, \dots, X_{i_{d+1}}}}\big]$. If $\left\{X_n\right\}_{n=1}^\infty$ are i.i.d., the two summands in~\eqref{eq: correction} are uncorrelated, and for the second summand $\Var(U_{2,n})=O(n^{-2})$, see e.g. \citet[Chapter~1.6]{Lee1990}. For the first summand in~\eqref{eq: correction}, with a short calculation, one obtains
    \begin{equation*}
    \Var\left(\frac{d+1}{m}\sum_{i=1}^m h_1(X_i)-\frac{d+1}{n}\sum_{i=1}^n h_1(X_i)\right)=\frac{1}{m}\frac{n-m}{n}(d+1)^2\Var(h(X_1)).
    \end{equation*}
Because  $m\rightarrow \infty$, the variances of $U_{2,n}$ and $U_{2,m}$ are thus of lower order. The variance of the difference $SD_m(x; \mathcal X_m)-SD_n(x; \mathcal X_n)$ is therefore asymptotically equivalent to the variance of $SD_{\widetilde{m}}$ with $\widetilde{m}=m\frac{n}{n-m}$. This leads to the bias-corrected estimator
    \begin{equation}    \label{eq: bc1}
    \widehat{\gamma}_{bc,1}=\frac{\log T_\ell - \log T_u}{\log (m_u)-\log(n-m_u)-\log (m_\ell)+\log(n-m_\ell)}.
    \end{equation}
Because this version of the estimator was derived under the assumption of non-degeneracy of the U-statistic $SD_n(x; \mathcal X_n)$, we refer to~\eqref{eq: bc1} also as to the \emph{bias-corrected estimator in the non-degenerate situation}.
    
A bias correction different from~\eqref{eq: bc1} can be derived under the assumption of degeneracy of $SD_n(x; \mathcal X_n)$. In this case, $\Var ( h_1(X_1))=0$, and we have to expand the statistic as
\begin{equation} \label{eq: correction 2}
SD_n(x; \mathcal X_n)=\frac{d(d+1)}{n(n-1)}\sum_{1\leq i_1<i_2\leq n} h_2(X_{i_1},X_{i_2})+U_{3,n},
\end{equation}
with $h_2(y,z)=\E\big[\I{x \in \Simplex{y, z, X_{i_3}, \dots, X_{i_{d+1}}}}\big]$. Note that under independence, the summands of the first part in~\eqref{eq: correction 2} are uncorrelated (see \citealp[Chapter~1.6]{Lee1990}) and thus, the variance of $\sum_{1\leq i_1<i_2\leq n} h_2(X_{i_1},X_{i_2})$ behaves like the variance of the sum of  $\binom{n}{2}$ independent summands. Furthermore,  $\Var(U_{3,n})=O(n^{-3})$, so the first part of the right-hand side in~\eqref{eq: correction 2} dominates the asymptotic behavior in~\eqref{eq: correction 2}. Using similar arguments as in the non-degenerate case, but replacing $m$ and $n$ with $m^2$ and $n^2$, one should use  $\widetilde{m}=m\sqrt{\frac{n^2}{n^2-m^2}}$ as a corrected subsample size, leading to the estimator
    \begin{equation}    \label{eq: bc2}
    \widehat{\gamma}_{bc,2}=\frac{\log T_\ell - \log T_u}{\log (m_u)-\log(n^2-m^2_u)/2-\log (m_\ell)+\log(n^2-m_\ell^2)/2}.
    \end{equation}
This estimator is also called the \emph{bias-corrected estimator in the degenerate situation}.

As the consistency properties of the estimators $\widehat \gamma$, $\widehat \gamma_\mathrm{t}$ and $\widehat \gamma_\mathrm{r}$ discussed in Theorem \ref{lem:gammaest} trivially transfer to their bias-corrected versions, we get the following corollary.

\begin{corollary}[Consistent bias-corrected estimation of $\gamma$]\label{cor:gammaest_bc} Under the assumptions of Theorem \ref{lem:gammaest}, we have the following:
    \begin{enumerate}[label=(\roman*)]
        \item If  $m_\ell<m_u<n$,  $m_\ell\rightarrow \infty$,  $n/m_u\rightarrow \infty$, $m_u/m_\ell \geq c_1n^{c_2}$ for some $c_1,c_2>0$ and $S\rightarrow \infty$, then $\frac{n^{\widehat{\gamma}_{bc,1}}}{n^\gamma}\xrightarrow{\mathcal{P}}1$ and $\frac{n^{\widehat{\gamma}_{bc,2}}}{n^\gamma}\xrightarrow{\mathcal{P}}1$.
        \item If additionally $\gamma\in[1/2,1]$, then $\frac{n^{\widehat{\gamma}_{\mathrm{t},bc,1}}}{n^\gamma}\xrightarrow{\mathcal{P}}1$ and $\frac{n^{\widehat{\gamma}_{\mathrm{t},bc,2}}}{n^\gamma}\xrightarrow{\mathcal{P}}1$ as well as, if $\gamma\in\{1/2,1\}$, then $\frac{n^{\widehat{\gamma}_{\mathrm{r},bc,1}}}{n^\gamma}\xrightarrow{\mathcal{P}}1$ and $\frac{n^{\widehat{\gamma}_{\mathrm{r},bc,2}}}{n^\gamma}\xrightarrow{\mathcal{P}}1$.
    \end{enumerate}  
\end{corollary}

As in practice it is not known whether the $U$-statistic $SD_n(x; \mathcal X_n)$ is degenerate or not, it is not clear which of the corrections~\eqref{eq: bc1} or~\eqref{eq: bc2} to use.  To achieve a conservative confidence interval, one should use the smaller value for an estimator of $\gamma$, because this leads to larger confidence intervals. The correction is stronger for the non-degenerate $U$-statistics in~\eqref{eq: bc1}, so $\widehat{\gamma}_{bc,1}$ leads to a higher coverage probability. We will see this phenomenon also in the simulations in Section~\ref{section: simulations}.

\subsection{Subsampling the distribution of \texorpdfstring{$SD_n$}{SD}}  \label{section:distributionsubsampling} 


To estimate the distribution of $ R_n = n^{\gamma}(SD_n(x; \mathcal X_n) - SD(x;P))$, one can use subsampling in the following way:

\begin{enumerate}
    \item Pick one of the estimators $\widehat{\gamma}$ for $\gamma$ (raw, trimmed, or rounded, with or without bias correction) described in Sections~\ref{section: gamma estimators}--\ref{section: bias correction}.
    \item Choose $m<n$, such that $m\rightarrow\infty$ and $n/m\rightarrow\infty$ as $n\rightarrow \infty$.
    \item For $b=1,\dots,B$:
    \begin{enumerate}
    \item Choose a subsample $\mathcal X_{m,b}^\star$ of size $m$ from $\mathcal X_n$ (drawing $m$  observations from the sample $\mathcal X_n$ uniformly and without replacement),
    \item calculate the sample depth $SD^\star_{m,b}(x;\mathcal X_n)=SD_m(x; \mathcal X_{m,b}^\star)$.
    \end{enumerate}
    \item Use the empirical distribution of the values     \begin{equation*}    
    R_{m,b}^\star = m^{\widehat{\gamma}}(SD_{m,b}^\star(x; \mathcal X_n) - SD_n(x; \mathcal X_n)) \qquad \mbox{for $b = 1, \dots, B$,}
    \end{equation*}
    as an estimator of the distribution of $R_n$.   
\end{enumerate}
By Theorem~\ref{lem:gammaest}, Corollary~\ref{cor:gammaest_bc}, and \citet[Theorem 2.2.1]{Politis_Romano1999}, this procedure leads to a consistent estimation of the distribution function of $ R_n = n^{\gamma}(SD_n(x; \mathcal X_n) - SD(x;P))$. 

\section{Simulation study} \label{section: simulations}

This section presents a numerical study of our estimators of the convergence rate of SD from Section~\ref{section: rate}, and the associated subsampling procedure for the distribution of the sample SD from Section~\ref{section:distributionsubsampling}. First, in Section~\ref{section: gamma simulations}, we assess the performance of the estimators of the rate parameter $\gamma$. Then, in Section~\ref{section: simulation distribution}, we use the estimated rates $\widehat{\gamma}$ and compare the resulting approximations of the asymptotic distribution of the sample simplicial depth $SD_n(x; \mathcal X_n)$ at individual points $x \in \R^2$.

\subsection{Estimating \texorpdfstring{$\gamma$}{gamma}} \label{section: gamma simulations}

In $1000$ independent runs, we generated samples of three sizes $n \in \left\{ 100, 1000, 10000 \right\}$ from $P \in \Prob[\R^2]$ the centered bivariate Gaussian distribution with covariance matrix with unit variances and $\rho = 0.8$ as the off-diagonal terms.\footnote{Since SD is invariant with respect to affine transforms of $\R^d$ \citep[formula~(1.9)]{Liu1990}, we could equally simulate from a standard normal distribution with $\rho = 0$. In our numerical study, however, the results appear to be more pronounced with a higher value of $\rho$.} The coefficient $\gamma$ was estimated in three scenarios:
    \begin{enumerate}[label=($x_{\arabic*}$), ref=($x_{\arabic*}$)]
        \item \label{x1} for $x = (0,0.5)^\mathsf{T}$ far away from the center of symmetry of $P$ (true value is $\gamma = 1/2$); 
        \item \label{x2} for $x = (0,0.2)^\mathsf{T}$ close to the center of symmetry of $P$ (true value is $\gamma = 1/2$); and
        \item \label{x3} for $x = c(0,0)^\mathsf{T}$ at the center of symmetry of $P$ (true value is $\gamma = 1$).
    \end{enumerate}
We used three setups for the choice of the subsampling rates $m_u$ and $m_\ell$:
    \begin{enumerate}[label=(\roman*)]
        \item \emph{small $m$}: $m_u = \lceil 1.25\, n^{1/2} \rceil$ and $m_\ell = \lceil 1.25\, n^{1/3} \rceil$;
        \item \emph{medium $m$}: $m_u = \lceil 1.25\, n^{3/4} \rceil$ and $m_\ell = \lceil 1.25\, n^{1/2} \rceil$; and
        \item \emph{large $m$}: $m_u = \lceil 1.25\, n^{4/5} \rceil$ and $m_\ell = \lceil 1.25\, n^{3/5} \rceil$.
    \end{enumerate}
The concrete values $m_u$ and $m_\ell$ used in each setup are in Table~\ref{table: m}.

\begin{table}[t]
    \centering
    \begin{tabular}{c|ccc}
    $\beta$ & $n = 100$ & $n = 1000$ & $n = 10000$ \\
    \hline
    1/3 & 6 & 13 & 27 \\
    1/2 & 13 & 40 & 125 \\
    3/5 & 20 & 79 & 314 \\
    3/4 & 40 & 223 & 1250 \\
    4/5 & 50 & 314 & 1982
    \end{tabular}
    \caption{Specific values of constants $m = \lceil 1.25\, n^\beta \rceil$ chosen in the subsampling procedures in the simulations in Section~\ref{section: simulations}.}
    \label{table: m}
\end{table}
    
For each sample, we use $S = 1000$ replicates to estimate $\gamma$, using three different methods:
    \begin{enumerate}[label=(\roman*)]
        \item the \emph{raw estimator} from~\eqref{eq: gamma hat}
            \[  \widehat{\gamma} = \frac{\log T_\ell - \log T_u}{\log m_u-\log m_\ell}; \]
        \item the \emph{trimmed estimator}
            \[  \widehat{\gamma}_\mathrm{t} = \max\left\{\min\left\{\frac{\log T_\ell - \log T_u}{\log m_u-\log m_\ell},1\right\},1/2\right\};  \]
        \item the \emph{rounded estimator}
            \[  \widehat{\gamma}_\mathrm{r} = 
            \begin{cases}
            1 & \mbox{if } \frac{\log T_\ell - \log T_u}{\log m_u-\log m_\ell} > 3/4, \\
            1/2 & \mbox{otherwise}.
            \end{cases}  
            \]
    \end{enumerate}
Each of these three estimators can be used in three versions:
    \begin{enumerate}[label=(\roman*)]
        \item without a bias correction, also denoted by a zero in the index, i.e., $\widehat{\gamma}_0$;
        \item with the bias correction for non-degenerate U-statistics from~\eqref{eq: bc1}, distinguished by an index one, i.e. $\widehat{\gamma}_1$; and
        \item with the bias correction for degenerate U-statistics from~\eqref{eq: bc2}, denoted by $\widehat{\gamma}_2$.
    \end{enumerate} 
In total, nine different estimators of $\gamma$ are considered. 

The results of the simulation study are presented in Figures~\ref{fig:box1}--\ref{fig:box3} and in Tables~\ref{Tab:gamma1}--\ref{Tab:gamma3} in the Supplementary Material. In the boxplots in Figures~\ref{fig:box1}--\ref{fig:box3}, the three versions of the bias-corrected estimators are determined by the color of the boxplot: 
    \begin{itemize}
        \item \textbf{\textcolor{gray}{gray color}} for estimators without bias correction $\widehat{\gamma}_0$,
        \item \textbf{\textcolor{red}{red color}} for bias-corrected estimators $\widehat{\gamma}_1$, and
        \item \textbf{\textcolor{yllw}{yellow color}} for bias-corrected estimators $\widehat{\gamma}_2$.
    \end{itemize} 
In Tables~\ref{Tab:gamma1}--\ref{Tab:gamma3} in Section~\ref{section: tables} in the Supplementary Material, we see the summary characteristics of the squared errors when estimating $\gamma$ using the three different bias-correction methods. For each combination of $x$ and $n$, the estimator with the smallest squared error corresponds to the value in bold in each table. 

We have the following conclusions to make:
    \begin{itemize}
    \item \textbf{Consistency.} All the considered estimators are consistent, meaning that with growing sample size $n$, all the estimators $\widehat{\gamma}$ appear to converge to the true values $\gamma$ ($1/2$ for the first two choices of $x$ in~\ref{x1} and~\ref{x2}, and $1$ for the last choice~\ref{x3}). This is clearly visible from the boxplots and their convergence to the thick horizontal lines representing the true value of $\gamma$. Nevertheless, as expected, the estimators are not very precise for smaller $n$.
    \item \textbf{Choice of $m$.} The subsampling with small $m$ is substantially poorer than the other two scenarios (medium and large $m$). This is seen especially with $n = 100$ (Figure~\ref{fig:box1}), where in scenario~\ref{x1}, the boxplots grossly overestimate the true value $\gamma = 1/2$, as well as in scenario~\ref{x3}, where the true $\gamma = 1$ is severely underestimated. This is the effect of choosing very small values $m_u = 13$ and $m_\ell = 6$, see Table~\ref{table: m}. The same effect is, however, seen also for larger $n$. From the remaining two setups (medium and large $m$), the medium $m$ regime seems to perform slightly better, although the differences are not very marked. 
    \item \textbf{Bias correction.} The estimators without bias correction $\widehat{\gamma}_0$ (gray boxplots) are systematically biased upwards, for all sample sizes $n$. In general, bias correction proposed in Section~\ref{section: bias correction} helps. The greater bias correction $\widehat{\gamma}_1$, made under the assumption of non-degeneracy (red boxplots), improves the estimators substantially, especially when $n$ is small, and the U-statistics are indeed non-degenerate (scenarios~\ref{x1} and~\ref{x2}). Even in the degenerate situation, however, the stronger bias correction $\widehat{\gamma}_1$ appears to improve the original estimators $\widehat{\gamma}_0$. The more conservative bias correction $\widehat{\gamma}_2$, made under the assumption of degeneracy, also helps to improve the original estimator. However, this improvement is only marginal in most cases. 
    \item \textbf{Effect of rounding and trimming.} Due to the systematic upward bias of most of the estimators, the effect of trimming and rounding of $\widehat{\gamma}$ can be profound. For the rounded estimators $\widehat{\gamma}_{\mathrm{r}}$, for example, it is enough that the raw estimator $\widehat{\gamma}$ lies below the horizontal line at $3/4$ (black horizontal lines in Figures~\ref{fig:box1}--\ref{fig:box3}) in order to estimate the asymptotic rate $\gamma$ correctly in scenarios~\ref{x1} and~\ref{x2}. In combination with an appropriate choice of $m$ and the bias correction, the trimmed/rounded estimators appear to perform quite well. 
    \end{itemize} 
As an overall conclusion, we see that the moderate value of medium $m$ in conjunction with bias corrections $\widehat{\gamma}_1$ (the more liberal one) or $\widehat{\gamma}_2$ (the more conservative one) and trimming or rounding can provide quite precise estimators of the rate $\gamma$. This is attested also in the detailed tables with results that are collected in Section~\ref{section: tables} in the Supplementary Material.

\begin{figure}[htpb]
    \centering
    \includegraphics[width=\linewidth]{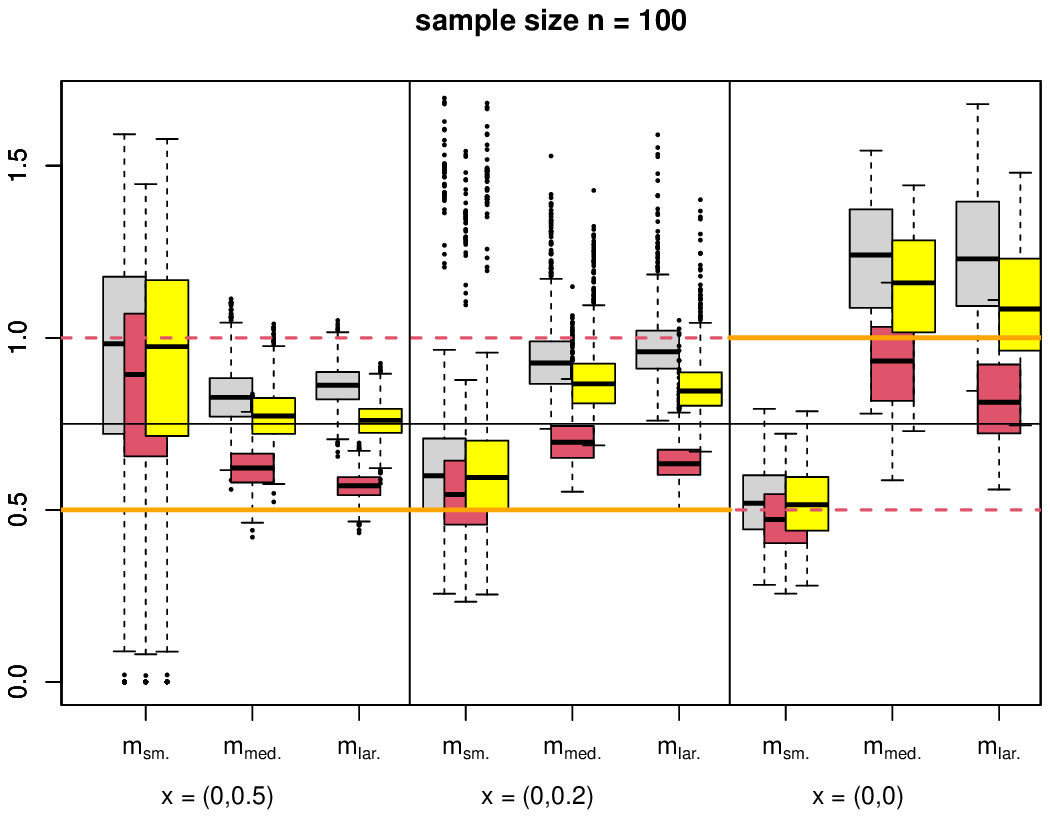}
    \caption{\textsf{Gaussian scenario:} Boxplots of the estimated raw (untrimmed, non-rounded) values of the exponent $\gamma$. The situation with sample size $n = 100$. The true value of $\gamma$ is $1/2$ for the first two scenarios, and $1$ for the third one (thick orange horizontal lines).}
    \label{fig:box1}
\end{figure}

\begin{figure}[htpb]
    \centering
    \includegraphics[width=\linewidth]{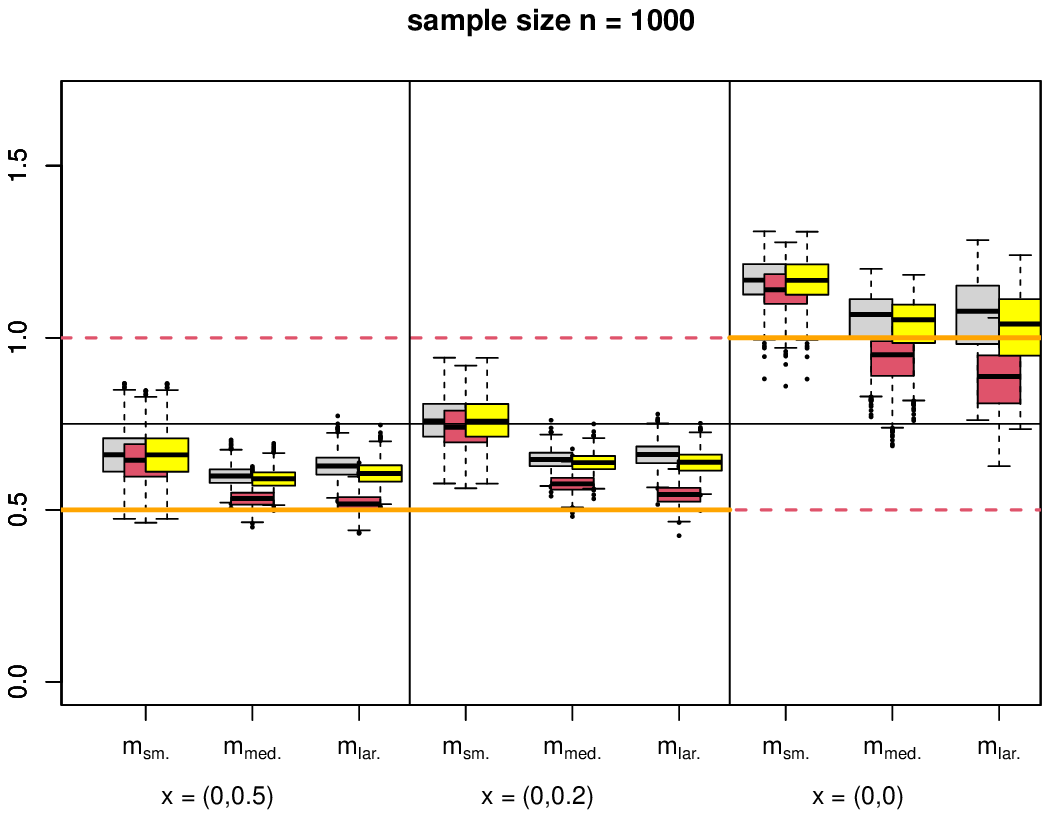}
    \caption{\textsf{Gaussian scenario:} Boxplots of the estimated raw (untrimmed, non-rounded) values of the exponent $\gamma$. The situation with sample size $n = 1000$. The true value of $\gamma$ is $1/2$ for the first two scenarios, and $1$ for the third one (thick orange horizontal lines).}
\end{figure}

\begin{figure}[htpb]
    \centering
    \includegraphics[width=\linewidth]{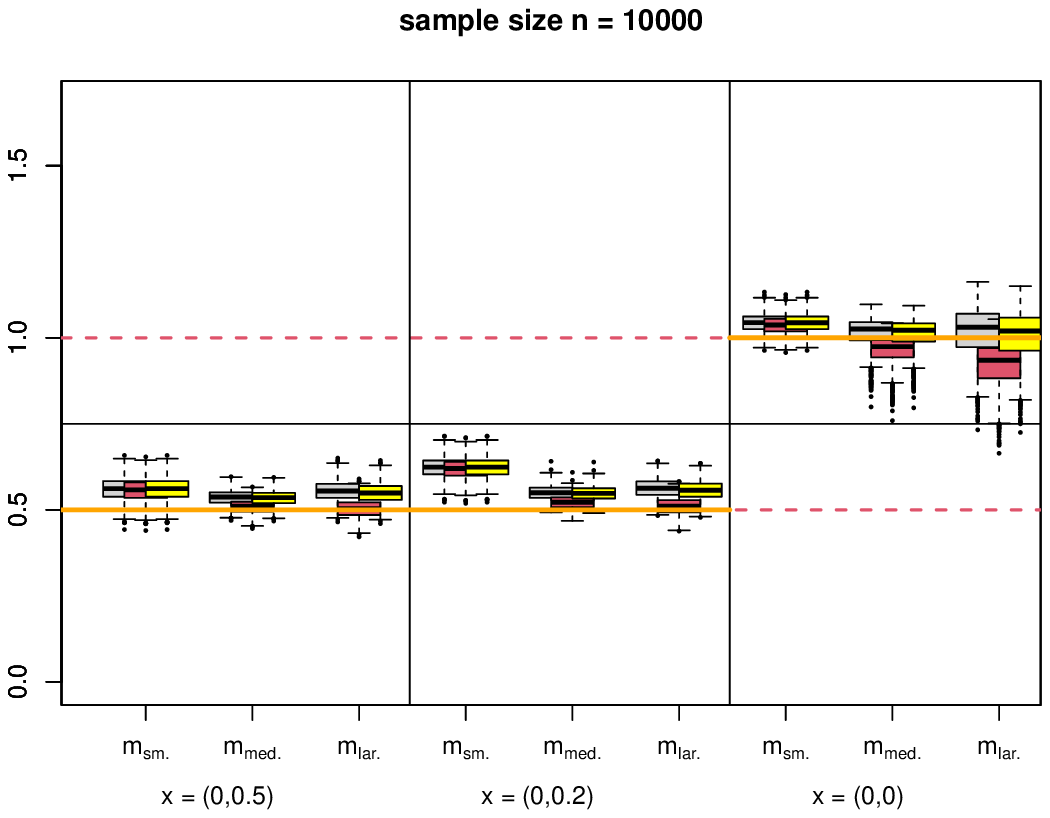}
    \caption{\textsf{Gaussian scenario:} Boxplots of the estimated raw (untrimmed, non-rounded) values of the exponent $\gamma$. The situation with sample size $n = 10000$. The true value of $\gamma$ is $1/2$ for the first two scenarios, and $1$ for the third one (thick orange horizontal lines).}
    \label{fig:box3}
\end{figure}

In addition to this situation with a Gaussian sample, we also performed an analogous simulation study with a heavy-tailed (Cauchy) distribution $P \in \Prob[\R^2]$. Its results are gathered in Section~\ref{section: Cauchy} in the Supplementary Material. The overall conclusions are, however, largely consistent with those from the \textsf{Gaussian scenario} considered here.

\subsection{Estimating the distribution of \texorpdfstring{SD}{SD}} \label{section: simulation distribution}

In the second step, we are interested in estimating the true distribution of
    \begin{equation*}  
    R_n = n^{\gamma}(SD_n(x; \mathcal X_n) - SD(x;P)),    \end{equation*}
which is estimated using our $m$-out-of-$n$ subsampling procedure from in Section~\ref{section:distributionsubsampling}, that is, by the empirical distribution of
    \begin{equation}   \label{eq: R*} 
    R_{m,b}^\star = m^{\widehat{\gamma}}(SD_{m,b}^\star(x; \mathcal X_n) - SD_n(x; \mathcal X_n)) \qquad \mbox{for $b = 1, \dots, B$.}
    \end{equation}
Here, $\widehat{\gamma}$ is one of the estimates of the rate parameter $\gamma$, and $SD_{m,b}^\star(x; \mathcal X_n)$ is the $b$-th independent (conditionally on the original sample $\mathcal X_n$) resample obtained from the data. 

We continue with the simulation study from Section~\ref{section: gamma simulations}. We use the same \textsf{Gaussian scenario}, the same settings for $x, n, m_u, m_\ell, S, B$, and the same nine estimators of $\gamma$. 

\begin{figure}[htpb]
    \centering
    \includegraphics[width=0.95\linewidth]{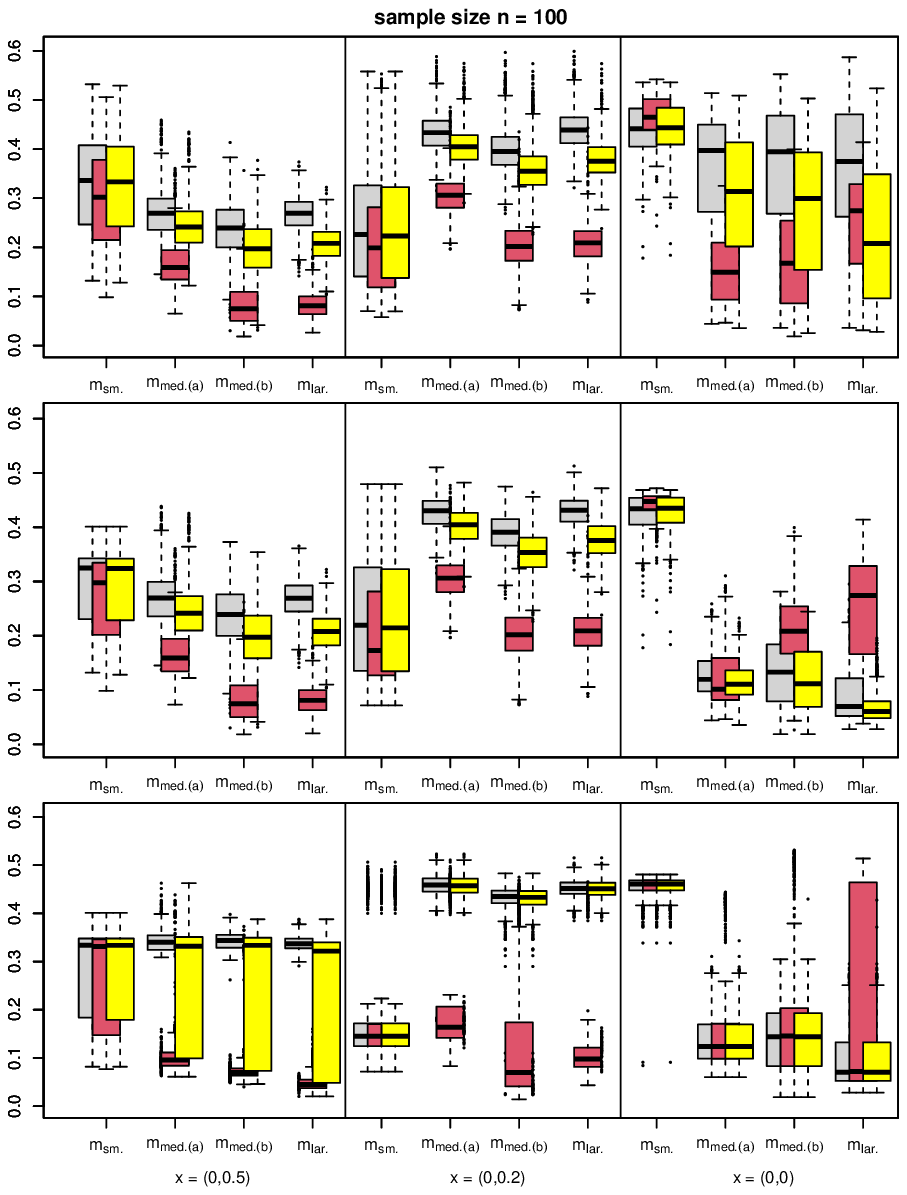}
    \caption{\textsf{Gaussian scenario,} \textbf{Kolmogorov-Smirnov dist., $n=100$:} Boxplots of the obtained Kolmogorov-Smirnov distances between the true and the resampled distribution of SD. The situation with sample size $n = 100$. The top panel corresponds to the raw estimator of $\widehat{\gamma}$, the middle panel to the trimmed estimator $\widehat{\gamma}_{\mathrm{t}}$, and the bottom panel to the rounded estimator $\widehat{\gamma}_{\mathrm{r}}$. The colors of boxplots are as in Figures~\ref{fig:box1}--\ref{fig:box3}.}
    \label{fig:KSbox1}
\end{figure}

\begin{figure}[htpb]
    \centering
    \includegraphics[width=0.95\linewidth]{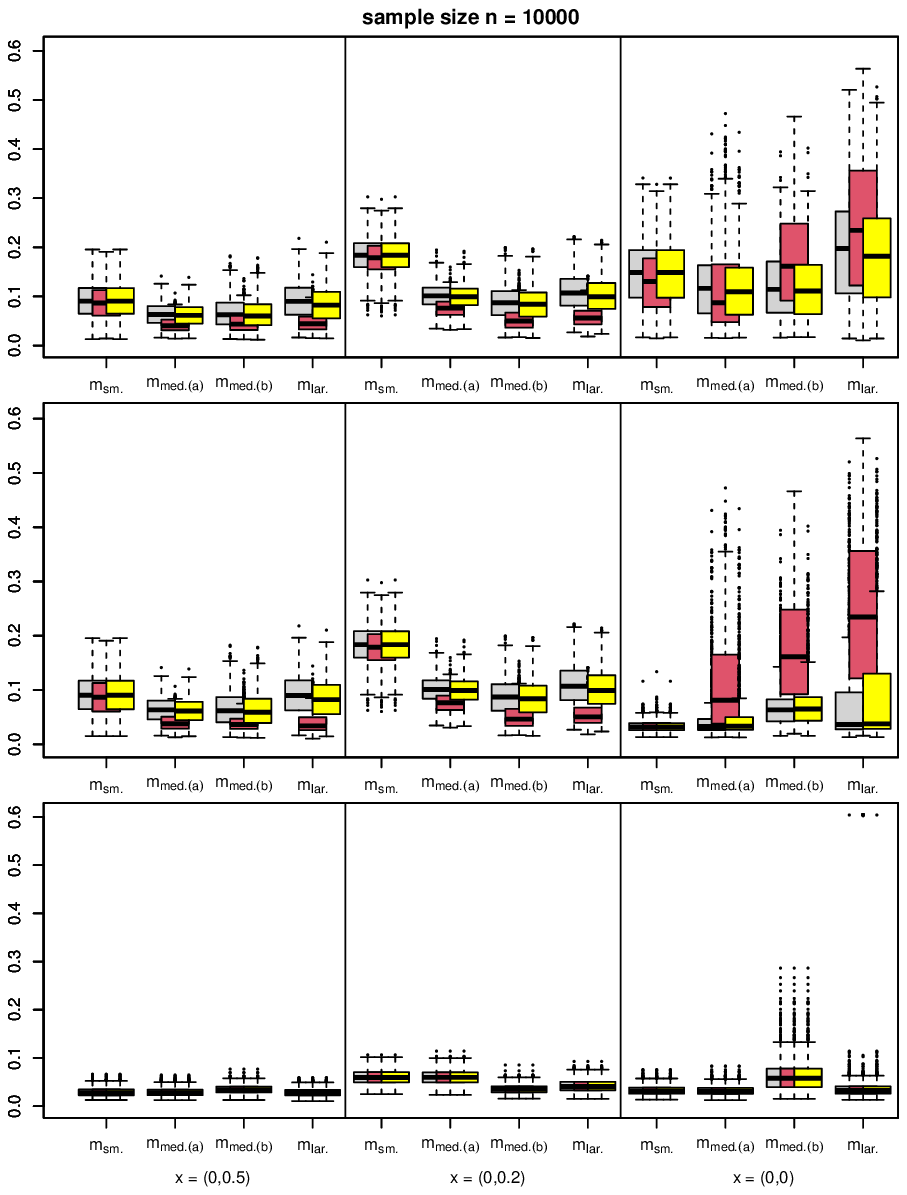}
    \caption{\textsf{Gaussian scenario,} \textbf{Kolmogorov-Smirnov dist., $n=10000$:} Boxplots of the obtained Kolmogorov-Smirnov distances between the true and the resampled distribution of SD. The situation with sample size $n = 10000$. The top panel corresponds to the raw estimator of $\widehat{\gamma}$, the middle panel to the trimmed estimator $\widehat{\gamma}_{\mathrm{t}}$, and the bottom panel to the rounded estimator $\widehat{\gamma}_{\mathrm{r}}$. The colors of boxplots are as in Figures~\ref{fig:box1}--\ref{fig:box3}.}
    \label{fig:KSbox3}
\end{figure}

The exact distribution of $R_n$ is approximated using a Monte Carlo simulation from the true sampling distribution $P$ (that is, multivariate normal), with the true value $SD(x; P)$ approximated using an average of additional $B_{\mathrm{MC}} = 10^5$ independent realizations of $SD_n(x; \mathcal X_n)$. This Monte Carlo approximate is compared to the distribution obtained by the subsampling procedure based on the estimated $\gamma$, where we consider four different choices of $m$, see also Table~\ref{table: m}:
    \begin{enumerate}
        \item \emph{small $m$}: $m = \lceil 1.25\, n^{1/2} \rceil$ used with $m_u = \lceil 1.25\, n^{1/2} \rceil$ and $m_\ell = \lceil 1.25\, n^{1/3} \rceil$;
        \item \emph{medium $m$(a)}: $m = \lceil 1.25\, n^{1/2} \rceil$ used with  $m_u = \lceil 1.25\, n^{3/4} \rceil$ and $m_\ell = \lceil 1.25\, n^{1/2} \rceil$;
        \item \emph{medium $m$(b)}: $m = \lceil 1.25\, n^{3/4} \rceil$ used with  $m_u = \lceil 1.25\, n^{3/4} \rceil$ and $m_\ell = \lceil 1.25\, n^{1/2} \rceil$; and       
        \item \emph{large $m$}: $m = \lceil 1.25\, n^{3/5} \rceil$ used with  $m_u = \lceil 1.25\, n^{4/5} \rceil$ and $m_\ell = \lceil 1.25\, n^{3/5} \rceil$.
    \end{enumerate}
The final Monte Carlo approximation and the bootstrap-estimated distributions of $R_n$ are compared using a Kolmogorov-Smirnov test statistic
    \begin{equation} \label{eq: KS} 
    KS(F_{\mathrm{MC}}, F_{\mathrm{Boot}}) = \sup_{y \in \R} \left\vert F_{\mathrm{MC}}(y) - F_{\mathrm{Boot}}(y) \right\vert, 
    \end{equation}
where 
    \begin{itemize}
        \item $F_{\mathrm{MC}}$ is the empirical approximation of the distribution function of $R_n$ with the true value of $\gamma$ (that is, $\gamma = 1/2$ in scenarios~\ref{x1} and~\ref{x2}, and $\gamma = 1$ for scenario~\ref{x3}) based on the $B_{\mathrm{MC}}$ samples from the true distributions, and 
        \item $F_{\mathrm{Boot}}$ is the empirical distribution function of $R_n$ with the estimated $\gamma$ (using one of the estimators $\widehat{\gamma}, \widehat{\gamma}_\mathrm{t}, \widehat{\gamma}_\mathrm{r}$ and with different bias-corrections), based on the $B$ subsamples $R_{m,b}^\star$ of size $m$ defined in~\eqref{eq: R*} taken from the data with sample size $n$.
    \end{itemize}
The complete results of this simulation study are in Tables~\ref{Tab:KS1}--\ref{Tab:KS3} in Section~\ref{section: tables} in the Supplementary Material. Those tables contain summaries of the resulting Kolmogorov-Smirnov distances; the same results are visualized in the boxplots in Figures~\ref{fig:KSbox1}--\ref{fig:KSbox3}.

In the interpretation of these results, we focus on the best estimators obtained in Section~\ref{section: gamma simulations} for simplicity. These were the scenarios of medium $m$, with bias-corrected and trimmed-or-rounded versions of the estimators. Among these, the overall performances are seen to be relatively similar. For non-degenerate cases~\ref{x1} and~\ref{x2}, the regime medium $m$(b) with the liberal bias-correction $\widehat{\gamma}_1$ (designed for the non-degenerate case) appears to be superior, while in the degenerate situation~\ref{x3} the regime medium $m$(b) and bias-correction $\widehat{\gamma}_2$ (designed for the degenerate case) performs better. Generally, lower squared errors are obtained using the estimators with rounding $\widehat{\gamma}_{\mathrm{r}}$ (bottom panels of Figures~\ref{fig:KSbox1}--\ref{fig:KSbox3}). Nevertheless, this result may be influenced by our choice of the performance criterion~\eqref{eq: KS}, where in the Monte Carlo approximation to the true distribution of $R_n$, the asymptotic value $\gamma \in \{1/2, 1\}$ is always considered. Necessarily, the squared errors will be smaller if also $\widehat{\gamma}$ is bound to lie in the same set $\{1/2,1\}$.

\begin{figure}[htpb]
    \centering
    \includegraphics[width=0.95\linewidth]{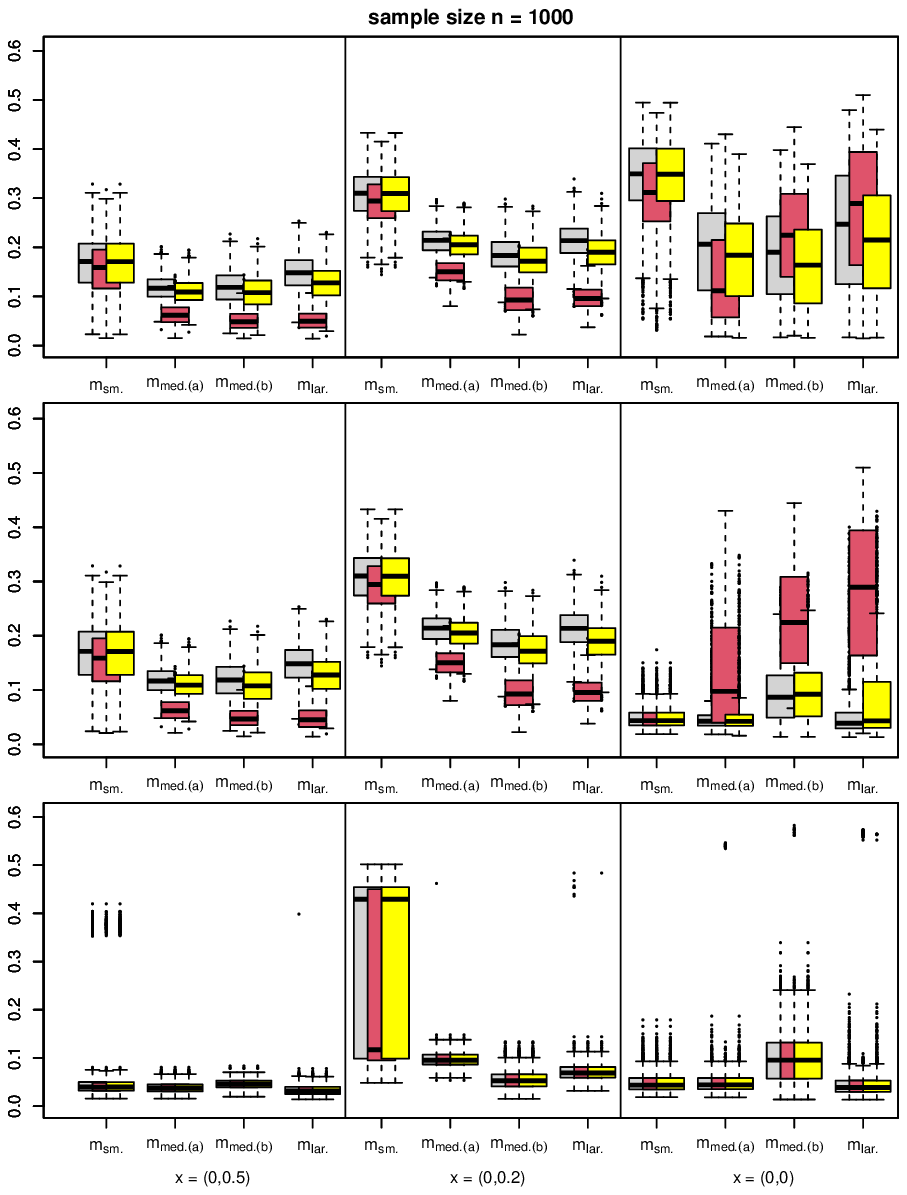}
    \caption{\textsf{Gaussian scenario,} \textbf{Kolmogorov-Smirnov dist., $n=1000$:} Boxplots of the obtained Kolmogorov-Smirnov distances between the true and the resampled distribution of SD. The situation with sample size $n = 1000$. The top panel corresponds to the raw estimator of $\widehat{\gamma}$, the middle panel to the trimmed estimator $\widehat{\gamma}_{\mathrm{t}}$, and the bottom panel to the rounded estimator $\widehat{\gamma}_{\mathrm{r}}$. The colors of boxplots are as in Figures~\ref{fig:box1}--\ref{fig:box3}.}
\end{figure}

Just as in Section~\ref{section: gamma simulations}, also in the present setup we run the whole simulation also with a heavy-tailed (Cauchy) base distribution $P \in \Prob[\R^2]$. The overall conclusions are the same as for the \textsf{Gaussian scenario}. The complete results are in Section~\ref{section: Cauchy} in the Supplementary Material.


%
%
%
%
%

\section{Applications to data analysis} \label{section: applications}

We now use the subsampling technique from Section~\ref{section: rate} in two applications: (i) the construction of confidence intervals for SD of a point in Section~\ref{section: application 1}, and (ii) for assessing the credibility of a depth-based classification method in Section~\ref{section: application 2}.

\subsection{Confidence intervals for SD}  \label{section: application 1}

For $x \in \R^d$ and $P \in \Prob$ given, our first task is to estimate $SD(x; P)$. We have a random sample $\mathcal X_n = \left\{ X_1, \dots, X_n \right\}$ from $P$. A natural point estimator of $SD(x; P)$ is the sample simplicial depth $SD_n(x; \mathcal X_n)$ from~\eqref{eq: SDn}. The subsampling principle from Section~\ref{section: rate} is used to find a confidence interval for the (unknown) true value $SD(x; P)$. First, we estimate the rate $\gamma$ in the asymptotic expression $R_n$ from~\eqref{eq: Rn}.
By Theorem~\ref{thm: Thm1}, we know that in the plane, $\gamma \in \{1/2, 1\}$, and \citet[Section~5.5]{Serfling1980} gives that the asymptotic distribution of $R_n$ must be either Gaussian (if $\gamma = 1/2$, that is if the associated U-statistic is non-degenerate), or a weighted sum of centered $\chi^2$-distributions (if $\gamma = 1$, that is if the U-statistic is degenerate). The asymptotic variance (in the non-degenerate case) or the whole shape (in the degenerate case) of the limiting distribution is, however, of a rather complicated form, and is difficult to estimate directly. In addition, for finite sample sizes, it might be sensible to consider also other estimated values $\widehat{\gamma} > 0$ (or $\widehat{\gamma} \in [1/2,1]$) to account for the finite sample behavior of $R_n$. This is because the leading terms in the asymptotic expansion of $R_n$ are either of order $\sqrt{n}$ or $n$, respectively, but other features that are asymptotically vanishing might be beneficial to capture as well by using an estimate $\widehat{\gamma} \in [1/2,1]$ for subsampling to potentially improve its finite sample performance. We will investigate this by simulations.

Estimating the exponent $\gamma$ in \eqref{eq: Rn} by $\widehat{\gamma}$ and taking $\widehat{\gamma}$ fixed, we derive a confidence interval for $SD(x; P)$ at level $\alpha \in (0,1)$ of the form
    \begin{equation}  \label{eq: confidence interval}
    \left[ SD_n(x; \mathcal X_n) - \frac{q(1-\alpha/2)}{n^{\widehat{\gamma}}}, SD_n(x; \mathcal X_n) - \frac{q(\alpha/2)}{n^{\widehat{\gamma}}} \right],  
    \end{equation}
where $q(\alpha)$ is the $\alpha$-quantile of the (asymptotic, or finite sample) distribution of $R_n$ from~\eqref{eq: Rn}. With the estimator $\widehat{\gamma}$ fixed, the latter quantiles are estimated using a $m$-out-of-$n$ subsampling procedure as in Section~\ref{section: simulation distribution}, that is, using an empirical $\alpha$-quantile of the resamples $R_{m,b}^\star$ from~\eqref{eq: R*}. Plugging in $R_{m,b}^\star$ into~\eqref{eq: confidence interval} and denoting by $q_{m,B}^\star(\alpha)$ the empirical $\alpha$-quantile of the set of resamples $SD_{m,b}^\star(x; \mathcal X_n)$, $b = 1, \dots, B$, we can rewrite the left-hand side of~\eqref{eq: confidence interval} into
    \[
    SD_n(x; \mathcal X_n) - \frac{q(1-\alpha/2)}{n^{\widehat{\gamma}}} = SD_n(x; \mathcal X_n) - \left(\frac{m}{n}\right)^{\widehat{\gamma}} \left( q_{m,B}^\star(1-\alpha/2) - SD_n(x; \mathcal X_n) \right),
    \]
and analogously for the right-hand side of~\eqref{eq: confidence interval}. The crucial role of the rate estimate $\widehat{\gamma}$ is obvious --- larger values $\widehat{\gamma}$ result in narrower confidence intervals.

In the simulation study, when evaluating the performance of our confidence intervals~\eqref{eq: confidence interval}, we consider $\alpha = 0.05$. For the \textsf{Gaussian scenario} and the setup with medium $m$ from Section~\ref{section: gamma simulations}, the empirical coverage of the confidence intervals is reported in the first part of Tables~\ref{Tab:1.0covlen_m2bc_2} (bias correction $\widehat{\gamma}_1$) and~\ref{Tab:1.0covlen_m2bc_3} (bias correction $\widehat{\gamma}_2$). The lengths of the same intervals are reported in the second part of Tables~\ref{Tab:1.0covlen_m2bc_2} (bias correction $\widehat{\gamma}_1$) and~\ref{Tab:1.0covlen_m2bc_3} (bias correction $\widehat{\gamma}_2$). 

From Tables~\ref{Tab:1.0covlen_m2bc_2} and~\ref{Tab:1.0covlen_m2bc_3}, we see that the overall best results are obtained by the rounded estimators $\widehat{\gamma}_{\mathrm{r}}$, whose results surpass the other two estimators by a wide margin. As for the bias correction, the liberal bias correction $\widehat{\gamma}_1$, designed under the assumption of non-degeneracy, gives better results than the conservative $\widehat{\gamma}_2$ in terms of coverage. This is somewhat compensated by the larger confidence intervals when using $\widehat{\gamma}_1$, as was to be expected due to the consistently smaller value of the exponent  $\widehat{\gamma}_1$ when compared to $\widehat{\gamma}_2$. In the non-degenerate situation (scenarios~\ref{x1} and~\ref{x2}), the confidence intervals tend to have a coverage smaller than the nominal value $0.95$; this effect, however, vanishes as the sample size grows. For the degenerate scenario~\ref{x3}, the confidence intervals are interestingly shorter and with higher coverage than $0.95$. Overall, we see that confidence intervals are consistent and well-behaved, especially when $\widehat{\gamma}_{\mathrm{r}}$ is used in conjunction with the bias correction $\widehat{\gamma}_1$. Their performance is, however, not very good with low sample sizes, as was to be expected because of the use of a subsampling procedure.

Just as in the previous simulation studies, we replicated the same numerical experiment also for the \textsf{Cauchy scenario} with a bivariate heavy-tailed distribution $P \in \Prob[\R^2]$. The complete results are in Section~\ref{section: Cauchy} in the Supplementary Material. The conclusions and recommendations remain the same as for the \textsf{Gaussian scenario}.

\input{1.0covlen_m2bc_2}

\input{1.0covlen_m2bc_3}

\subsection{Assessing confidence in classification} \label{section: application 2}

There are two distributions $P \ne Q \in \Prob$, and we are given independent random samples $\mathcal X_{n_1} = \left\{ X_1, \dots, X_{n_1} \right\}$ from $P$, and $\mathcal Y_{n_2} = \left\{ Y_1, \dots, Y_{n_2} \right\}$ from $Q$. For simplicity, assume that $n_1 = n_2 = n$. 
We are engaged in a classification task, that is, for a given new point $x \in \R^d$, the problem is to decide from which of the distributions $P, Q$ the new observation $x$ came, and to quantify the certainty that $x$ comes from that distribution. 

A simple yet well-performing classifier in this setting is the maximum depth classifier \citep{Ghosh_Chaudhuri2005}, which assigns $x$ to $P$ if and only if $SD_n(x; \mathcal X_{n}) > SD_n(x; \mathcal Y_n)$.\footnote{The use of the simple maximum depth classifier is known to be suboptimal in many scenarios, and more sophisticated depth-based classification techniques such as the classifiers based on $DD$-plots \citep{Li_etal2012} or their modifications \citep{Mozharovskyi_etal2015, Cuesta_etal2017, Hubert_etal2017} perform better. We use the maximum depth classifier for simplicity; our ideas are straightforward to extend to other depth-based classifiers.} The idea is to use subsampling to assess the degree of certainty in correctly classifying $x$ into a group.

For both $\mathcal X_n$ and $\mathcal Y_n$, we estimate the distributions of $R_n$ from~\eqref{eq: Rn} using $R_{m,b}^\star$ from~\eqref{eq: R*}. Write $\gamma_X$ for the exponent $\gamma$ in~\eqref{eq: Rn}, and let $\gamma_Y$ be the exponent in~\eqref{eq: Rn} when $SD_n(x; \mathcal X_n)$ is replaced by $SD_n(x; \mathcal Y_n)$ and $SD(x; P)$ is replaced by $SD(x; Q)$. First, we estimate $\gamma_X$ (or $\gamma_Y$) as in Section~\ref{section: gamma simulations} by $\widehat{\gamma}_X$ (or $\widehat{\gamma}_Y$). Then, taking $\widehat{\gamma}_X$ and $\widehat{\gamma}_Y$ fixed (as we did in Section~\ref{section: simulation distribution}), we use our subsampling procedure to obtain a large number $B = 10^4$ of bootstrap realizations
    \begin{equation} \label{eq: R star}  
    R_{m,b}^\star = m^{\gamma}(SD_{m,b}^\star(x) - SD_n(x)), \quad \mbox{for }b=1, \dots, B, 
    \end{equation}
of $R_n$ given by
    \begin{equation} \label{eq: R2}  R_n = n^{\gamma}(SD_n(x) - SD(x)).    \end{equation}
The quantities~\eqref{eq: R star} and~\eqref{eq: R2} with $\gamma = \widehat{\gamma}_X$, $SD_n(x) = SD_n(x; \mathcal X_n)$, $SD(x) = SD(x; P)$ and $SD_{m,b}^\star(x) = SD_{m,b}^\star(x; \mathcal X_n)$ are denoted by $R_{m,b,X}^\star$, $b = 1, \dots, B$ and $R_{n,X}$, respectively. In an analogous way, we obtain $B$ bootstrap realizations~\eqref{eq: R star} of $R_n$ from~\eqref{eq: R2} with $\gamma = \widehat{\gamma}_Y$, $SD_n(x) = SD_n(x; \mathcal Y_n)$, $SD(x) = SD(x; Q)$ and $SD_{m,b}^\star(x) = SD_{m,b}^\star(x; \mathcal Y_n)$, and these are denoted by $R_{m,b,Y}^\star$, $b = 1, \dots, B$ and $R_{n,Y}$, respectively. 

To assess the degree of uncertainty of whether $SD_n(x; \mathcal X_n) > SD_n(x; \mathcal Y_n)$ does imply $SD(x; P) > SD(x; Q)$, we propose to use approximation
    \begin{equation} \label{eq: estimate X}
    \begin{aligned}
    SD(x; P) & = SD_n(x; \mathcal X_n) - \frac{R_{n,X}}{n^{\widehat{\gamma}_X}}  \approx SD_n(x; \mathcal X_n) - \frac{R_{m,b,X}^\star}{n^{\widehat{\gamma}_X}} \\
    & = SD_n(x; \mathcal X_n) - \frac{m^{\widehat{\gamma}_X}(SD_{m,b}^\star(x; \mathcal X_n) - SD_n(x; \mathcal X_n))}{n^{\widehat{\gamma}_X}} \\
    & = SD_n(x; \mathcal X_n)(1+\lambda_X) - \lambda_X \, SD_{m,b}^\star(x; \mathcal X_n),
    \end{aligned}
    \end{equation}
where $\lambda_X = (m/n)^{\widehat{\gamma}_X}$. Analogously, for $SD(x; Q)$ we obtain
    \begin{equation} \label{eq: estimate Y}
    SD(x; Q) \approx SD_n(x; \mathcal Y_n)(1+\lambda_Y) - \lambda_Y \, SD_{m,b}^\star(x; \mathcal Y_n)
    \end{equation}
with $\lambda_Y = (m/n)^{\widehat{\gamma}_Y}$. Denote the right-hand side approximations in~\eqref{eq: estimate X} and~\eqref{eq: estimate Y} by $\widehat{SD}^\star_{m,b}(x; \mathcal X_n)$ and $\widehat{SD}^\star_{m,b}(x; \mathcal Y_n)$, respectively.
Our intention is to estimate the probability 
    \[  p_X(x) = \PP\left( SD_n(x; \mathcal X_n) > SD_n(x; \mathcal Y_n) \right)  \]
of assigning $x$ into the group $\mathcal X_n$. 
Due to the discrete and finite set of values that $SD_n(x; \mathcal X_n)$ can take, it may happen that also $SD_n(x; \mathcal X_n) = SD_n(x; \mathcal Y_n)$ with non-zero probability. Thus, in what follows we modify $p_X$ slightly, and symmetrize $p_X(x)$ and its complementary probability $p_Y(x) = \PP\left( SD_n(x; \mathcal X_n) < SD_n(x; \mathcal Y_n) \right)$ by considering
    \[  p(x) = p_X(x) + \frac{1}{2} \, \PP\left( SD_n(x; \mathcal X_n) = SD_n(x; \mathcal Y_n) \right). \]
This probability can be estimated in several ways. First, using the approximations~\eqref{eq: estimate X} and~\eqref{eq: estimate Y}, the probability $p(x)$ can be estimated by the ``Hall-type bootstrap estimator''
that is associated with the two-sample U-statistic
    \begin{equation}    \label{eq: p4}
    \begin{aligned}
    \widehat{p}_{H}(x) & = \frac{1}{B^{2}} \sum_{b =1}^B \sum_{b' =1}^B \I{\widehat{SD}^\star_{m,b}(x; \mathcal X_n) >  \widehat{SD}^\star_{m,b'}(x; \mathcal Y_n)} \\
    & \phantom{=} + \frac{1}{2 B^2} \sum_{b =1}^B \sum_{b' =1}^B \I{\widehat{SD}^\star_{m,b}(x; \mathcal X_n) =  \widehat{SD}^\star_{m,b'}(x; \mathcal Y_n)}.
    \end{aligned}
    \end{equation}
A particular estimator of this type is the one with the ``naive'', trivial estimators $\widehat{\gamma}_X = \widehat{\gamma}_Y = 0$. In that case, $\lambda_X = \lambda_Y = 1$, and we obtain the ``naive Hall-type bootstrap estimator'' in the form
    \begin{equation}    \label{eq: p4 naive}
    \begin{aligned}
    & \widehat{p}_{H,\mathrm{naive}}(x) \\
    & = \frac{1}{B^{2}} \sum_{b =1}^B \sum_{b' =1}^B \I{2 SD_n(x; \mathcal X_n) - SD_{m,b}^\star(x; \mathcal X_n) > 2 SD_n(x; \mathcal Y_n) - SD_{m,b'}^\star(x; \mathcal Y_n)} \\
    & \phantom{=} + \frac{1}{2 B^2} \sum_{b =1}^B \sum_{b' =1}^B \I{2 SD_n(x; \mathcal X_n) - SD_{m,b}^\star(x; \mathcal X_n) = 2 SD_n(x; \mathcal Y_n) - SD_{m,b'}^\star(x; \mathcal Y_n)}.
    \end{aligned}
    \end{equation}
The estimator~\eqref{eq: p4 naive} is not particularly useful or well-performing in practice, but it will be useful to gain insights into the performance of the original Hall-type quantity $p_H$ from~\eqref{eq: p4}. 

Another natural estimator of the probability $p(x)$ is the ``percentile-like'' estimator given by
the two-sample U-statistic
    \begin{equation}    \label{eq: p3}
    \begin{aligned}
    \widehat{p}_{P}(x) & = \frac{1}{B^{2}} \sum_{b=1}^B \sum_{b'=1}^B \I{SD_{m,b}^\star(x; \mathcal X_n) > SD_{m, b'}^\star(x; \mathcal Y_n)} \\
    & \phantom{=} + \frac{1}{2 B^2} \sum_{b=1}^B \sum_{b'=1}^B \I{SD_{m,b}^\star(x; \mathcal X_n) = SD_{m,b'}^\star(x; \mathcal Y_n)}, 
    \end{aligned}
    \end{equation}
which is obtained directly from the subsampled values of the depths.
    
An interesting feature of the simpler estimators~\eqref{eq: p4 naive} and~\eqref{eq: p3} is that the convergence rates $\gamma_X$ and $\gamma_Y$ do not need to be estimated. The estimator~\eqref{eq: p3} is, in fact, a U-statistic that corresponds to the test statistic of the Mann-Whitney two-sample U-test of the hypothesis that $SD(x; P) > SD(x; Q)$. 

To gain insights into the difference between the Hall-type estimators~\eqref{eq: p4} and~\eqref{eq: p4 naive}, and the percentile-like estimator~\eqref{eq: p3}, we can in $\widehat{p}_{H,\mathrm{naive}}(x)$  rewrite
    \begin{equation} \label{eq: symmetrizing}
    2 SD_n(x; \mathcal X_n) - SD^\star_{m,b}(x; \mathcal X_n) = SD_n(x; \mathcal X_n) - (SD^\star_{m,b}(x; \mathcal X_n) - SD_n(x; \mathcal X_n)),
    \end{equation}
which can be interpreted as ``flipping" the estimated bootstrap distribution given by $SD_{m,b}^\star(x; \mathcal X_n)$ around the point estimator $SD_n(x; \mathcal X_n)$ of the simplicial depth, and analogously for the sample $\mathcal Y_n$. In the same way, also $\widehat{SD}_{m,b}^\star(x; \mathcal X_n)$ encountered in $\widehat{p}_H(x)$ can be expressed as 
    \begin{equation} \label{eq: symmetrizing 2}
    \widehat{SD}_{m,b}^\star(x; \mathcal X_n) = SD_n(x; \mathcal X_n) - \lambda_X(SD_{m,b}^\star(x; \mathcal X_n) - SD_n(x; \mathcal X_n)),
    \end{equation}
which is again a similar ``flipping'' of the bootstrap distribution around a central point in $SD_n(x; \mathcal X_n)$, this time coupled with a scaling of the transformed distribution by a factor of $\lambda_X$.


\begin{figure}[htpb]
    \centering   
    \includegraphics[width=0.475\linewidth]{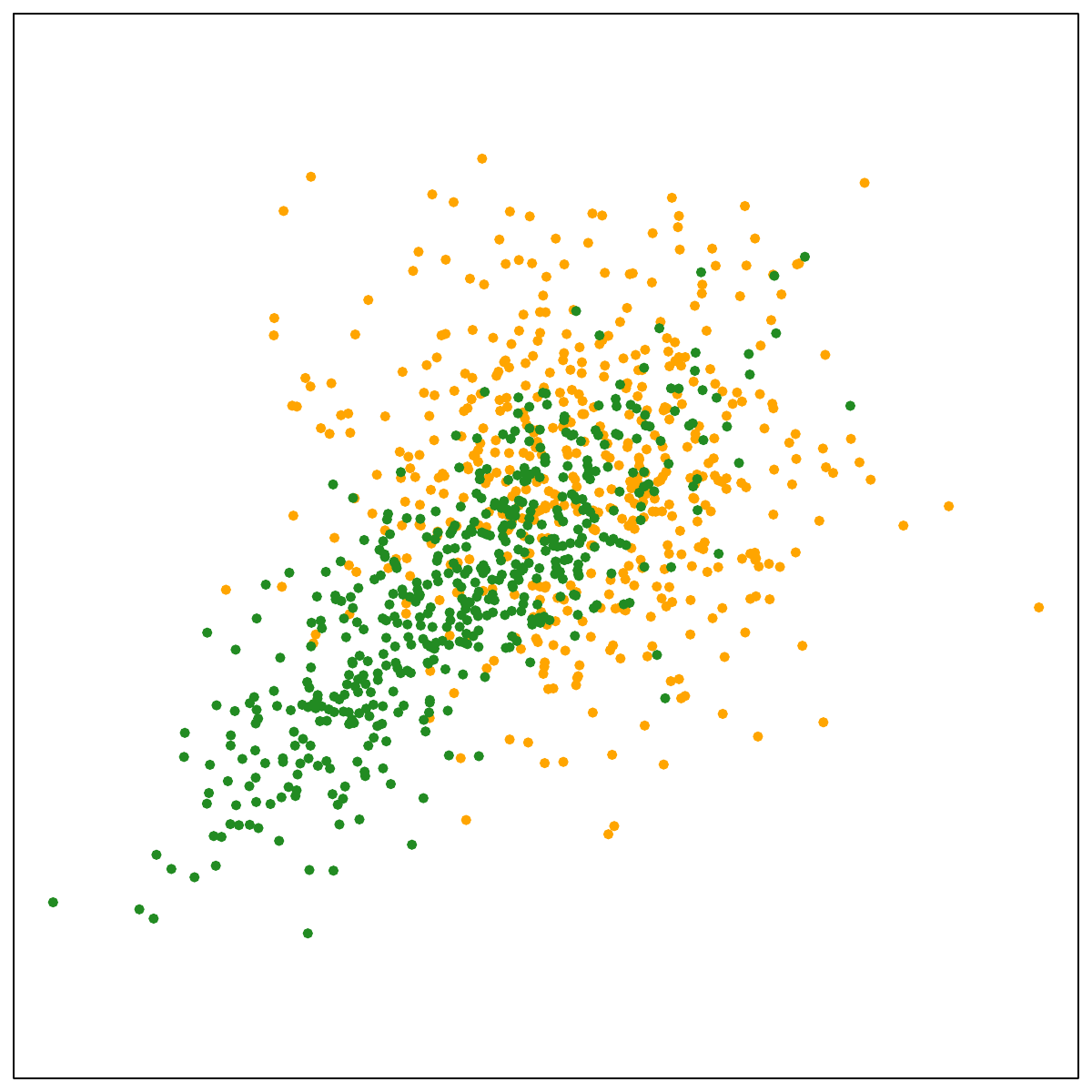} 
    \includegraphics[width=0.475\linewidth]{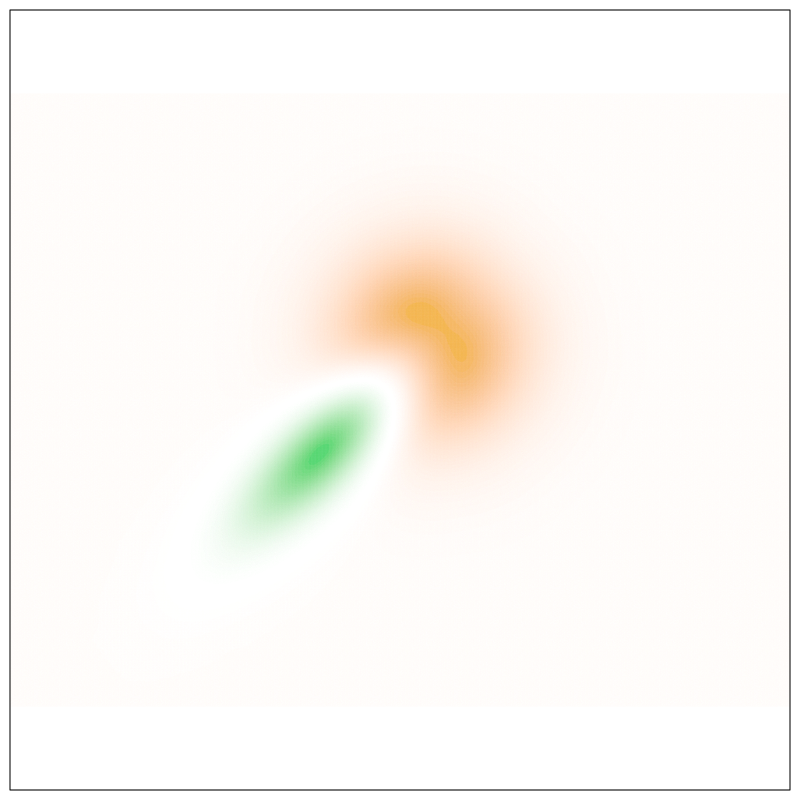} 
    \includegraphics[width=0.475\linewidth]{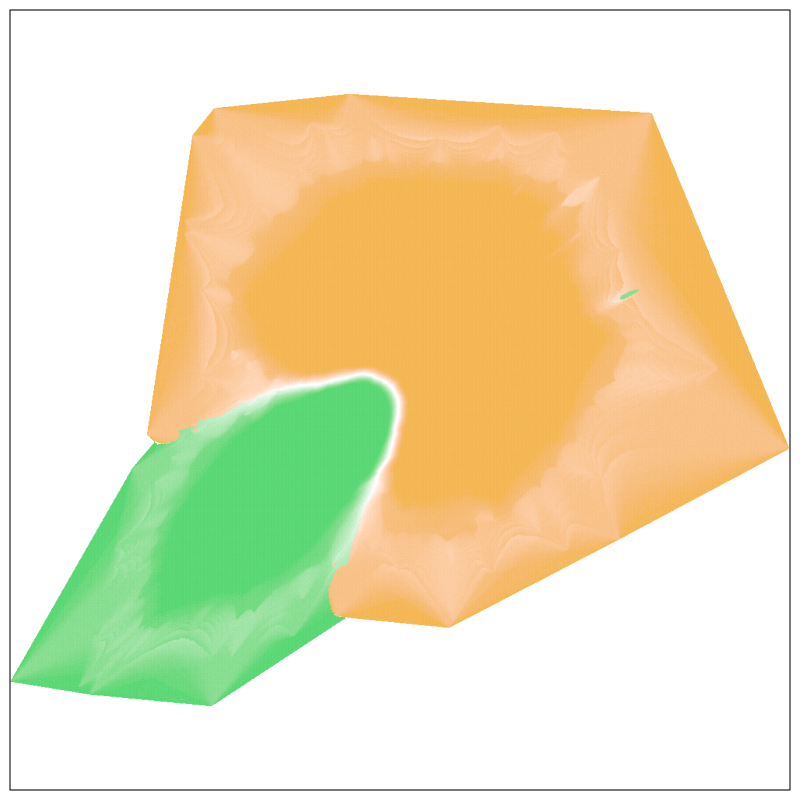} 
    \includegraphics[width=0.475\linewidth]{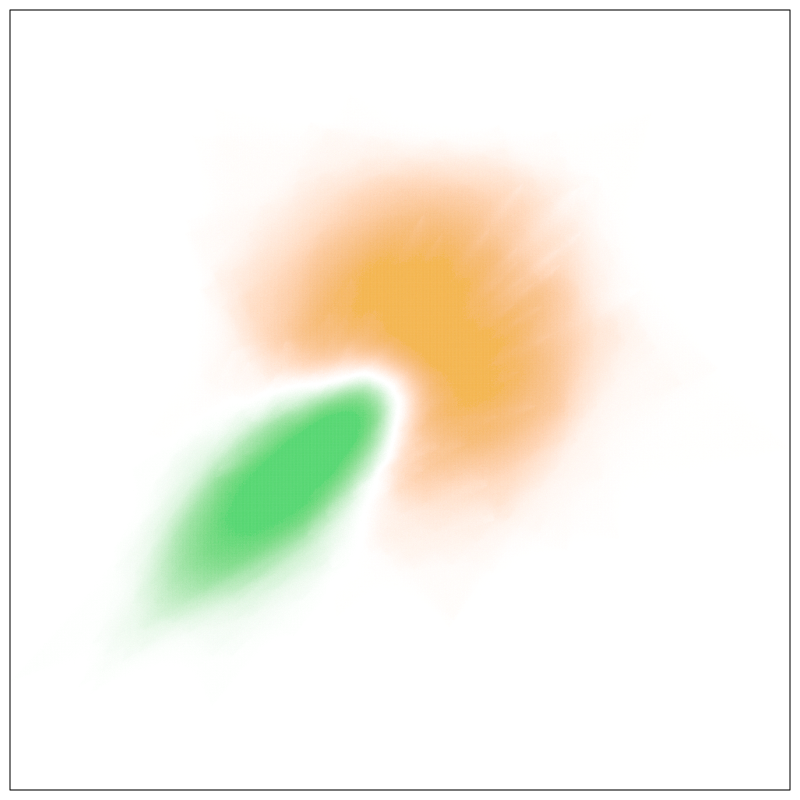} 
    \caption{A simulated dataset. Top left: two samples from Gaussian distributions $P$ (orange) and $Q$ (green), respectively, each of sample size $n_1 = n_2 = 500$. Top right: The difference of the true simplicial depths $SD(x; P) - SD(x; Q)$, approximated using random samples of sizes $10^5$ from both $P$ and $Q$. Bottom panels: Estimated contours of $\widehat{p}_{H}$ (left) and $\widehat{p}_{P}$ (right). The colors of the contours used are orange for $\widehat{p}>1/2$ and green for $\widehat{p}<1/2$. Darker tones are used for values closer to $0$ or $1$; white color corresponds to $\widehat{p} \approx 1/2$.}
    \label{fig: classification normal}
\end{figure}

\begin{figure}[htpb]
    \centering
    \includegraphics[width=0.475\linewidth]{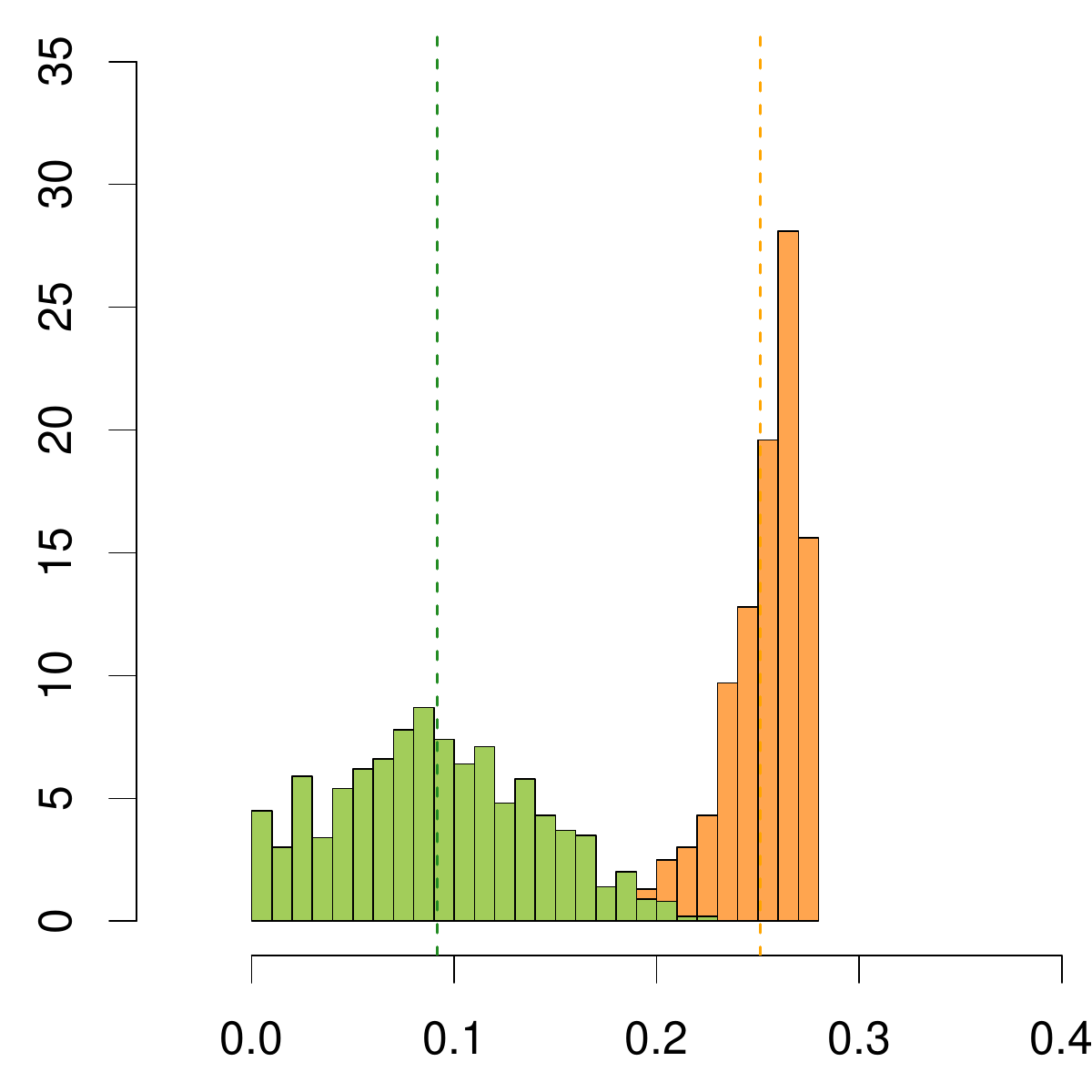}
    \includegraphics[width=0.475\linewidth]{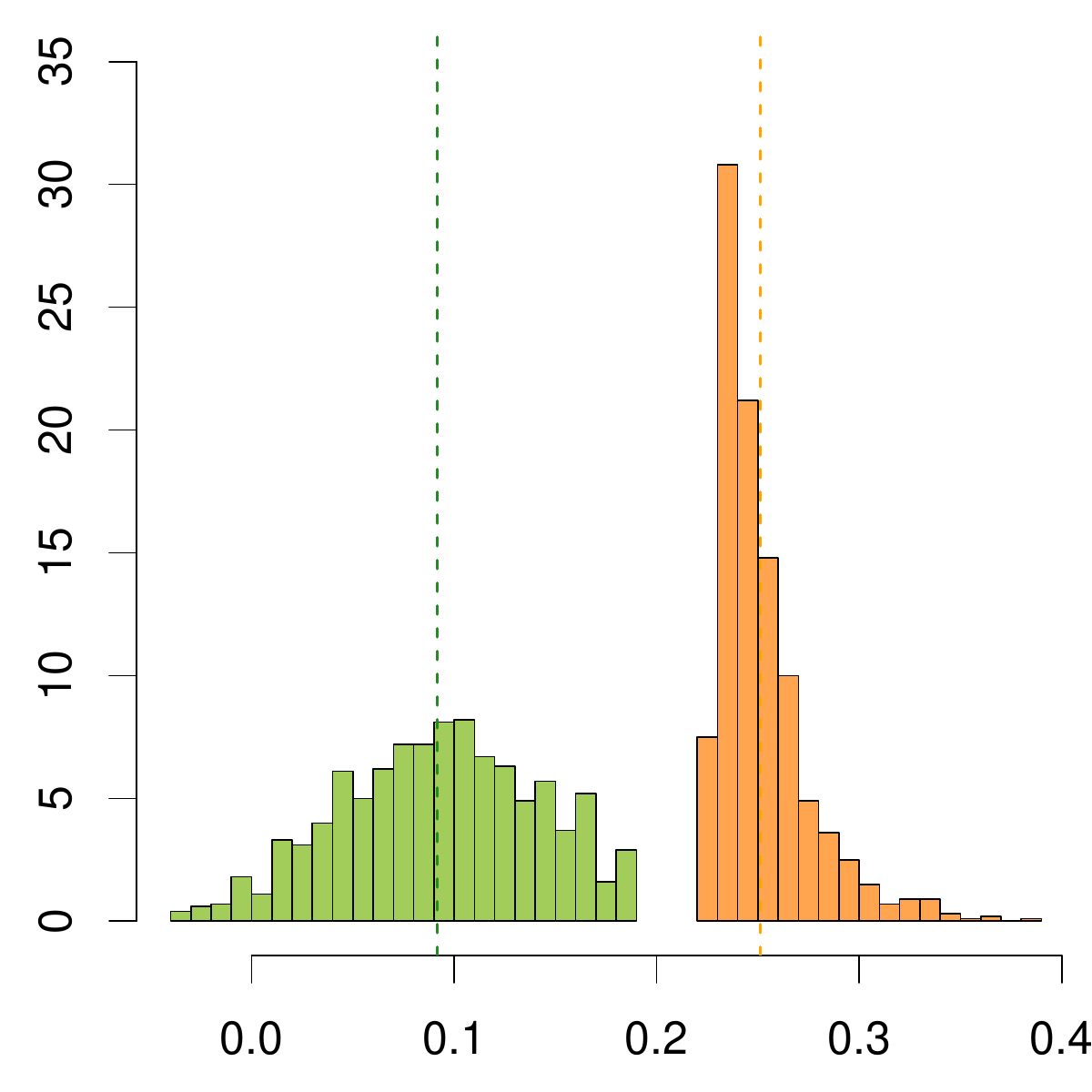}
    \caption{Two simulated datasets $\mathcal X_n$ and $\mathcal Y_n$ from Gaussian distributions, each of sample size $n = n_1 = n_2 = 500$, and a point $x = (1,1)^\mathsf{T}$ at the center of $\mathcal X_n$. Left-hand panel: Histograms of the resampled values $SD^\star_{m,b}(x; \mathcal X_n)$ (orange) and $SD^\star_{m,b'}(x; \mathcal Y_n)$ (green), with dashed vertical lines standing for the point estimators $SD_n(x; \mathcal X_n)$ (orange) and $SD_n(x; \mathcal Y_n)$ (green). The two distributions on the left-hand panel are not perfectly separated, meaning that $\widehat{p}_P(x) < 1$. Right-hand panel: Histograms of the two symmetrized distributions~\eqref{eq: symmetrizing}. Here, the separation of the two distributions is obvious, and hence $\widehat{p}_H(x) = \widehat{p}_{H,\mathrm{naive}}(x) = 1$.}
    \label{figure: normal symmetrize}
\end{figure}

\begin{figure}[htpb]
    \centering   
    \includegraphics[width=0.475\linewidth]{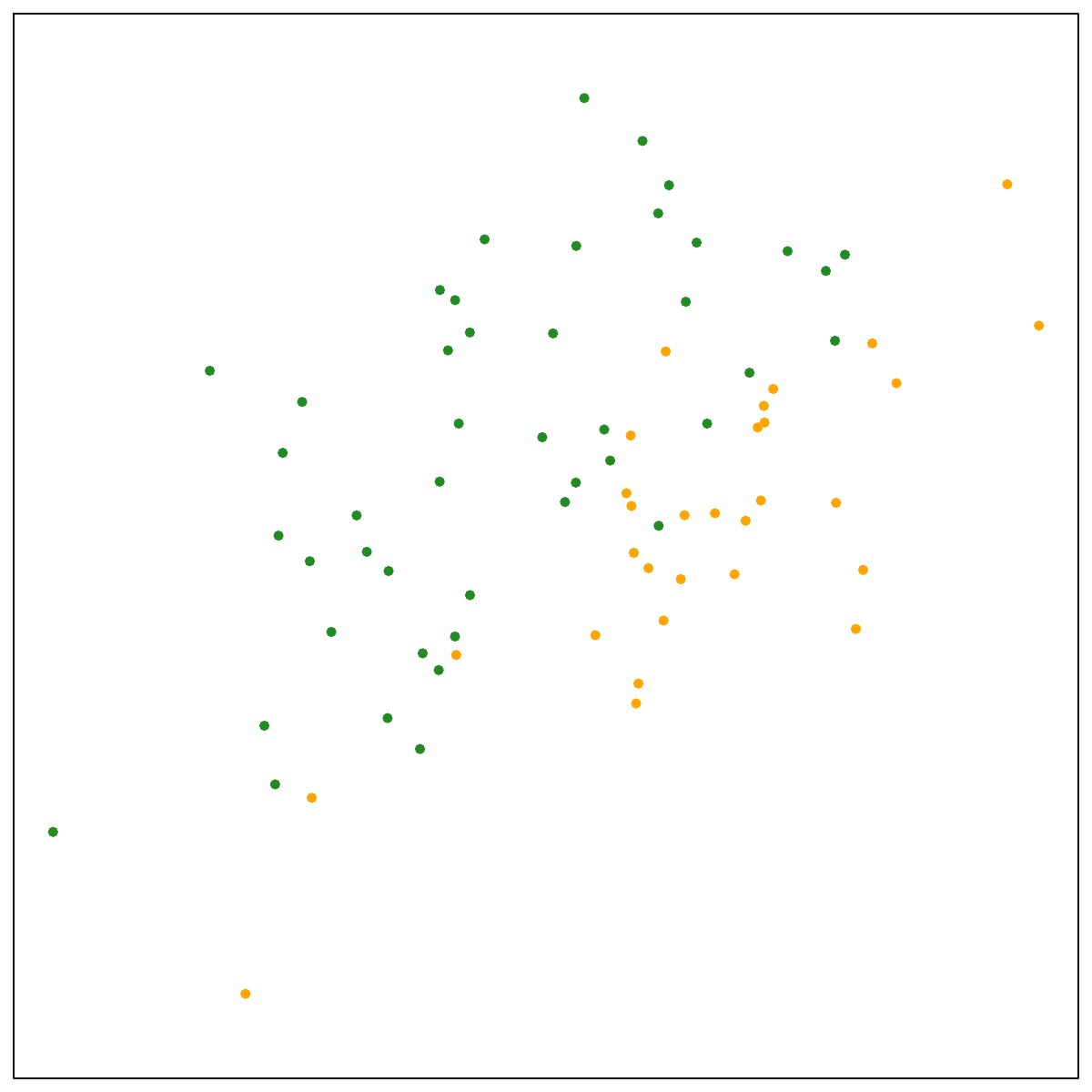} 
    \includegraphics[width=0.475\linewidth]{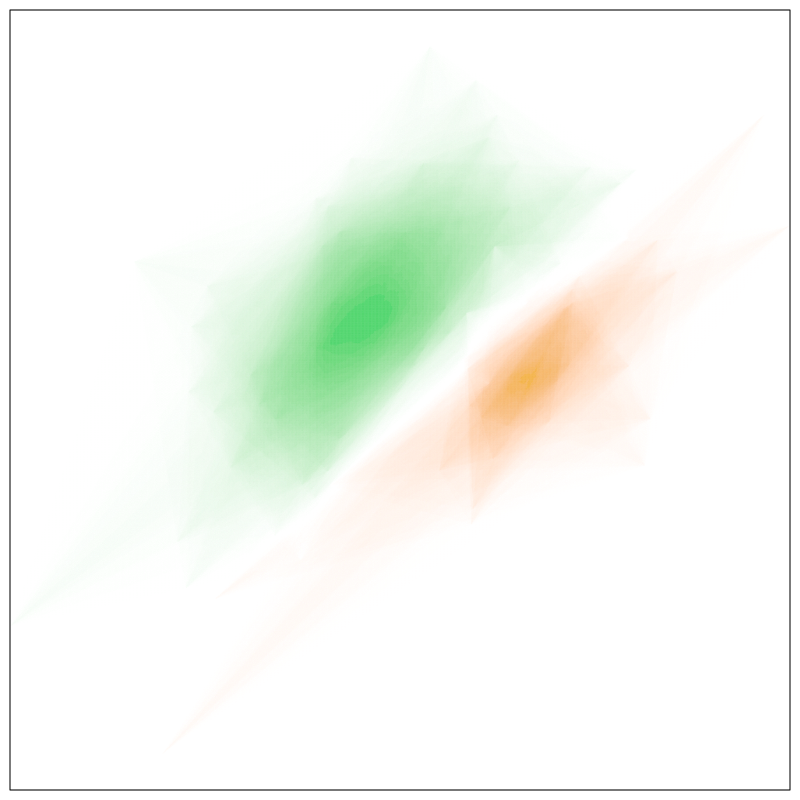} 
    \includegraphics[width=0.475\linewidth]{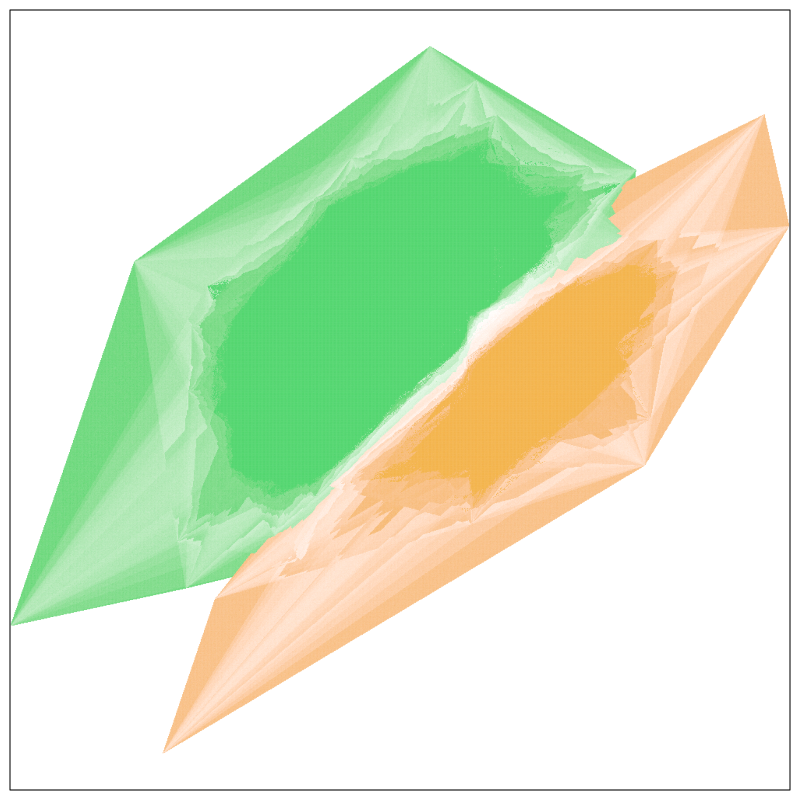} 
    \includegraphics[width=0.475\linewidth]{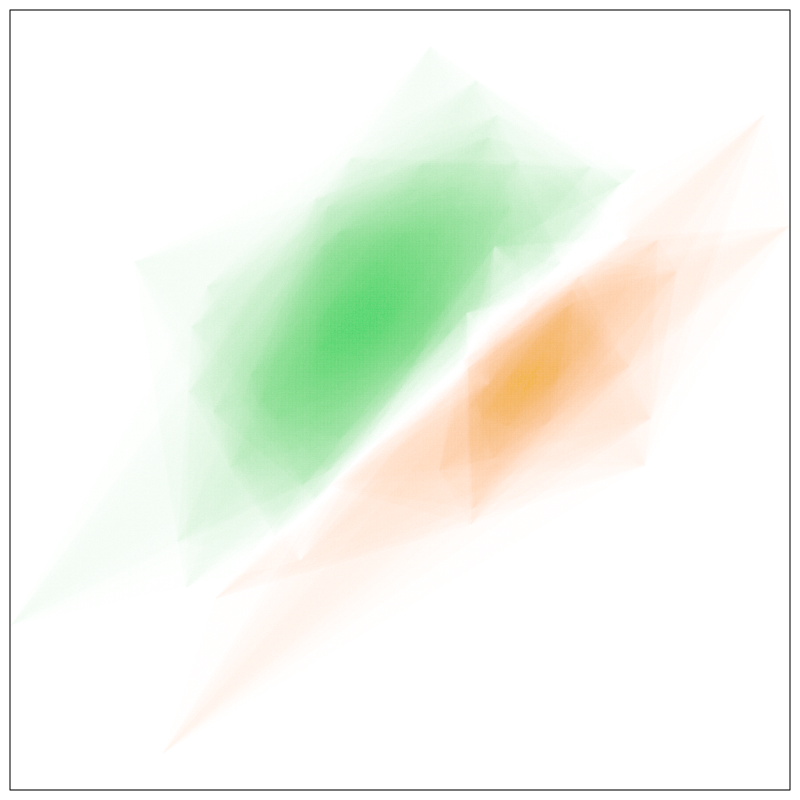} 
    \caption{A real dataset: the \textsf{hemophilia} data with sample sizes $n_1 = 30$ for dataset $\mathcal X_{n_1}$ (orange) and $n_2 = 45$ for dataset $\mathcal Y_{n_2}$ (green). Top left: The original dataset. Top right: The difference of the exact sample simplicial depths $SD_{n_1}(x; \mathcal X_{n_1}) - SD(x; \mathcal Y_{n_2})$. Bottom panels: Estimated contours of $\widehat{p}_{H}$ (left) and $\widehat{p}_{P}$ (right). The colors of the contours used are orange for $\widehat{p}>1/2$ and green for $\widehat{p}<1/2$. Darker tones are used for values closer to $0$ or $1$; white color corresponds to $\widehat{p} \approx 1/2$.}
    \label{fig: classification hemophilia}
\end{figure}

To empirically evaluate the performance of the estimators~\eqref{eq: p4}--\eqref{eq: p3}, we consider numerical examples. According to the conclusions of the simulations in Section~\ref{section: simulations}, we consider only estimators $\widehat{\gamma}$ in the medium(a) scenario, the rounded estimators $\widehat{\gamma}_{\mathrm{r}}$, and the bias correction for non-degenerate U-statistics $\widehat{\gamma}_1$. With other reasonable estimators $\widehat{\gamma}$, the results turned out to be very similar (a comparison of the results for the Hall-type estimators with choices $\widehat{\gamma} = 0$ as in $\widehat{p}_{H, \mathrm{naive}}(x)$, the rounded estimator $\widehat{\gamma}_{\mathrm{r}}$ and the trimmed estimator $\widehat{\gamma}_{\mathrm{t}}$ can be found in Figure~\ref{fig: classification H} in Supplementary Material~\ref{section: Gaussian classification 2}).

In each example, we plot the estimated contours of the functions $\widehat{p}_H$ and $\widehat{p}_P$ in the plane. We consider two simulated Gaussian scenarios (a) and (b), whose results can be found in Figure~\ref{fig: classification normal} and in Section~\ref{section: Gaussian classification 2} in the Supplementary Material, and (c) a real dataset \textsf{hemophilia} from \textsf{R} package \textsf{ddalpha} \citep{ddalpha}, as can be seen in Figure~\ref{fig: classification hemophilia}.

In the simulated Gaussian scenario (a), $X$ is bivariate Gaussian with mean $(1,1)^\top$ and unit covariance matrix (orange points in the top left panel of Figure~\ref{fig: classification normal}) and $Y$ is the bivariate Gaussian with mean $(0,0)^\top$ and the covariance matrix with unit diagonal terms and $\rho = 0.8$ as the off-diagonal terms (green points in Figure~\ref{fig: classification normal}). Several conclusions about the performance of both estimators can be drawn from Figure~\ref{fig: classification normal}:
    \begin{itemize}
        \item \textbf{Overall performance.} Especially in the central parts of the distributions, both~\eqref{eq: p4} and~\eqref{eq: p3} perform well, and properly distinguish the two samples, assigning higher values of $\widehat{p}_H(x)$ and $\widehat{p}_P(x)$ to points $x$ close to the center of $\mathcal X_n$. 
        \item \textbf{Outsider problem.} Outside the convex hull of $\mathcal X_n$ we have $SD_n(x; \mathcal X_n) = SD_{m,b}^\star(x; \mathcal X_n) = 0$ for all $b = 1,\dots, B$, and the same is true for the other sample $\mathcal Y_n$. Thus, for $x$ outside the convex hulls of both $\mathcal X_n$ and $\mathcal Y_n$, we obtain trivially $\widehat{p}_H(x) = \widehat{p}_P(x) = 1/2$. The latter phenomenon is well known in the depth literature as the outsider problem \citep{Mosler_Hoberg2006, Mozharovskyi_etal2015}. It is tied to the intrinsic property of SD to attach zero value to all points outside the convex hull of the data.
        \item \textbf{Comparison at the border.} Inspecting the extreme levels of the estimated functions $\widehat{p}_H$ and $\widehat{p}_P$ near $0$ and $1$, we observe a difference in behavior. While the extreme contours of $\widehat{p}_H$ are indeed copying the shape of the convex hulls of the datasets, the contours of the percentile estimator $\widehat{p}_P$ do not perform so well. This can be accounted for as follows. Take, for instance, a point $x$ inside the convex hull of $\mathcal X_n$, but near its boundary, so that $SD_n(x; \mathcal X_n)$ is positive but small. Suppose that $x$ is outside the convex hull of $\mathcal Y_n$, i.e. $SD_n(x; \mathcal Y_n) = 0$ and also $SD^\star_{m,b}(x; \mathcal Y_n) = 0$ almost surely. Since $SD_n(x; \mathcal X_n)$ is small, there are only very few $\mathcal X_n$-determined simplices that contain $x$, and thus in the resampling procedure, we will see $SD^\star_{m,b}(x; \mathcal X_n) = 0$ with very high probability. Naturally, for the percentile estimator we then have $0 = SD^\star_{m,b}(x; \mathcal X) = SD^\star_{m,b}(x; \mathcal Y_n)$, and $\widehat{p}_P(x) = 1/2$ with high probability. Thus, even though the point $x$ lies inside the convex hull of $\mathcal X_n$ but outside of the convex hull of $\mathcal Y_n$, the percentile estimator will very likely claim that it is impossible to decide whether $x$ came from $\mathcal X_n$ or $\mathcal Y_n$. On the other hand, for the Hall estimator, the small but positive value of $SD_n(x; \mathcal X_n)$ in~\eqref{eq: p4} combined with $SD^\star_{m,b}(x; \mathcal X_n) = 0$ with high probability gives $(1+\lambda_X) SD_n(x; \mathcal X_n) - \lambda_X\, SD^\star_{m,b}(x; \mathcal X_n) > (1+\lambda_Y) SD_n(x; \mathcal Y_n) - \lambda_Y SD^\star_{m,b'}(x; \mathcal Y_n) = 0$, and $\widehat{p}_H(x) = 1$ with high probability. Thus, in this situation, the effect of symmetrizing the depth distribution in~\eqref{eq: symmetrizing} and~\eqref{eq: symmetrizing 2} helps in distinguishing points near the boundary of the convex hulls of the data.
        \item \textbf{Comparison at the center.} A further interesting effect of symmetrization in~\eqref{eq: symmetrizing} appears for $x$ near the center of one of the distributions, where the behavior of $SD_n(x; \mathcal X_n)$ and $SD_n(x; \mathcal Y_n)$ is much different. We have seen in Theorem~\ref{thm: Thm1} that for $x$ at the center of symmetry of $P$ (corresponding to the sample $\mathcal X_n$) we have $SD_n(x; \mathcal X_n) \approx 1/4$ and the asymptotic distribution of $SD_n(x; \mathcal X_n)$ is heavily skewed to the left. For $\mathcal Y_n$, we have that $SD_n(x; \mathcal Y_n)$ is typically positive but relatively low, and its asymptotic distribution is a symmetric Gaussian distribution. The large difference between $SD_n(x; \mathcal X_n)$ and $SD_n(x; \mathcal Y_n)$ should make it easy to decide that $p(x)$ should be very high. On the other hand, due to the skewness of the distribution of $SD^\star_{m,b}(x; \mathcal X_n)$, there are typically pairs $b, b'$ such that $SD^\star_{m,b}(x; \mathcal X_n) < SD^\star_{m,b'}(x; \mathcal Y_n)$, and thus $\widehat{p}_P(x)$ is not going to be equal to $1$. For the Hall-type estimators, however, the symmetrization in~\eqref{eq: symmetrizing} and~\eqref{eq: symmetrizing 2} makes the two clusters of the resampled depths much better separated, and typically $\widehat{p}_H(x) = 1$. This effect can be clearly seen in Figure~\ref{figure: normal symmetrize}, where the resampled depths $SD^\star_{m,b}(x; \mathcal X_n)$ and $SD^\star_{m,b'}(x; \mathcal Y_n)$ are displayed.
    \end{itemize}
%
%
The results for the same simulation with a real dataset \textsf{hemophilia} in task (c) can be found in Figure~\ref{fig: classification hemophilia}. Finally, the simulation exercise (b) was performed with two Gaussian distributions also in the situation when the difference between the two clusters is only in location, i.e. by choosing $\rho = 0$ in the covariance matrix of $Y$. The results analogous to those in Figure~\ref{fig: classification normal} are in Figure~\ref{fig: classification normal0} in Section~\ref{section: Gaussian classification 2} in the Supplementary Material. 

As an overall conclusion, from our numerical examples it appears that the Hall-type estimator $\widehat{p}_H(x)$ from~\eqref{eq: p4} is preferable to the percentile estimator $\widehat{p}_P(x)$ from~\eqref{eq: p3} or the naive estimator $\widehat{p}_{H,\mathrm{naive}}(x)$ from~\eqref{eq: p4 naive}. As such, estimating $p(x)$ using our resampling procedure significantly enhances the performance and the interpretability of supervised depth-based classification.

\subsection*{Acknowledgments} The research of S. Nagy was supported by the Czech Science Foundation (project n.~24-10822S) and the ERC~CZ grant LL2407 of the Ministry of Education, Youth and Sport of the Czech Republic. The research of C. Jentsch was partially supported by the Deutsche Forschungsgemeinschaft (DFG, German Research Foundation; Project-ID 520388526; TRR 391: Spatio-temporal Statistics for the Transition of Energy and Transport).

\subsection*{Disclosure statement} The authors report there are no competing interests to declare. 

\bibliographystyle{apalike}
\bibliography{SDarXiv.bbl}

\newpage
\FloatBarrier

%
%
%
%

\title{Resampling simplicial depth: Supplementary Material}


\setcounter{section}{0}
\setcounter{table}{0}
\setcounter{figure}{0}
\renewcommand{\thesection} {S.\arabic{section}} 
\renewcommand{\theequation}{S.\arabic{equation}}
\renewcommand{\thefigure}{S.\arabic{figure}}
\renewcommand{\thetable}{S.\arabic{table}}
\renewcommand{\thetheorem}{S.\arabic{theorem}}
\renewcommand{\theexample}{S.\arabic{example}}

\maketitle

This is the Supplementary Material accompanying the paper \emph{Resampling simplicial depth}. This document gathers the proofs of our theoretical results and the complete results of the full simulation study performed in the main paper.

\section{Proofs of the theoretical results} \label{section: proofs}

\subsection*{Proof of Theorem~\ref{thm: Thm1}}

The first part of the theorem follows from \citet[Remark~B]{Liu1990}.

We prove the non-trivial implication in the second part of the theorem in the plane. That is, we want to show that if $x \in \R^2$ is not a center of symmetry and at the same time $SD(x; P) > 0$, then~\eqref{eq: SDn} is non-degenerate.

Using the affine invariance of $SD_n$, we may assume that $x$ is the origin $0 \in \R^2$. The kernel of the U-statistic~\eqref{eq: SDn} is $h(x,y,z) = \I{ 0 \in \Simplex{x, y, z}}$. Write
    \[
    h_1(x) = \int_{\R^2} \int_{\R^2} h(x, y, z) \dd P(y) \dd P(z) = \int_{\R^2} \int_{\R^2} \I{ 0 \in \Simplex{x, y, z}} \dd P(y) \dd P(z) 
    \]
for $x \in \R^2$. The U-statistic is degenerate of order 1 if and only if $h_1(x) = SD_n(0; \mathcal X_n)$ for $P$-almost all $x \in \R^2$ \citep[Section~5.2.1]{Serfling1980}. We will evaluate $h_1(x)$ explicitly in terms of the distribution function of a transform of $P$. To do so, first, observe that the simplicial depth at the origin $0 \in \R^d$ is invariant w.r.t. the radial projection $\xi \colon \R^d\setminus\{0\} \to \Sph \colon x \mapsto x/\left\Vert x \right\Vert$, where $\Sph$ stands for the unit sphere in $\R^d$. That is, 
    \[  SD(0; P) = SD(0; P_{\xi(X)})    \]
where $P_{\xi(X)}$ is the distribution of $\xi(X)$ with $X \sim P$ that lives in the unit sphere $\Sph$. This identity is valid for any $P$ that has no mass at $0$. Thus, it is enough to assume that $P \in \Prob[\R^2]$ is supported on the unit circle $\Sph[1]$. 

\begin{figure}[htpb]
    \centering
    \includegraphics[width=0.45\linewidth]{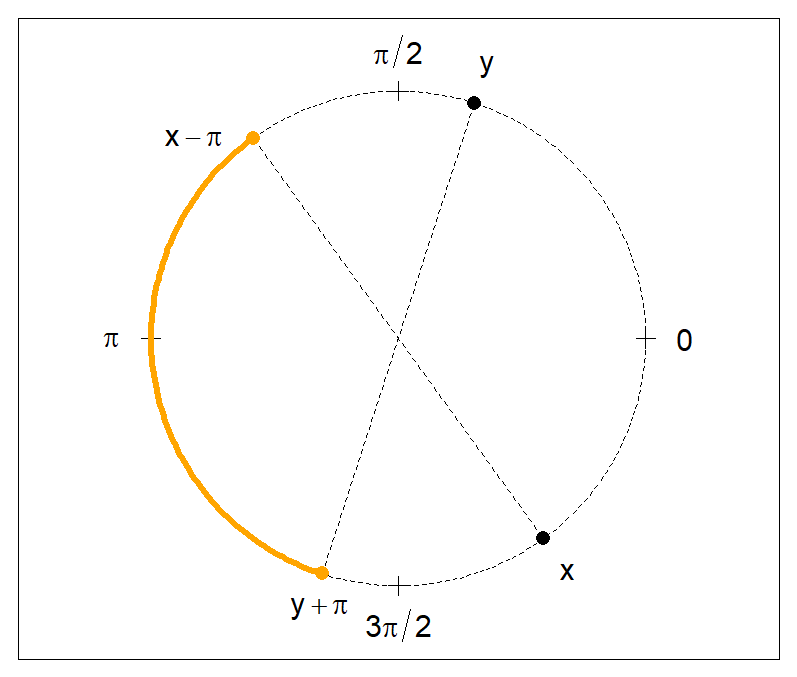}
    \includegraphics[width=0.45\linewidth]{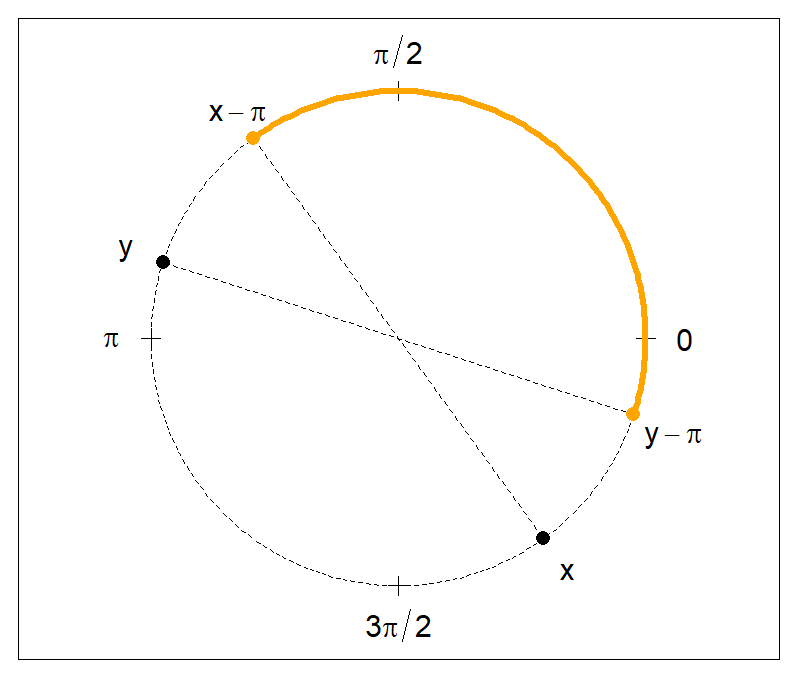} \\
    \includegraphics[width=0.45\linewidth]{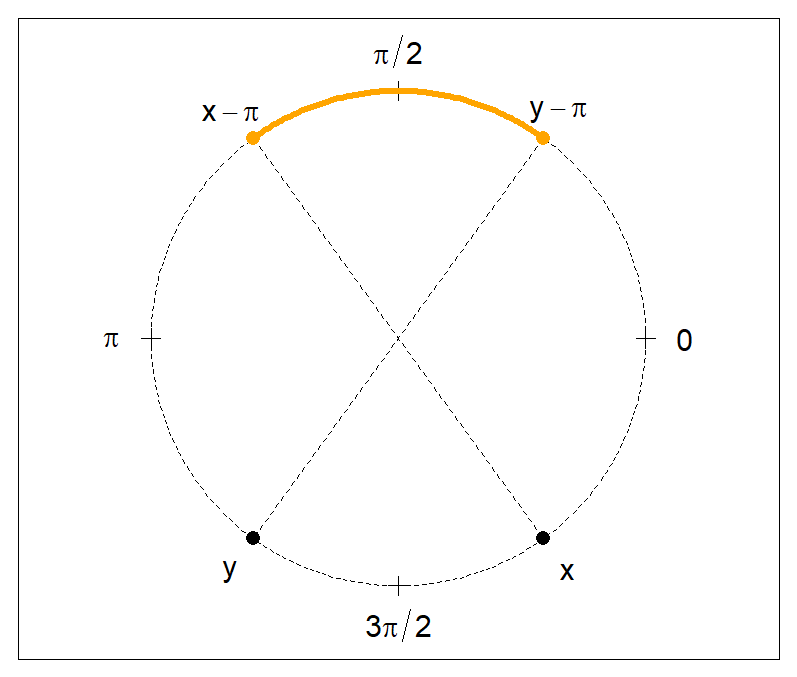}
    \includegraphics[width=0.45\linewidth]{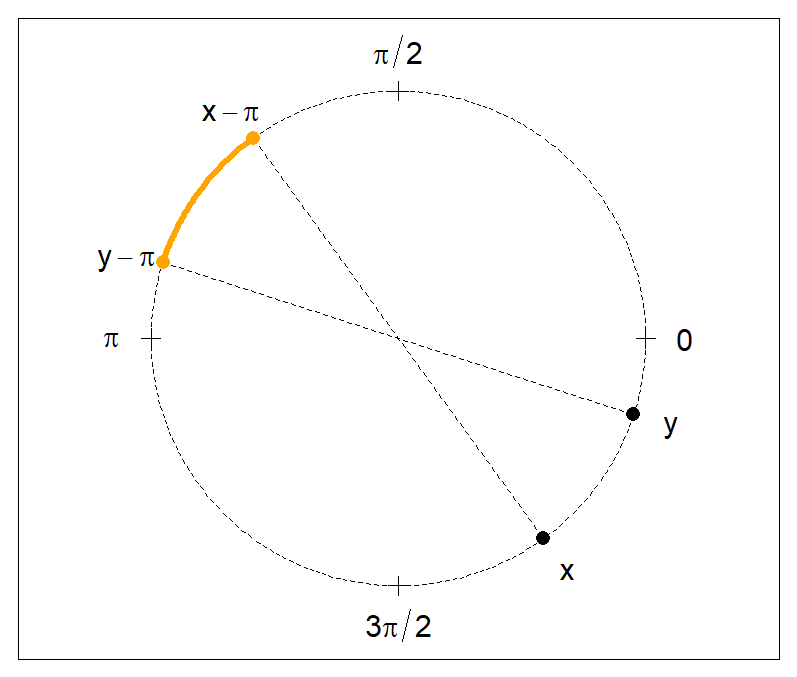}
    \caption{Proof of Theorem~\ref{thm: Thm1}: Four situations to be considered when evaluating $h_1(x)$. In each case, the orange arc represents the set where $z \in \Sph[1]$ must lie in order to $0 \in \Simplex{x,y,z}$.}
    \label{fig: Thm1}
\end{figure}

Consider the angular distribution function $F$ of $X \sim P \in \Prob[{\Sph[1]}]$ around the origin defined as
    \[  F(t) = \PP\left( T \leq t \right) \quad \mbox{for }t \in [0,2\,\pi),    \]
where $T$ is the random angle of $X = \left( \cos(T), \sin(T) \right)^\top \in \R^2$ when expressed in polar coordinates. In the same way, also other points on the unit circle will be canonically identified with their angles in $[0,2\pi)$. Let $x \in \Sph[1]$ be given, and without loss of generality, suppose that $x > \pi$. Denote $a = F(x-\pi)$. Then, separating the event $y \in [0,2\pi)$ into four cases $y \in [0, x - \pi)$, $y \in (x-\pi, \pi]$, $y \in (\pi, x)$, $y \in (x, 2\pi)$ as in Figure~\ref{fig: Thm1} and neglecting events of null probability, we have
    \[  
    \begin{aligned}
    h_1(x) & = \int_{\R^2} \int_{\R^2} \I{0 \in \Simplex{x, y, z}} \dd P(z) \dd P(y) \\
    & = \int_{0}^{x - \pi} F(y+\pi) - F(x-\pi) \dd F(y) + \int_{x-\pi}^\pi 1 - F(y+\pi) + F(x-\pi) \dd F(y) \\
    & \phantom{=} + \int_\pi^x F(x-\pi) - F(y-\pi) \dd F(y) + \int_{x}^{2\pi} F(y-\pi) - F(x-\pi) \dd F(y) \\
    & = -a^2 + (1+a)(F(\pi) - a) + a(F(x) - F(\pi)) - a(1 - F(x)) \\
    & \phantom{=} + \int_{0}^{x - \pi} F(y+\pi) \dd F(y) - \int_{x-\pi}^\pi F(y+\pi) \dd F(y) \\
    & \phantom{=} - \int_\pi^x F(y-\pi) \dd F(y) + \int_{x}^{2\pi} F(y-\pi) \dd F(y) \\
    & = 2 a(F(x) - 1 - a) + F(\pi) + \int_{0}^{x - \pi} F(y+\pi) \dd F(y) - \int_{x-\pi}^\pi F(y+\pi) \dd F(y) \\
    & \phantom{=} - \int_\pi^x F(y-\pi) \dd F(y) + \int_{x}^{2\pi} F(y-\pi) \dd F(y). \\
    \end{aligned}
    \]
Take a derivative of the last expression w.r.t. $x$. Writing $f = F'$ for the angular density of $F$, we have that the derivative of $a$ is $f(x-\pi)$, and the fundamental theorem of calculus \citep[Theorem~7.2.1]{Dudley2002} gives
    \[
    \begin{aligned}
    \frac{\partial h_1(x)}{\partial x} & = 2 f(x-\pi)(F(x) - 1 - F(x-\pi)) + 2 F(x-\pi)(f(x) - f(x-\pi)) \\
    & \phantom{=} + 2 F(x) f(x-\pi) - 2 F(x-\pi) f(x) \\
    & = 2 f(x-\pi) (2 F(x) - 2 F(x-\pi) - 1). 
    \end{aligned}
    \]
This expression is equal to zero if and only if either $f(x-\pi) = 0$, or if $F(x) = F(x-\pi) + 1/2$. 

First, consider the situation when the density $f$ is non-vanishing on $[0,2\pi)$. Then, the latter condition $F(x) = F(x-\pi) + 1/2$ for each $x$ is equivalent to the distribution of $X$ being halfspace symmetric around the origin, which is by \citet[Theorem~2]{Rousseeuw_Struyf2004} equivalent with angular symmetry of $X \sim P$ around $0 \in \R^2$.

Now, let $f$ be any density, and let $x$ be a point such that $f(x-\pi) = 0$. Because of our assumption on the existence of the density $f$, the distribution function $F$ must be continuous everywhere on $[0,2\pi)$. Consider a sequence of points $x_n \to x$ on $\Sph[1]$.  For each such point $x_n \in \Sph[1]$, either $f(x_n - \pi) = 0$ or $F(x_n) = F(x_n-\pi) + 1/2$. Suppose first that the second condition is true for infinitely many points in the sequence $\left\{ x_n \right\}_{n=1}^\infty$. Then, by the continuity of $F$, also $F(x) = F(x-\pi) + 1/2$, and the halfspace whose boundary line passes through the origin and $x$ has $P$-mass $1/2$. On the other hand, suppose that $f(x_n-\pi) = 0$ for all but a finite number of indices $n$, for all sequences $x_n \to x$. Then, necessarily, there is a (circular) interval $I \subset \Sph[1]$ containing $x$ such that $f(y-\pi) = 0$ for all $y \in I$. Expand this interval so that $I$ is the largest circular interval containing $x$ such that for almost all $y \in I$, $f(y-\pi)=0$. Because we assumed that $SD(0; P) > 0$, the interval $I$ must be contained in some semi-circle, and its boundary points $y_1, y_2 \in \Sph[1]$ can be approached by sequences $y_{1,n} \to y_1$ and $y_{2,n} \to y_2$ in $\Sph[1]$ such that $F(y_{i,n}) = F(y_{i,n}-\pi) + 1/2$ for all $n$ and $i = 1, 2$. The continuity of $F$ again gives $F(y_i) = F(y_i - \pi) + 1/2$ for $i=1,2$. Rotating the whole setup if necessary, we can assume that $I$ is contained in the lower semi-circle of angles greater than $\pi$. Simultaneously, by our construction of $I$, $F$ must be a constant function on $I$. In particular, $F(y_1 - \pi) + 1/2 = F(y_1) = F(y_2) = F(y_2 - \pi) + 1/2$, and consequently also $F(y_1 - \pi) = F(y_2 - \pi)$. We obtain that just as the cone of angles in $I \subset \Sph[1]$, also the antipodal cone $\left\{ y \in \Sph[1] \colon y - \pi \in I \right\}$ has zero $P$-mass (recall that this mass is determined by the angle $T$ of $X$ giving the distribution function $F$). Thus, $F(y) = F(y - \pi) + 1/2$ for all $y \in I$, and this is true in particular also for $x \in I$. 

Overall, we have shown that for any $x \in \Sph[1]$ in the lower semi-circle, we must have $F(x) = F(x-\pi) + 1/2$. Analogously, the same can be shown also for $x \in \Sph[1]$ in the upper semi-circle. Altogether, we obtain that if $SD(0; P)> 0$ and if $SD(0; P)$ is degenerate, then $P$ must be halfspace symmetric around the origin.

%
%

\subsection*{Proof of Theorem~\ref{lem:gammaest}} 

As $M$ is the median of the limit distribution of $n^{\gamma}|\widehat{\theta}_n-\theta|$,  we have by \citet[Theorem 2.2.1]{Politis_Romano1999} $m_u^\gamma \,T_u\xrightarrow{\mathcal{P}} M$ and $m_\ell^\gamma\, T_\ell\xrightarrow{\mathcal{P}} M$. Thus, we get the convergence
    \[
    \frac{m_u^\gamma \, T_u}{m_\ell^\gamma \, T_\ell}\xrightarrow{\mathcal{P}}1, \quad \mbox{ which implies } \quad \gamma (\log m_u-\log m_\ell)-(\log T_\ell - \log T_u)\xrightarrow{\mathcal{P}}0.
    \]
By our assumptions, $\log m_u-\log m_\ell=\log(m_u/m_{\ell})\geq \log(c_1)+c_2\log(n)$, so for $\widehat{\gamma}=(\log T_\ell - \log T_u) /(\log m_u-\log m_\ell)$, we get
    \[
    \begin{aligned}
    \log\left(\frac{n^{\gamma}}{n^{\widehat{\gamma}}}\right) & =\log(n)\big(\gamma-\widehat{\gamma}\big)\\
    & =\frac{\log(n)}{\log m_u-\log m_\ell}\left(\gamma (\log m_u-\log m_\ell)-(\log T_\ell - \log T_u)\right)\xrightarrow{\mathcal{P}}0.
    \end{aligned}
    \]
If $\gamma\in[1/2,1]$, then $|1-n^{\gamma} / n^{\widehat{\gamma}_\mathrm{t}}|\leq |1-n^{\gamma} / n^{\widehat{\gamma}}|\xrightarrow{\mathcal{P}}0$, and if $\gamma\in\{1/2,1\}$, we have that
\begin{equation*}
\PP\left(\frac{n^{\widehat{\gamma}_\mathrm{r}}}{n^\gamma}\neq 1\right)\leq \PP\left(\frac{n^{\widehat{\gamma}}}{n^\gamma}\geq n^{1/4} \ \text{or} \ \frac{n^{\widehat{\gamma}}}{n^\gamma}\leq n^{-1/4}\right)\xrightarrow[n \to \infty]{} 0.
\end{equation*}
The theorem is proved.

\section{Tables with detailed simulation results} \label{section: tables}

This section contains additional tables with complete results of the simulation studies performed in Sections~\ref{section: gamma simulations} and~\ref{section: simulation distribution} for the \textsf{Gaussian scenario}. These tables correspond to the boxplots displayed in Sections~\ref{section: gamma simulations} and~\ref{section: simulation distribution}.

\input{1.0gamma1.tex}

\input{1.0gamma2.tex}

\input{1.0gamma3.tex}

\input{1.0KS1.tex}

\input{1.0KS2.tex}

\input{1.0KS3.tex}

\FloatBarrier

\section{Simulation study: Cauchy scenario}   \label{section: Cauchy}

\begin{figure}[htpb]
    \centering
    \includegraphics[width=\linewidth]{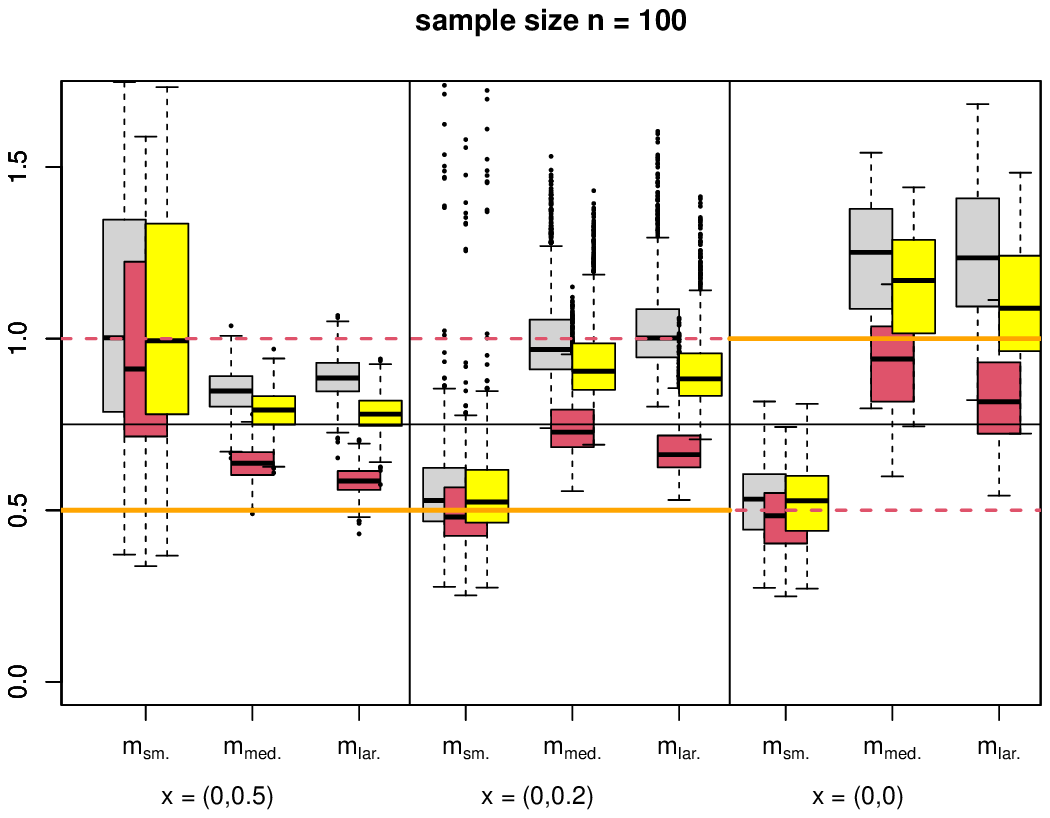}
    \caption{\textsf{Cauchy scenario:} Boxplots of the estimated raw (untrimmed, non-rounded) values of the exponent $\gamma$. The situation with sample size $n = 100$. The true value of $\gamma$ is $1/2$ for the first two scenarios, and $1$ for the third one.}
    \label{fig:box1Cauchy}
\end{figure}

\begin{figure}[htpb]
    \centering
    \includegraphics[width=\linewidth]{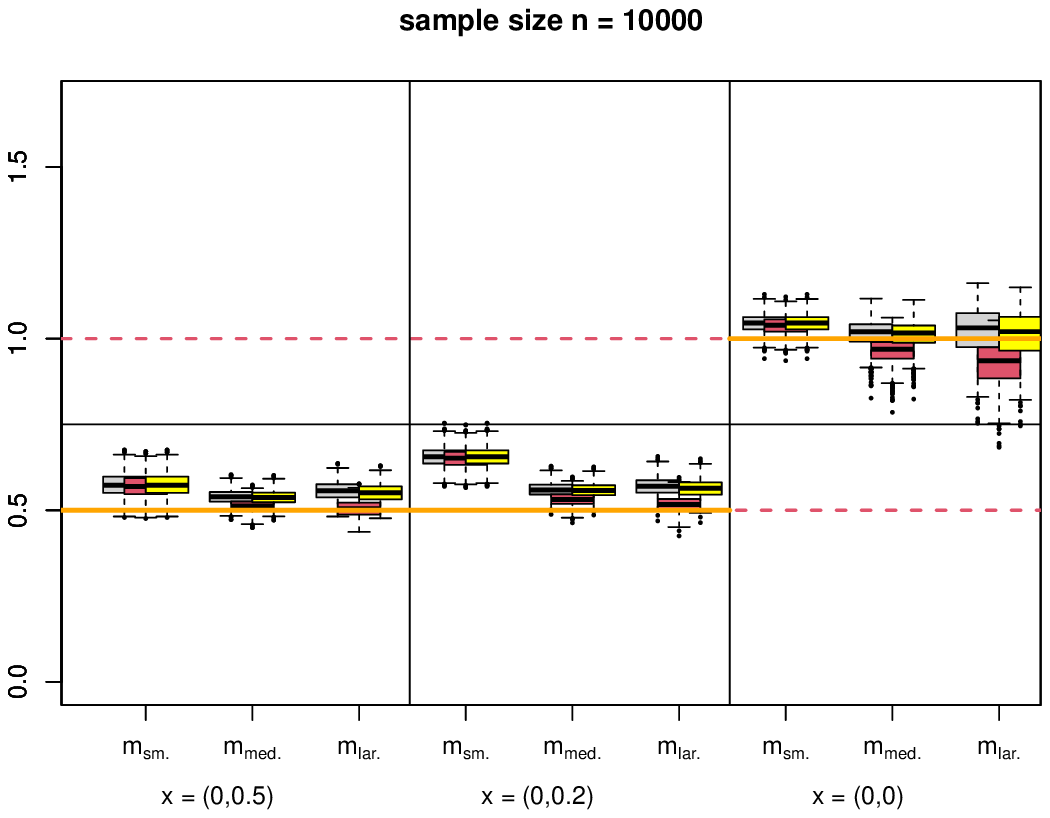}
    \caption{\textsf{Cauchy scenario:} Boxplots of the estimated raw (untrimmed, non-rounded) values of the exponent $\gamma$. The situation with sample size $n = 10000$. The true value of $\gamma$ is $1/2$ for the first two scenarios, and $1$ for the third one.}
    \label{fig:box3Cauchy}
\end{figure}

The same simulations and under the same setups as in Sections~\ref{section: gamma simulations} and~\ref{section: simulation distribution} were also performed with a heavy-tailed distribution $P$. In this case, $P$ was taken to be the centered bivariate elliptical Cauchy distribution with scatter matrix with one on its diagonal and $0.8$ as its off-diagonal terms. The boxplots with the three bias-corrected estimators of the rate $\gamma$ can be found in Figures~\ref{fig:box1Cauchy}--\ref{fig:box3Cauchy}; the boxplots with the Kolmogorov-Smirnov distances analogous to those from Section~\ref{section: simulation distribution} are in Figures~\ref{fig:KSbox1Cauchy}--\ref{fig:KSbox3Cauchy}.

\begin{figure}[htpb]
    \centering
    \includegraphics[width=\linewidth]{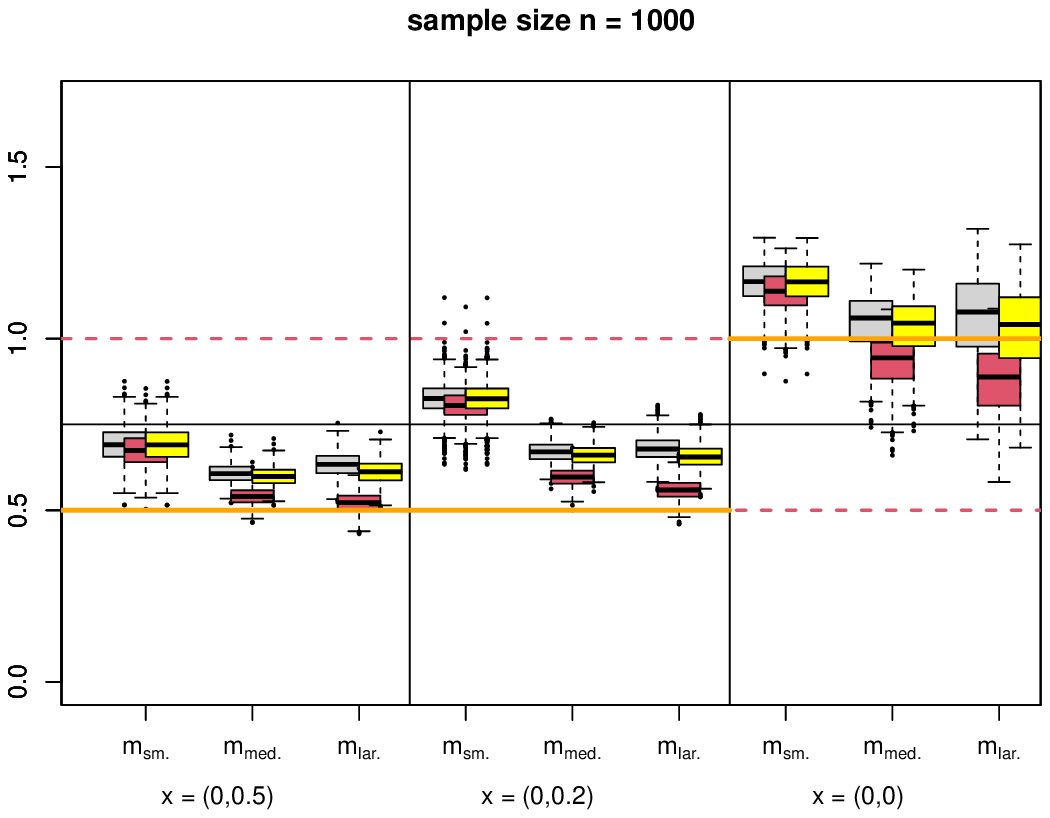}
    \caption{\textsf{Cauchy scenario:} Boxplots of the estimated raw (untrimmed, non-rounded) values of the exponent $\gamma$. The situation with sample size $n = 1000$. The true value of $\gamma$ is $1/2$ for the first two scenarios, and $1$ for the third one.}
\end{figure}

\begin{figure}[htpb]
    \centering
    \includegraphics[width=0.95\linewidth]{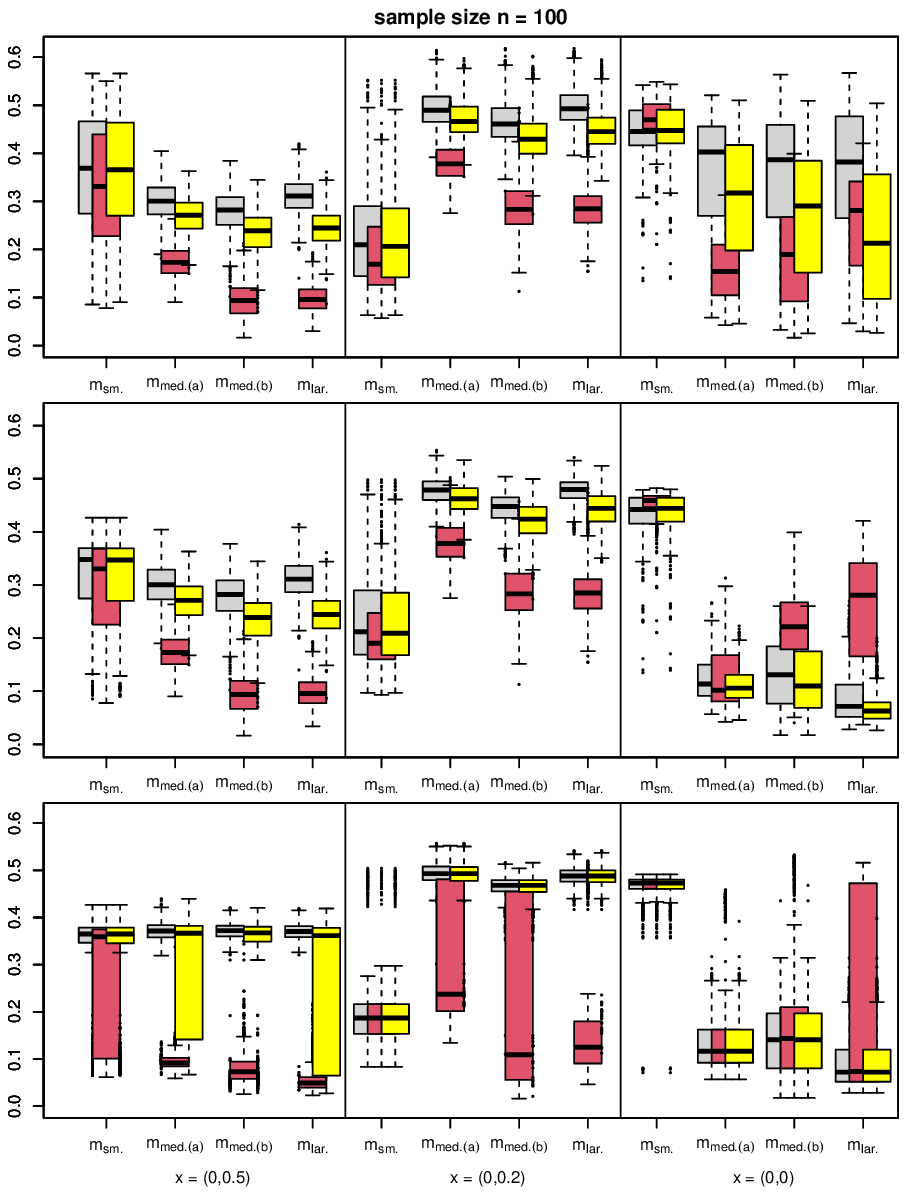}
    \caption{\textsf{Cauchy scenario,} \textbf{Kolmogorov-Smirnov dist., $n=100$:} Boxplots of the obtained Kolmogorov-Smirnov distances between the true and the resampled distribution of SD. The situation with sample size $n = 100$. The top panel corresponds to the raw estimator of $\widehat{\gamma}$, the middle panel to the trimmed estimator $\widehat{\gamma}_{\mathrm{t}}$, and the bottom panel to the rounded estimator $\widehat{\gamma}_{\mathrm{r}}$. The colors of boxplots are as in Figures~\ref{fig:box1Cauchy}--\ref{fig:box3Cauchy}.}
    \label{fig:KSbox1Cauchy}
\end{figure}

\begin{figure}[htpb]
    \centering
    \includegraphics[width=0.95\linewidth]{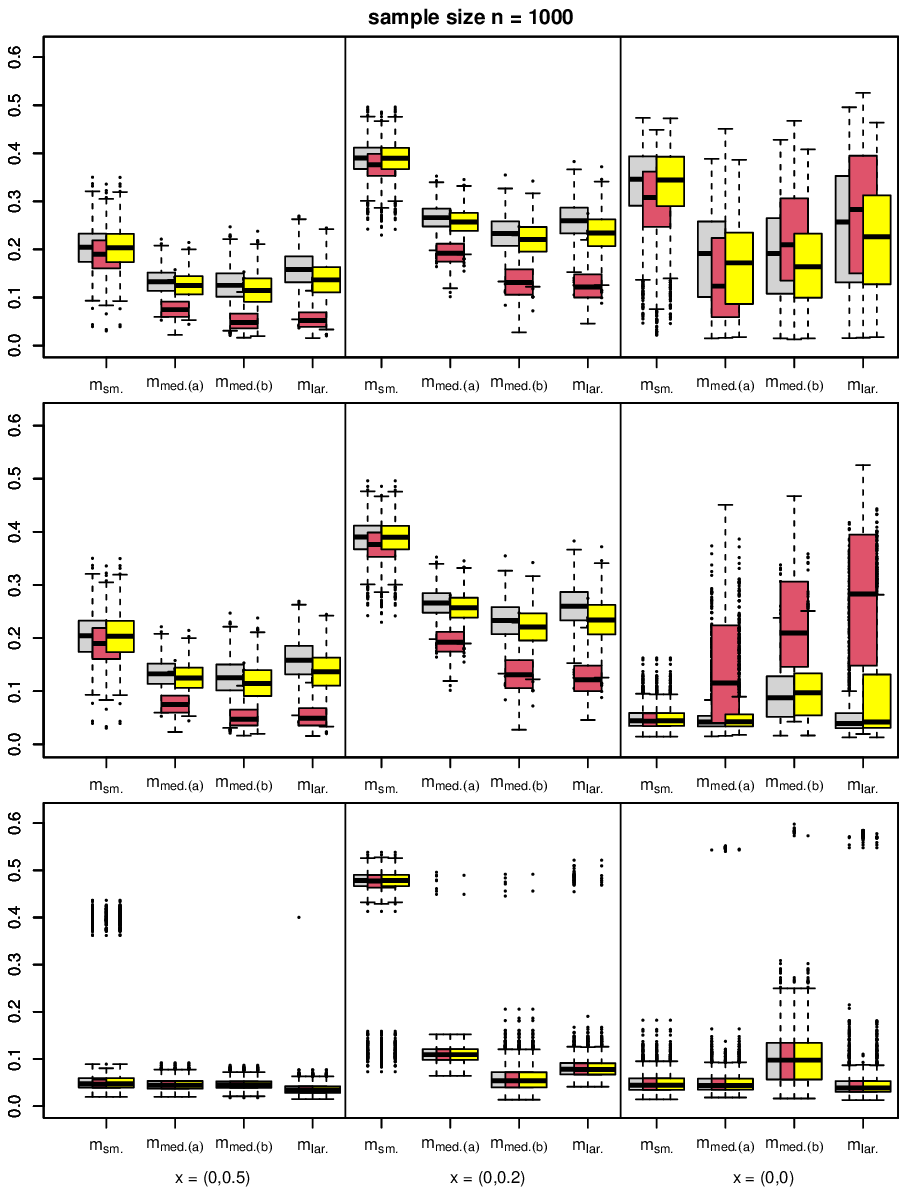}
    \caption{\textsf{Cauchy scenario,} \textbf{Kolmogorov-Smirnov dist., $n=1000$:} Boxplots of the obtained Kolmogorov-Smirnov distances between the true and the resampled distribution of SD. The situation with sample size $n = 1000$. The top panel corresponds to the raw estimator of $\widehat{\gamma}$, the middle panel to the trimmed estimator $\widehat{\gamma}_{\mathrm{t}}$, and the bottom panel to the rounded estimator $\widehat{\gamma}_{\mathrm{r}}$. The colors of boxplots are as in Figures~\ref{fig:box1Cauchy}--\ref{fig:box3Cauchy}.}
\end{figure}

\begin{figure}[htpb]
    \centering
    \includegraphics[width=0.95\linewidth]{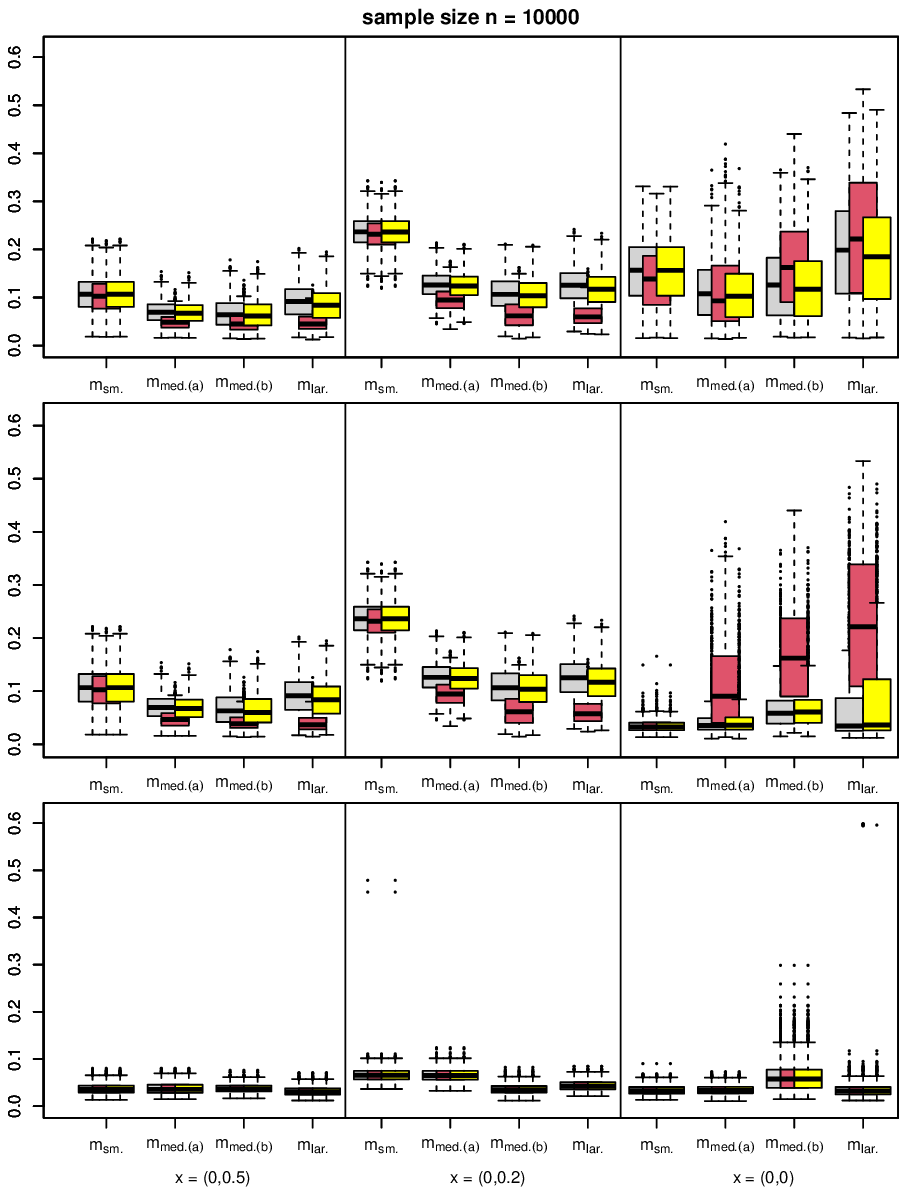}
    \caption{\textsf{Cauchy scenario,} \textbf{Kolmogorov-Smirnov dist., $n=10000$:} Boxplots of the obtained Kolmogorov-Smirnov distances between the true and the resampled distribution of SD. The situation with sample size $n = 10000$. The top panel corresponds to the raw estimator of $\widehat{\gamma}$, the middle panel to the trimmed estimator $\widehat{\gamma}_{\mathrm{t}}$, and the bottom panel to the rounded estimator $\widehat{\gamma}_{\mathrm{r}}$. The colors of boxplots are as in Figures~\ref{fig:box1Cauchy}--\ref{fig:box3Cauchy}.}
    \label{fig:KSbox3Cauchy}
\end{figure}

\input{2.0covlen_m2bc_2}

\input{2.0covlen_m2bc_3}

\FloatBarrier

\section{Simulations: Additional figures in classification}   \label{section: Gaussian classification 2}


\begin{figure}[htpb]
    \centering   
    \includegraphics[width=0.475\linewidth]{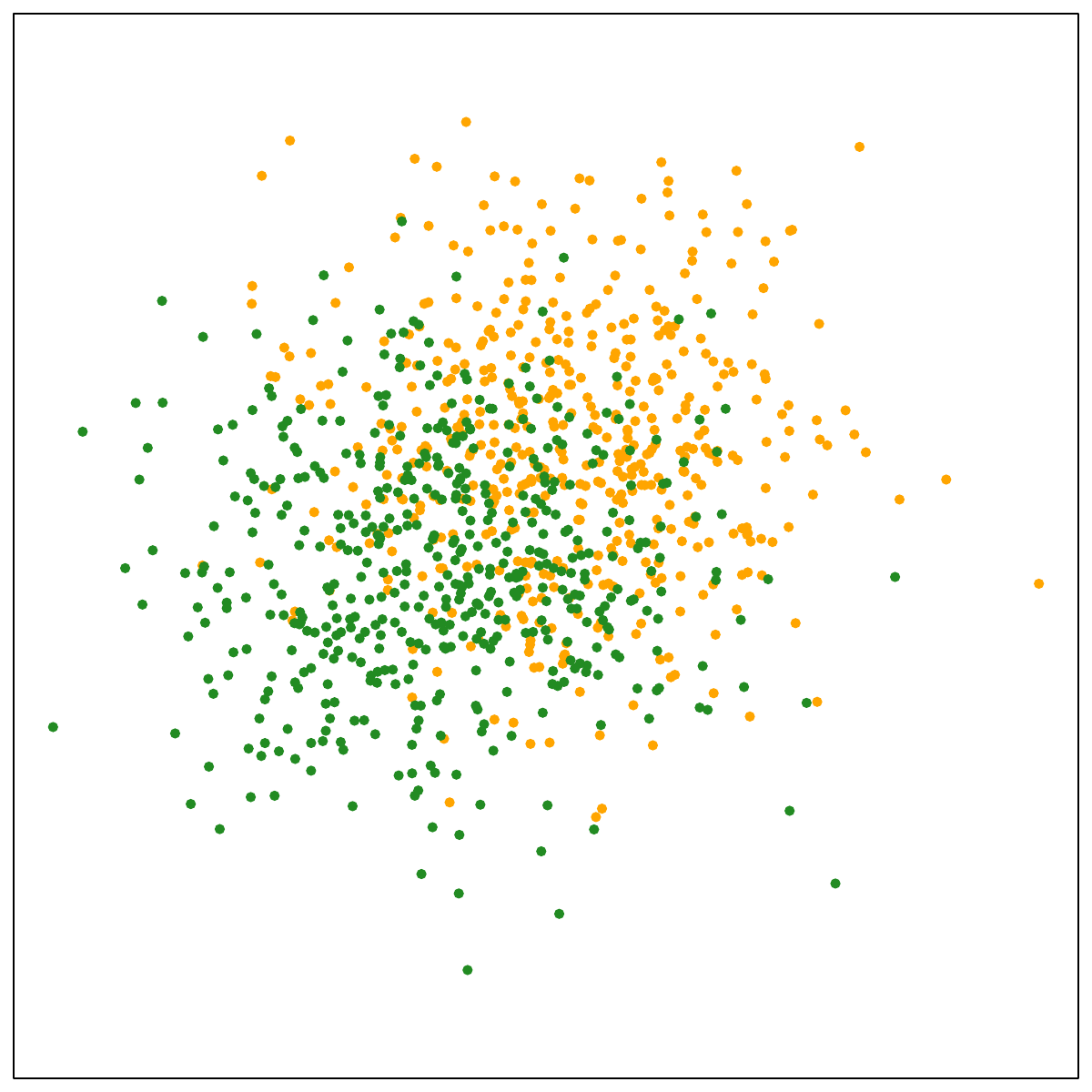} 
    \includegraphics[width=0.475\linewidth]{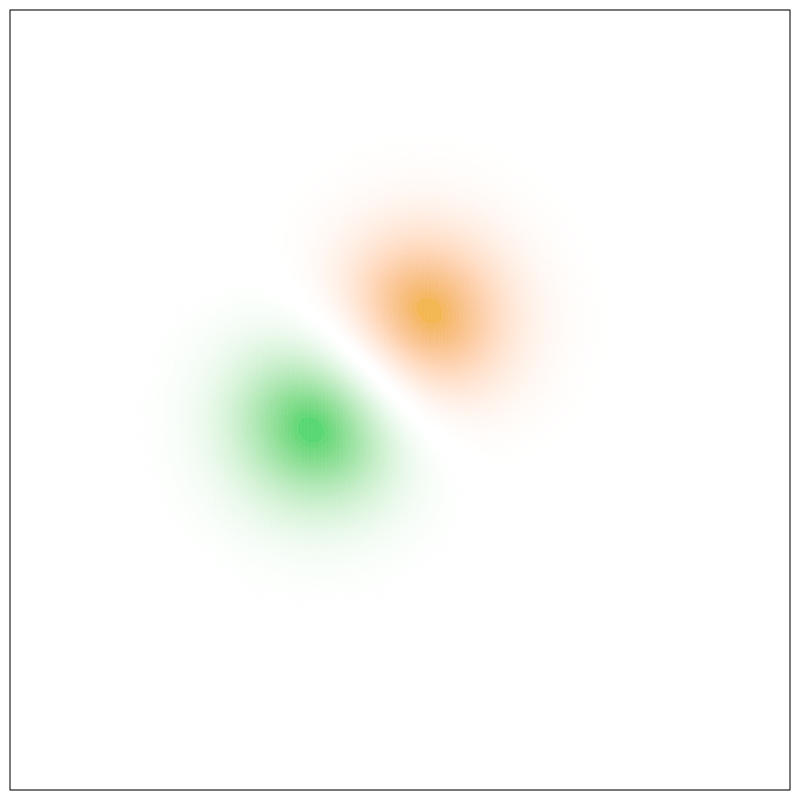} 
    \includegraphics[width=0.475\linewidth]{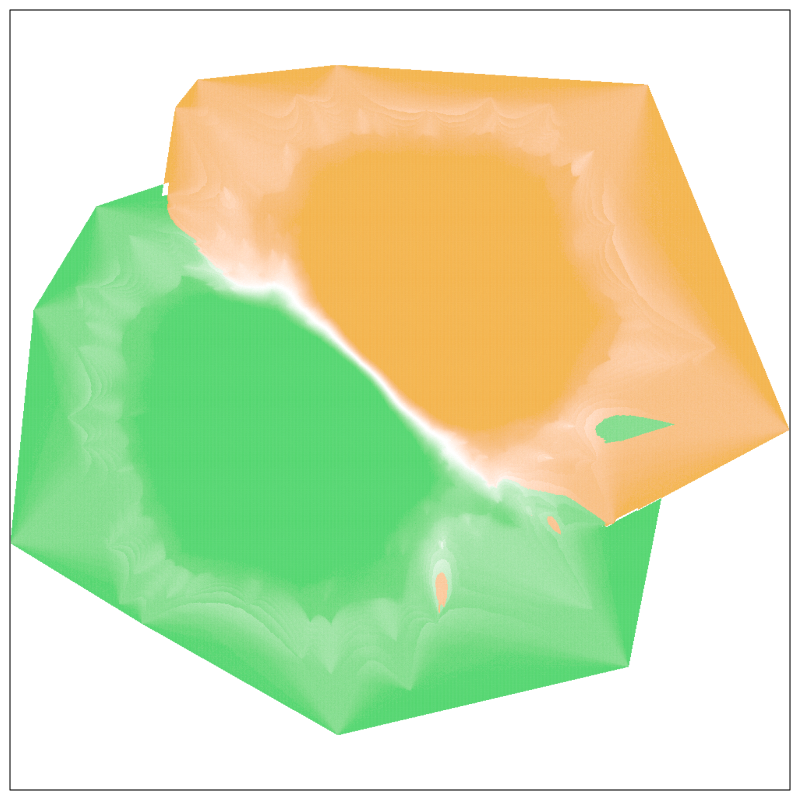} 
    \includegraphics[width=0.475\linewidth]{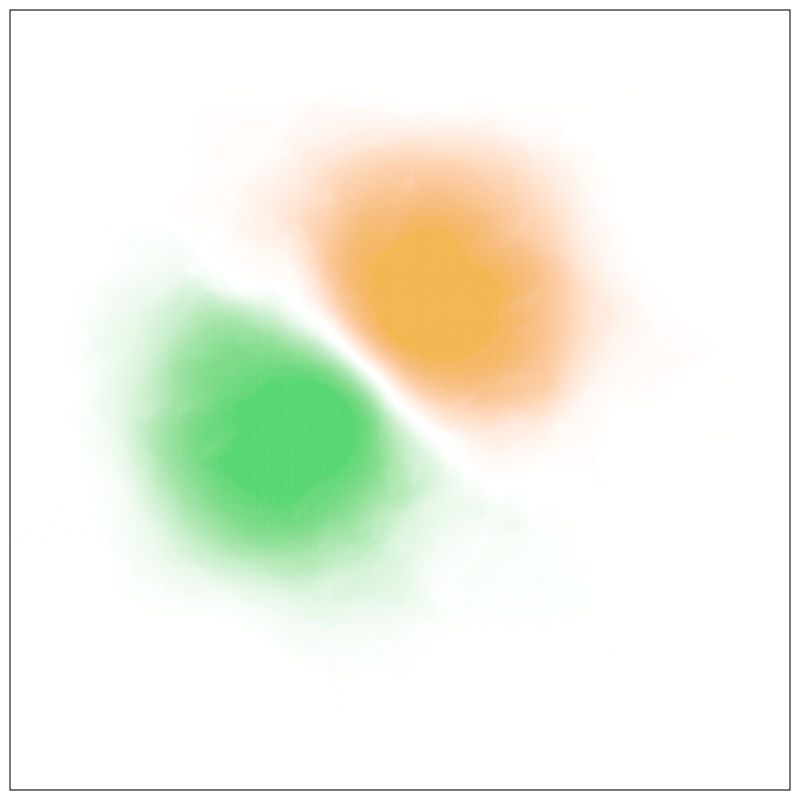} 
    \caption{A simulated dataset. Top left: two samples from Gaussian distributions $P$ (orange) and $Q$ (green), respectively, each of sample size $n_1 = n_2 = 500$. Top right: The difference of the true simplicial depths $SD(x; P) - SD(x; Q)$, approximated using random samples of sizes $10^5$ from both $P$ and $Q$. Bottom panels: Estimated contours of $\widehat{p}_{H}$ (left) and $\widehat{p}_{P}$ (right). The colors of the contours used are orange for $\widehat{p}>1/2$ and green for $\widehat{p}<1/2$. Darker tones are used for values closer to $0$ or $1$; white color corresponds to $\widehat{p} \approx 1/2$.}
    \label{fig: classification normal0}
\end{figure}

\begin{figure}[htpb]
    \centering   
    \includegraphics[width=0.32\linewidth]{normal_H_b1r_light.png} 
    \includegraphics[width=0.32\linewidth]{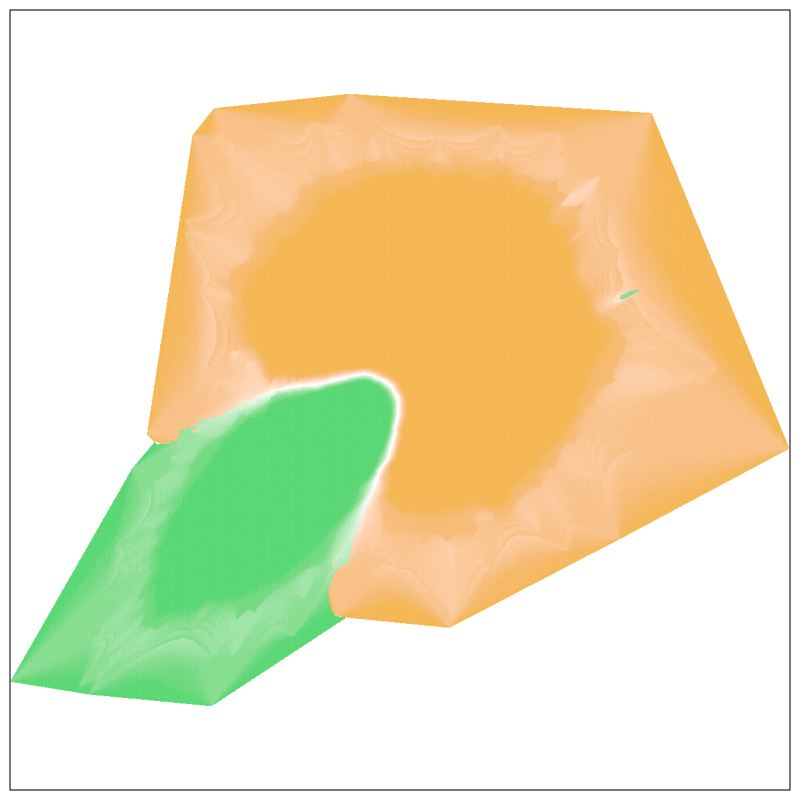} 
    \includegraphics[width=0.32\linewidth]{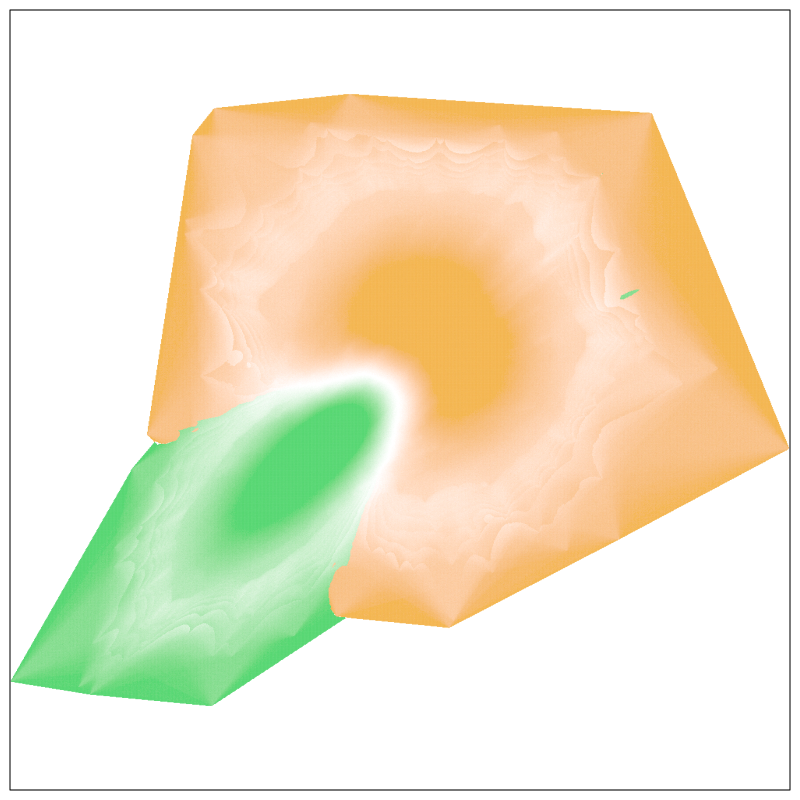} 
    \includegraphics[width=0.32\linewidth]{normal0_H_b1r_light.png} 
    \includegraphics[width=0.32\linewidth]{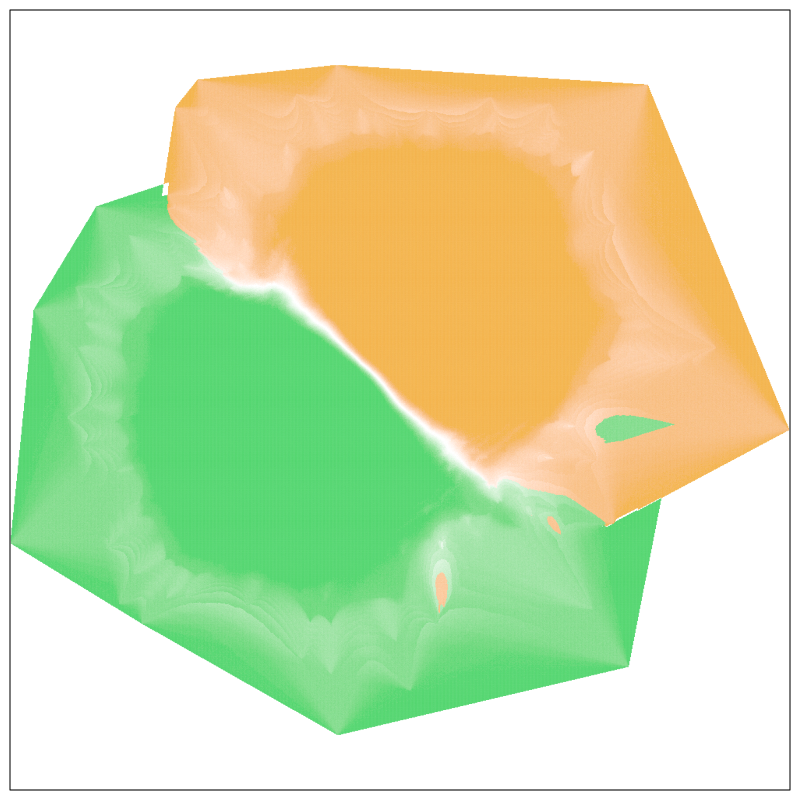} 
    \includegraphics[width=0.32\linewidth]{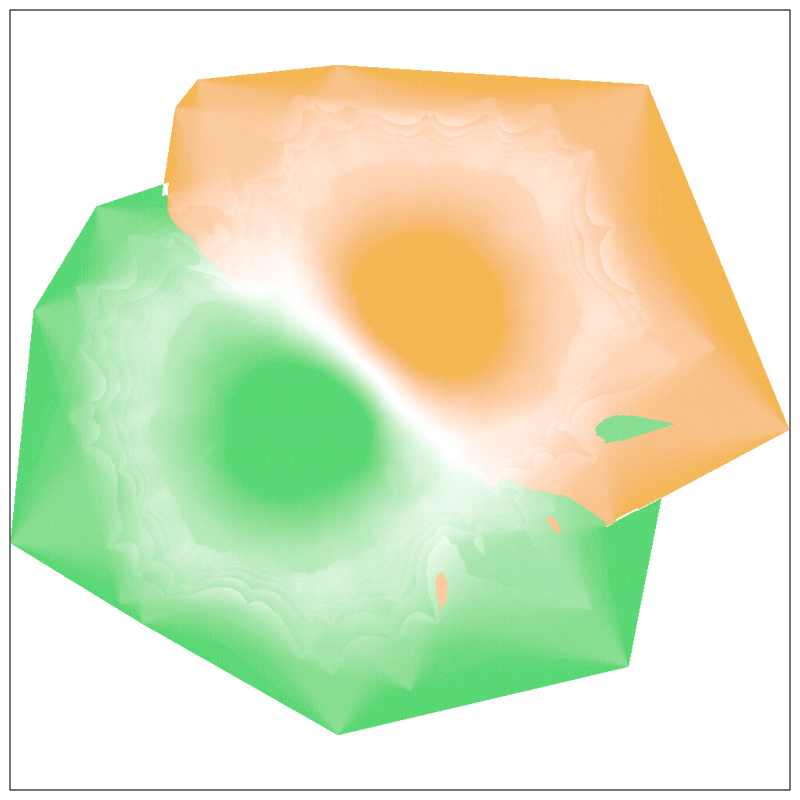}
    \includegraphics[width=0.32\linewidth]{hemo_H_b1r_light.png} 
    \includegraphics[width=0.32\linewidth]{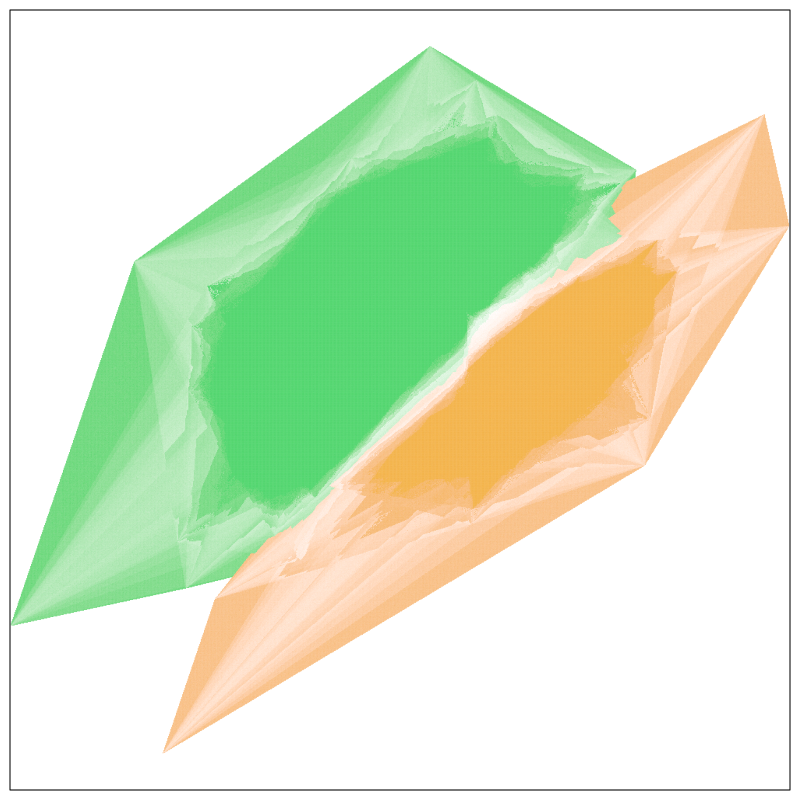} 
    \includegraphics[width=0.32\linewidth]{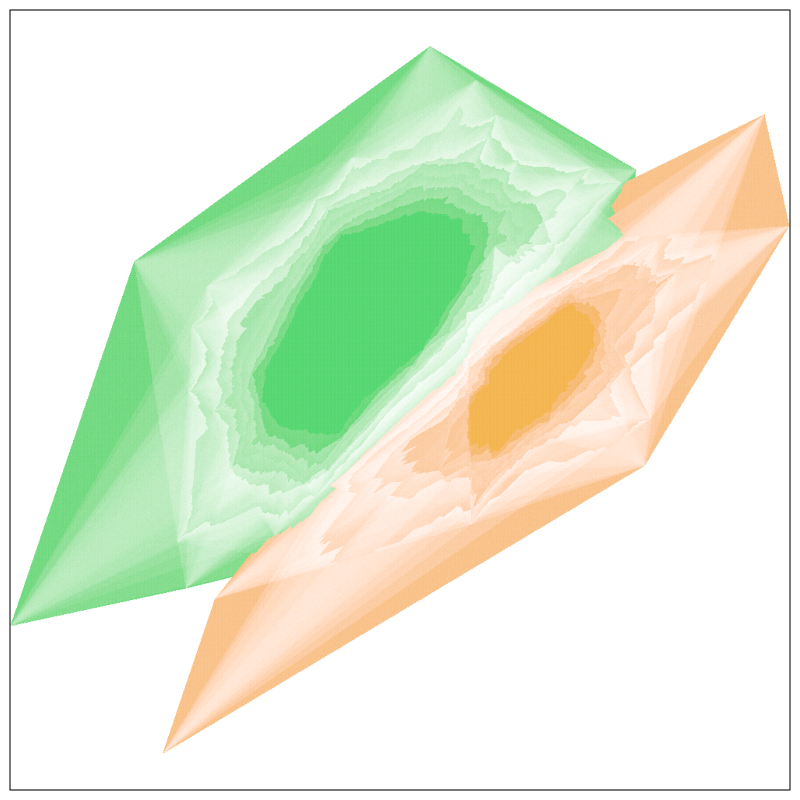}
    \caption{Three datasets from the simulation study in Section~\ref{section: application 2}  (in rows) and the estimated contours of $\widehat{p}_{H}$ using three different estimators $\widehat{\gamma}$. The left-hand panels correspond to the rounded estimators $\widehat{\gamma}_{\mathrm{r}}$ (the same plots as in Figures~\ref{fig: classification normal}, \ref{fig: classification hemophilia}, and \ref{fig: classification normal0}), the middle panels are for the trimmed estimators $\widehat{\gamma}_{\mathrm{t}}$, and the right-hand panels to the trivial estimator $\widehat{\gamma} = 0$ (that is, $\widehat{p}_{H,\mathrm{naive}}(x)$). The colors of the contours used are orange for $\widehat{p}>1/2$ and green for $\widehat{p}<1/2$. Darker tones are used for values closer to $0$ or $1$; white color corresponds to $\widehat{p} \approx 1/2$.}
    \label{fig: classification H}
\end{figure}


\end{document}

%% file: 1.0covlen_m2bc_2.tex
\begin{table}[t]
\centering
\caption{\textsf{Gaussian case: }\textbf{coverage and length,} regime medium $m$(a), bias correction $\widehat{\gamma}_1$, means and standard deviations (in brackets) of the coverage of the confidence intervals for the simplicial depth.} 
\label{Tab:1.0covlen_m2bc_2}
\begin{tabular}{cc|c|ccc}
  \hline
                                                       &                                                                                 &                    $x$              & \multicolumn{1}{l}{                          $n=100$} & \multicolumn{1}{l}{                         $n=1000$} & \multicolumn{1}{l}{                        $n=10000$} \\ 
   \hline
\multirow{9}{*}{\rotatebox[origin=c]{90}{coverage}} & \multirow{3}{*}{$\widehat{\gamma}$}               & $\left(0,0.5\right)$ & 0.74500  \scriptsize{(0.43608)} & 0.90400  \scriptsize{(0.29474)} & 0.93300  \scriptsize{(0.25015)} \\ 
                                                        &                                                                                 & $\left(0,0.2\right)$ & 0.78100  \scriptsize{(0.41378)} & 0.84400  \scriptsize{(0.36304)} & 0.92200  \scriptsize{(0.26831)} \\ 
                                                        &                                                                                 & $\left(0,0\right)$   & 0.90300  \scriptsize{(0.29611)} & 0.95400  \scriptsize{(0.20959)} & 0.97700  \scriptsize{(0.14998)} \\  \cdashline{2-6}
                                                        & \multirow{3}{*}{$\widehat{\gamma}_{\mathrm{t}}$} & $\left(0,0.5\right)$ & 0.74500  \scriptsize{(0.43608)} & 0.90400  \scriptsize{(0.29474)} & 0.92700  \scriptsize{(0.26027)} \\ 
                                                        &                                                                                 & $\left(0,0.2\right)$ & 0.78100  \scriptsize{(0.41378)} & 0.84400  \scriptsize{(0.36304)} & 0.92000  \scriptsize{(0.27143)} \\ 
                                                        &                                                                                 & $\left(0,0\right)$   & 0.98900  \scriptsize{(0.10435)} & 0.98500  \scriptsize{(0.12161)} & 0.99500  \scriptsize{(0.07057)} \\  \cdashline{2-6}
                                                        & \multirow{3}{*}{$\widehat{\gamma}_{\mathrm{r}}$} & $\left(0,0.5\right)$ & 0.85100  \scriptsize{(0.35627)} & 0.93400  \scriptsize{(0.24841)} & 0.94900  \scriptsize{(0.22011)} \\ 
                                                        &                                                                                 & $\left(0,0.2\right)$ & 0.80800  \scriptsize{(0.39407)} & 0.91800  \scriptsize{(0.27450)} & 0.94900  \scriptsize{(0.22011)} \\ 
                                                        &                                                                                 & $\left(0,0\right)$   & 0.98900  \scriptsize{(0.10435)} & 0.97400  \scriptsize{(0.15921)} & 0.96500  \scriptsize{(0.18387)} \\  \hline
  \multirow{9}{*}{\rotatebox[origin=c]{90}{length}}   & \multirow{3}{*}{$\widehat{\gamma}$}               & $\left(0,0.5\right)$ & 0.07691  \scriptsize{(0.01273)} & 0.02955  \scriptsize{(0.00265)} & 0.00993  \scriptsize{(0.00094)} \\ 
                                                        &                                                                                 & $\left(0,0.2\right)$ & 0.05861  \scriptsize{(0.01469)} & 0.01737  \scriptsize{(0.00222)} & 0.00597  \scriptsize{(0.00059)} \\ 
                                                        &                                                                                 & $\left(0,0\right)$   & 0.03125  \scriptsize{(0.01073)} & 0.00298  \scriptsize{(0.00091)} & 0.00027  \scriptsize{(0.00006)} \\  \cdashline{2-6}
                                                        & \multirow{3}{*}{$\widehat{\gamma}_{\mathrm{t}}$} & $\left(0,0.5\right)$ & 0.07682  \scriptsize{(0.01256)} & 0.02939  \scriptsize{(0.00238)} & 0.00972  \scriptsize{(0.00066)} \\ 
                                                        &                                                                                 & $\left(0,0.2\right)$ & 0.05863  \scriptsize{(0.01463)} & 0.01737  \scriptsize{(0.00222)} & 0.00593  \scriptsize{(0.00052)} \\ 
                                                        &                                                                                 & $\left(0,0\right)$   & 0.03220  \scriptsize{(0.00984)} & 0.00301  \scriptsize{(0.00088)} & 0.00028  \scriptsize{(0.00006)} \\  \cdashline{2-6}
                                                        & \multirow{3}{*}{$\widehat{\gamma}_{\mathrm{r}}$} & $\left(0,0.5\right)$ & 0.09586  \scriptsize{(0.01415)} & 0.03269  \scriptsize{(0.00091)} & 0.01030  \scriptsize{(0.00029)} \\ 
                                                        &                                                                                 & $\left(0,0.2\right)$ & 0.07636  \scriptsize{(0.02796)} & 0.02204  \scriptsize{(0.00168)} & 0.00656  \scriptsize{(0.00028)} \\ 
                                                        &                                                                                 & $\left(0,0\right)$   & 0.03126  \scriptsize{(0.01772)} & 0.00253  \scriptsize{(0.00144)} & 0.00023  \scriptsize{(0.00001)} \\ 
   \hline
\end{tabular}
\end{table}

%% file: 1.0covlen_m2bc_3.tex
\begin{table}[t]
\centering
\caption{\textsf{Gaussian case: }\textbf{coverage and length,} regime medium $m$(a), bias correction $\widehat{\gamma}_2$, means and standard deviations (in brackets) of the coverage of the confidence intervals for the simplicial depth.} 
\label{Tab:1.0covlen_m2bc_3}
\begin{tabular}{cc|c|ccc}
  \hline
                                                       &                                                                                 &                 $x$                 & \multicolumn{1}{l}{                          $n=100$} & \multicolumn{1}{l}{                         $n=1000$} & \multicolumn{1}{l}{                        $n=10000$} \\ 
   \hline
\multirow{9}{*}{\rotatebox[origin=c]{90}{coverage}} & \multirow{3}{*}{$\widehat{\gamma}$}               & $\left(0,0.5\right)$ & 0.61900  \scriptsize{(0.48588)} & 0.81800  \scriptsize{(0.38604)} & 0.89600  \scriptsize{(0.30541)} \\ 
                                                        &                                                                                 & $\left(0,0.2\right)$ & 0.67600  \scriptsize{(0.46823)} & 0.78200  \scriptsize{(0.41309)} & 0.88400  \scriptsize{(0.32039)} \\ 
                                                        &                                                                                 & $\left(0,0\right)$   & 0.72500  \scriptsize{(0.44674)} & 0.77600  \scriptsize{(0.41713)} & 0.88500  \scriptsize{(0.31918)} \\  \cdashline{2-6}
                                                        & \multirow{3}{*}{$\widehat{\gamma}_{\mathrm{t}}$} & $\left(0,0.5\right)$ & 0.61900  \scriptsize{(0.48588)} & 0.81800  \scriptsize{(0.38604)} & 0.89600  \scriptsize{(0.30541)} \\ 
                                                        &                                                                                 & $\left(0,0.2\right)$ & 0.67600  \scriptsize{(0.46823)} & 0.78200  \scriptsize{(0.41309)} & 0.88400  \scriptsize{(0.32039)} \\ 
                                                        &                                                                                 & $\left(0,0\right)$   & 0.98600  \scriptsize{(0.11755)} & 0.98100  \scriptsize{(0.13659)} & 0.99100  \scriptsize{(0.09449)} \\  \cdashline{2-6}
                                                        & \multirow{3}{*}{$\widehat{\gamma}_{\mathrm{r}}$} & $\left(0,0.5\right)$ & 0.59200  \scriptsize{(0.49171)} & 0.93400  \scriptsize{(0.24841)} & 0.94900  \scriptsize{(0.22011)} \\ 
                                                        &                                                                                 & $\left(0,0.2\right)$ & 0.58700  \scriptsize{(0.49262)} & 0.91800  \scriptsize{(0.27450)} & 0.94900  \scriptsize{(0.22011)} \\ 
                                                        &                                                                                 & $\left(0,0\right)$   & 0.97600  \scriptsize{(0.15313)} & 0.96000  \scriptsize{(0.19606)} & 0.96500  \scriptsize{(0.18387)} \\  \hline
  \multirow{9}{*}{\rotatebox[origin=c]{90}{length}}   & \multirow{3}{*}{$\widehat{\gamma}$}               & $\left(0,0.5\right)$ & 0.05673  \scriptsize{(0.01096)} & 0.02461  \scriptsize{(0.00242)} & 0.00891  \scriptsize{(0.00088)} \\ 
                                                        &                                                                                 & $\left(0,0.2\right)$ & 0.04164  \scriptsize{(0.01166)} & 0.01426  \scriptsize{(0.00193)} & 0.00534  \scriptsize{(0.00055)} \\ 
                                                        &                                                                                 & $\left(0,0\right)$   & 0.02022  \scriptsize{(0.00834)} & 0.00218  \scriptsize{(0.00073)} & 0.00022  \scriptsize{(0.00005)} \\  \cdashline{2-6}
                                                        & \multirow{3}{*}{$\widehat{\gamma}_{\mathrm{t}}$} & $\left(0,0.5\right)$ & 0.05674  \scriptsize{(0.01092)} & 0.02461  \scriptsize{(0.00242)} & 0.00888  \scriptsize{(0.00082)} \\ 
                                                        &                                                                                 & $\left(0,0.2\right)$ & 0.04211  \scriptsize{(0.01079)} & 0.01426  \scriptsize{(0.00193)} & 0.00534  \scriptsize{(0.00054)} \\ 
                                                        &                                                                                 & $\left(0,0\right)$   & 0.02653  \scriptsize{(0.00455)} & 0.00253  \scriptsize{(0.00053)} & 0.00024  \scriptsize{(0.00004)} \\  \cdashline{2-6}
                                                        & \multirow{3}{*}{$\widehat{\gamma}_{\mathrm{r}}$} & $\left(0,0.5\right)$ & 0.05970  \scriptsize{(0.03181)} & 0.03269  \scriptsize{(0.00091)} & 0.01030  \scriptsize{(0.00029)} \\ 
                                                        &                                                                                 & $\left(0,0.2\right)$ & 0.03447  \scriptsize{(0.01480)} & 0.02204  \scriptsize{(0.00168)} & 0.00656  \scriptsize{(0.00028)} \\ 
                                                        &                                                                                 & $\left(0,0\right)$   & 0.02545  \scriptsize{(0.00301)} & 0.00234  \scriptsize{(0.00014)} & 0.00023  \scriptsize{(0.00001)} \\ 
   \hline
\end{tabular}
\end{table}

%% file: 1.0gamma1.tex
\begin{table}[htpb]
\centering
\caption{\textbf{Estimation of $\gamma$,} \textbf{no bias correction:} Means and standard deviations (first number in brackets) of squared errors between the estimated parameter $\widehat{\gamma}$ and the true exponent $\gamma$. The second number in brackets is the proportion of precisely estimated $\gamma$.} 
\label{Tab:gamma1}
\begin{tabular}{ccl|ccc}
   \hline
                                                                              &                                                      &           & \multicolumn{1}{l}{                                      $n=100$} & \multicolumn{1}{l}{                                     $n=1000$} & \multicolumn{1}{l}{                                    $n=10000$} \\ 
   \hline
\multirow{9}{*}{\rotatebox[origin=c]{90}{$x = \left(0,0.5\right)$}} & \multirow{3}{*}{$\widehat{\gamma}$}               & small $m$ & 0.310  \scriptsize{(0.261/0.000)}           & 0.029  \scriptsize{(0.022/0.000)}           & 0.005  \scriptsize{(0.005/0.000)}           \\ 
                                                                                &                                                      & med. $m$  & 0.115  \scriptsize{(0.058/0.000)}           & 0.011  \scriptsize{(0.006/0.000)}           & 0.002  \scriptsize{(0.002/0.000)}           \\ 
                                                                                &                                                      & large $m$ & 0.131  \scriptsize{(0.043/0.000)}           & 0.017  \scriptsize{(0.009/0.000)}           & 0.004  \scriptsize{(0.003/0.000)}           \\  \cdashline{2-6}
                                                                                & \multirow{3}{*}{$\widehat{\gamma}_{\mathrm{t}}$} & small $m$ & 0.172  \scriptsize{(0.100/0.051)}           & 0.029  \scriptsize{(0.022/0.003)}           & 0.005  \scriptsize{(0.005/0.027)}           \\ 
                                                                                &                                                      & med. $m$  & \textbf{0.113}  \scriptsize{(0.053/0.000)} & 0.011  \scriptsize{(0.006/0.001)}           & 0.002  \scriptsize{(0.002/0.048)}           \\
\hline
  \end{tabular}
\end{table}

%% file: 1.0gamma2.tex
\begin{table}[htpb]
\centering
\caption{\textbf{Estimation of $\gamma$,} \textbf{bias correction 1:} Means and standard deviations (first number in brackets) of squared errors between the estimated parameter $\widehat{\gamma}$ and the true exponent $\gamma$. The second number in brackets is the proportion of precisely estimated $\gamma$.} 
\begin{tabular}{ccl|ccc}
   \hline
                                                                              &                                                      &           & \multicolumn{1}{l}{                                      $n=100$} & \multicolumn{1}{l}{                                     $n=1000$} & \multicolumn{1}{l}{                                    $n=10000$} \\ 
   \hline
\multirow{9}{*}{\rotatebox[origin=c]{90}{$x = \left(0,0.5\right)$}} & \multirow{3}{*}{$\widehat{\gamma}$}               & small $m$ & 0.220  \scriptsize{(0.195/0.000)}           & 0.024  \scriptsize{(0.020/0.000)}           & 0.005  \scriptsize{(0.004/0.000)}           \\ 
                                                                                &                                                      & med. $m$  & 0.019  \scriptsize{(0.018/0.000)}           & 0.002  \scriptsize{(0.002/0.000)}           & 0.001  \scriptsize{(0.001/0.000)}           \\ 
                                                                                &                                                      & large $m$ & 0.006  \scriptsize{(0.006/0.000)}           & 0.001  \scriptsize{(0.002/0.000)}           & 0.001  \scriptsize{(0.001/0.000)}           \\  \cdashline{2-6}
                                                                                & \multirow{3}{*}{$\widehat{\gamma}_{\mathrm{t}}$} & small $m$ & 0.145  \scriptsize{(0.099/0.060)}           & 0.024  \scriptsize{(0.020/0.008)}           & 0.005  \scriptsize{(0.004/0.035)}           \\ 
                                                                                &                                                      & med. $m$  & 0.019  \scriptsize{(0.018/0.017)}           & 0.002  \scriptsize{(0.002/0.112)}           & \textbf{0.000}  \scriptsize{(0.001/0.329)} \\
\hline
  \end{tabular}
\end{table}

%% file: 1.0gamma3.tex
\begin{table}[htpb]
\centering
\caption{\textbf{Estimation of $\gamma$,} \textbf{bias correction 2:} Means and standard deviations (first number in brackets) of squared errors between the estimated parameter $\widehat{\gamma}$ and the true exponent $\gamma$. The second number in brackets is the proportion of precisely estimated $\gamma$.} 
\label{Tab:gamma3}
\begin{tabular}{ccl|ccc}
   \hline
                                                                              &                                                      &           & \multicolumn{1}{l}{                                      $n=100$} & \multicolumn{1}{l}{                                     $n=1000$} & \multicolumn{1}{l}{                                    $n=10000$} \\ 
   \hline
\multirow{9}{*}{\rotatebox[origin=c]{90}{$x = \left(0,0.5\right)$}} & \multirow{3}{*}{$\widehat{\gamma}$}               & small $m$ & 0.301  \scriptsize{(0.255/0.000)}           & 0.029  \scriptsize{(0.022/0.000)}           & 0.005  \scriptsize{(0.005/0.000)}           \\ 
                                                                                &                                                      & med. $m$  & 0.081  \scriptsize{(0.046/0.000)}           & 0.009  \scriptsize{(0.005/0.000)}           & 0.002  \scriptsize{(0.002/0.000)}           \\ 
                                                                                &                                                      & large $m$ & \textbf{0.068}  \scriptsize{(0.027/0.000)} & 0.012  \scriptsize{(0.007/0.000)}           & 0.003  \scriptsize{(0.003/0.000)}           \\  \cdashline{2-6}
                                                                                & \multirow{3}{*}{$\widehat{\gamma}_{\mathrm{t}}$} & small $m$ & 0.170  \scriptsize{(0.100/0.052)}           & 0.029  \scriptsize{(0.022/0.003)}           & 0.005  \scriptsize{(0.005/0.027)}           \\ 
                                                                                &                                                      & med. $m$  & 0.081  \scriptsize{(0.044/0.000)}           & 0.009  \scriptsize{(0.005/0.002)}           & 0.002  \scriptsize{(0.002/0.058)}           \\
\hline
  \end{tabular}
\end{table}

%% file: 1.0KS1.tex
\begin{table}[htpb]
\centering
\caption{\textbf{Kolmogorov-Smirnov dist.,} \textbf{no bias correction:} Means and standard deviations (in brackets) of the Kolmogorov-Smirnov distance between the estimated distribution and the true distribution.} 
\label{Tab:KS1}
\begin{tabular}{ccl|ccc}
  \hline
                                                                                &                                                      &             & \multicolumn{1}{l}{                                $n=100$} & \multicolumn{1}{l}{                               $n=1000$} & \multicolumn{1}{l}{                              $n=10000$} \\ 
   \hline
\multirow{12}{*}{\rotatebox[origin=c]{90}{$x = \left(0,0.5\right)$}} & \multirow{4}{*}{$\widehat{\gamma}$}               & small $m$   & 0.337  \scriptsize{(0.098)}           & 0.158  \scriptsize{(0.055)}           & 0.091  \scriptsize{(0.036)}           \\ 
                                                                                 &                                                      & med. $m$(a) & 0.276  \scriptsize{(0.047)}           & 0.110  \scriptsize{(0.027)}           & 0.062  \scriptsize{(0.023)}           \\ 
                                                                                 &                                                      & med. $m$(b) & \textbf{0.243}  \scriptsize{(0.055)} & 0.110  \scriptsize{(0.037)}           & 0.065  \scriptsize{(0.031)}           \\ 
                                                                                 &                                                      & large $m$   & 0.273  \scriptsize{(0.037)}           & 0.138  \scriptsize{(0.036)}           & 0.090  \scriptsize{(0.036)}           \\  \cdashline{2-6}
                                                                                 & \multirow{4}{*}{$\widehat{\gamma}_{\mathrm{t}}$} & small $m$   & 0.299  \scriptsize{(0.072)}           & 0.158  \scriptsize{(0.055)}           & 0.091  \scriptsize{(0.036)}           \\
\hline
\end{tabular}
\end{table}

%% file: 1.0KS2.tex
\begin{table}[htpb]
\centering
\caption{\textbf{Kolmogorov-Smirnov dist.,} \textbf{bias correction 1:} Means and standard deviations (in brackets) of the Kolmogorov-Smirnov distance between the estimated distribution and the true distribution.} 
\begin{tabular}{ccl|ccc}
  \hline
                                                                                &                                                      &             & \multicolumn{1}{l}{                                $n=100$} & \multicolumn{1}{l}{                               $n=1000$} & \multicolumn{1}{l}{                              $n=10000$} \\ 
   \hline
\multirow{12}{*}{\rotatebox[origin=c]{90}{$x = \left(0,0.5\right)$}} & \multirow{4}{*}{$\widehat{\gamma}$}               & small $m$   & 0.307  \scriptsize{(0.098)}           & 0.145  \scriptsize{(0.054)}           & 0.088  \scriptsize{(0.035)}           \\ 
                                                                                 &                                                      & med. $m$(a) & 0.167  \scriptsize{(0.047)}           & 0.057  \scriptsize{(0.021)}           & 0.043  \scriptsize{(0.016)}           \\ 
                                                                                 &                                                      & med. $m$(b) & 0.086  \scriptsize{(0.040)}           & 0.049  \scriptsize{(0.020)}           & 0.049  \scriptsize{(0.021)}           \\ 
                                                                                 &                                                      & large $m$   & 0.086  \scriptsize{(0.028)}           & 0.046  \scriptsize{(0.019)}           & 0.047  \scriptsize{(0.020)}           \\  \cdashline{2-6}
                                                                                 & \multirow{4}{*}{$\widehat{\gamma}_{\mathrm{t}}$} & small $m$   & 0.281  \scriptsize{(0.078)}           & 0.145  \scriptsize{(0.054)}           & 0.088  \scriptsize{(0.036)}           \\
\hline
\end{tabular}
\end{table}

%% file: 1.0KS3.tex
\begin{table}[htpb]
\centering
\caption{\textbf{Kolmogorov-Smirnov dist.,} \textbf{bias correction 2:} Means and standard deviations (in brackets) of the Kolmogorov-Smirnov distance between the estimated distribution and the true distribution.} 
\label{Tab:KS3}
\begin{tabular}{ccl|ccc}
  \hline
                                                                                &                                                      &             & \multicolumn{1}{l}{                                $n=100$} & \multicolumn{1}{l}{                               $n=1000$} & \multicolumn{1}{l}{                              $n=10000$} \\ 
   \hline
\multirow{12}{*}{\rotatebox[origin=c]{90}{$x = \left(0,0.5\right)$}} & \multirow{4}{*}{$\widehat{\gamma}$}               & small $m$   & 0.334  \scriptsize{(0.098)}           & 0.158  \scriptsize{(0.055)}           & 0.091  \scriptsize{(0.036)}           \\ 
                                                                                 &                                                      & med. $m$(a) & 0.248  \scriptsize{(0.047)}           & 0.103  \scriptsize{(0.027)}           & 0.061  \scriptsize{(0.023)}           \\ 
                                                                                 &                                                      & med. $m$(b) & \textbf{0.203}  \scriptsize{(0.055)} & 0.100  \scriptsize{(0.036)}           & 0.063  \scriptsize{(0.031)}           \\ 
                                                                                 &                                                      & large $m$   & 0.212  \scriptsize{(0.037)}           & 0.117  \scriptsize{(0.035)}           & 0.083  \scriptsize{(0.035)}           \\  \cdashline{2-6}
                                                                                 & \multirow{4}{*}{$\widehat{\gamma}_{\mathrm{t}}$} & small $m$   & 0.298  \scriptsize{(0.073)}           & 0.158  \scriptsize{(0.055)}           & 0.091  \scriptsize{(0.036)}           \\
\hline
\end{tabular}
\end{table}

%% file: 2.0covlen_m2bc_2.tex
\begin{table}[htpb]
\centering
\caption{\textsf{Cauchy case: }\textbf{coverage and length,} regime medium $m$(a), bias correction $\widehat{\gamma}_1$, means and standard deviations (in brackets) of the coverage of the confidence intervals for the simplicial depth.} 
\begin{tabular}{cc|c|ccc}
  \hline
                                                       &                                                                                 &           $x$                       & \multicolumn{1}{l}{                          $n=100$} & \multicolumn{1}{l}{                         $n=1000$} & \multicolumn{1}{l}{                        $n=10000$} \\ 
   \hline
\multirow{9}{*}{\rotatebox[origin=c]{90}{coverage}} & \multirow{3}{*}{\rotatebox[origin=c]{90}{$\widehat{\gamma}$}}               & $\left(0,0.5\right)$ & 0.78100  \scriptsize{(0.41378)} & 0.90900  \scriptsize{(0.28775)} & 0.92600  \scriptsize{(0.26190)} \\ 
                                                        &                                                                                 & $\left(0,0.2\right)$ & 0.76700  \scriptsize{(0.42295)} & 0.80100  \scriptsize{(0.39945)} & 0.89400  \scriptsize{(0.30799)} \\ 
                                                        &                                                                                 & $\left(0,0\right)$   & 0.89100  \scriptsize{(0.31180)} & 0.95800  \scriptsize{(0.20069)} & 0.98800  \scriptsize{(0.10894)} \\  \cdashline{2-6}
                                                        & \multirow{3}{*}{\rotatebox[origin=c]{90}{$\widehat{\gamma}_{\mathrm{t}}$}} & $\left(0,0.5\right)$ & 0.78100  \scriptsize{(0.41378)} & 0.90800  \scriptsize{(0.28917)} & 0.92200  \scriptsize{(0.26831)} \\ 
                                                        &                                                                                 & $\left(0,0.2\right)$ & 0.76700  \scriptsize{(0.42295)} & 0.80100  \scriptsize{(0.39945)} & 0.89400  \scriptsize{(0.30799)} \\ 
                                                        &                                                                                 & $\left(0,0\right)$   & 0.99000  \scriptsize{(0.09955)} & 0.98900  \scriptsize{(0.10435)} & 1.00000  \scriptsize{(0.00000)} \\  \cdashline{2-6}
                                                        & \multirow{3}{*}{\rotatebox[origin=c]{90}{$\widehat{\gamma}_{\mathrm{r}}$}} & $\left(0,0.5\right)$ & 0.88000  \scriptsize{(0.32512)} & 0.94600  \scriptsize{(0.22613)} & 0.94500  \scriptsize{(0.22809)} \\ 
                                                        &                                                                                 & $\left(0,0.2\right)$ & 0.71300  \scriptsize{(0.45259)} & 0.89600  \scriptsize{(0.30541)} & 0.93200  \scriptsize{(0.25187)} \\ 
                                                        &                                                                                 & $\left(0,0\right)$   & 0.99100  \scriptsize{(0.09449)} & 0.97700  \scriptsize{(0.14998)} & 0.98700  \scriptsize{(0.11333)} \\  \hline
  \multirow{9}{*}{\rotatebox[origin=c]{90}{length}}   & \multirow{3}{*}{\rotatebox[origin=c]{90}{$\widehat{\gamma}$}}               & $\left(0,0.5\right)$ & 0.07866  \scriptsize{(0.00928)} & 0.02681  \scriptsize{(0.00253)} & 0.00904  \scriptsize{(0.00087)} \\ 
                                                        &                                                                                 & $\left(0,0.2\right)$ & 0.05066  \scriptsize{(0.01472)} & 0.01400  \scriptsize{(0.00220)} & 0.00471  \scriptsize{(0.00051)} \\ 
                                                        &                                                                                 & $\left(0,0\right)$   & 0.03090  \scriptsize{(0.01042)} & 0.00306  \scriptsize{(0.00095)} & 0.00027  \scriptsize{(0.00006)} \\  \cdashline{2-6}
                                                        & \multirow{3}{*}{\rotatebox[origin=c]{90}{$\widehat{\gamma}_{\mathrm{t}}$}} & $\left(0,0.5\right)$ & 0.07866  \scriptsize{(0.00927)} & 0.02676  \scriptsize{(0.00242)} & 0.00889  \scriptsize{(0.00067)} \\ 
                                                        &                                                                                 & $\left(0,0.2\right)$ & 0.05075  \scriptsize{(0.01455)} & 0.01400  \scriptsize{(0.00220)} & 0.00470  \scriptsize{(0.00049)} \\ 
                                                        &                                                                                 & $\left(0,0\right)$   & 0.03192  \scriptsize{(0.00946)} & 0.00309  \scriptsize{(0.00092)} & 0.00027  \scriptsize{(0.00005)} \\  \cdashline{2-6}
                                                        & \multirow{3}{*}{\rotatebox[origin=c]{90}{$\widehat{\gamma}_{\mathrm{r}}$}} & $\left(0,0.5\right)$ & 0.10307  \scriptsize{(0.00716)} & 0.03047  \scriptsize{(0.00118)} & 0.00950  \scriptsize{(0.00030)} \\ 
                                                        &                                                                                 & $\left(0,0.2\right)$ & 0.06351  \scriptsize{(0.03022)} & 0.01898  \scriptsize{(0.00172)} & 0.00540  \scriptsize{(0.00027)} \\ 
                                                        &                                                                                 & $\left(0,0\right)$   & 0.03090  \scriptsize{(0.01713)} & 0.00258  \scriptsize{(0.00158)} & 0.00023  \scriptsize{(0.00001)} \\ 
   \hline
\end{tabular}
\end{table}

%% file: 2.0covlen_m2bc_3.tex
\begin{table}[htpb]
\centering
\caption{\textsf{Cauchy case: }\textbf{coverage and length,} regime medium $m$(a), bias correction $\widehat{\gamma}_2$, means and standard deviations (in brackets) of the coverage of the confidence intervals for the simplicial depth.} 
\begin{tabular}{cc|c|ccc}
  \hline
                                                       &                                                                                 &         $x$                         & \multicolumn{1}{l}{                          $n=100$} & \multicolumn{1}{l}{                         $n=1000$} & \multicolumn{1}{l}{                        $n=10000$} \\ 
   \hline
\multirow{9}{*}{\rotatebox[origin=c]{90}{coverage}} & \multirow{3}{*}{\rotatebox[origin=c]{90}{$\widehat{\gamma}$}}               & $\left(0,0.5\right)$ & 0.69200  \scriptsize{(0.46190)} & 0.83500  \scriptsize{(0.37137)} & 0.88700  \scriptsize{(0.31675)} \\ 
                                                        &                                                                                 & $\left(0,0.2\right)$ & 0.68400  \scriptsize{(0.46515)} & 0.72500  \scriptsize{(0.44674)} & 0.85900  \scriptsize{(0.34820)} \\ 
                                                        &                                                                                 & $\left(0,0\right)$   & 0.70200  \scriptsize{(0.45761)} & 0.80700  \scriptsize{(0.39485)} & 0.90800  \scriptsize{(0.28917)} \\  \cdashline{2-6}
                                                        & \multirow{3}{*}{\rotatebox[origin=c]{90}{$\widehat{\gamma}_{\mathrm{t}}$}} & $\left(0,0.5\right)$ & 0.69200  \scriptsize{(0.46190)} & 0.83500  \scriptsize{(0.37137)} & 0.88700  \scriptsize{(0.31675)} \\ 
                                                        &                                                                                 & $\left(0,0.2\right)$ & 0.68600  \scriptsize{(0.46435)} & 0.72500  \scriptsize{(0.44674)} & 0.85900  \scriptsize{(0.34820)} \\ 
                                                        &                                                                                 & $\left(0,0\right)$   & 0.98900  \scriptsize{(0.10435)} & 0.98800  \scriptsize{(0.10894)} & 0.99900  \scriptsize{(0.03162)} \\  \cdashline{2-6}
                                                        & \multirow{3}{*}{\rotatebox[origin=c]{90}{$\widehat{\gamma}_{\mathrm{r}}$}} & $\left(0,0.5\right)$ & 0.61400  \scriptsize{(0.48707)} & 0.94600  \scriptsize{(0.22613)} & 0.94500  \scriptsize{(0.22809)} \\ 
                                                        &                                                                                 & $\left(0,0.2\right)$ & 0.60600  \scriptsize{(0.48888)} & 0.89500  \scriptsize{(0.30671)} & 0.93200  \scriptsize{(0.25187)} \\ 
                                                        &                                                                                 & $\left(0,0\right)$   & 0.97700  \scriptsize{(0.14998)} & 0.96100  \scriptsize{(0.19369)} & 0.98700  \scriptsize{(0.11333)} \\  \hline
  \multirow{9}{*}{\rotatebox[origin=c]{90}{length}}   & \multirow{3}{*}{\rotatebox[origin=c]{90}{$\widehat{\gamma}$}}               & $\left(0,0.5\right)$ & 0.05750  \scriptsize{(0.00802)} & 0.02227  \scriptsize{(0.00228)} & 0.00810  \scriptsize{(0.00082)} \\ 
                                                        &                                                                                 & $\left(0,0.2\right)$ & 0.03537  \scriptsize{(0.01164)} & 0.01141  \scriptsize{(0.00189)} & 0.00420  \scriptsize{(0.00047)} \\ 
                                                        &                                                                                 & $\left(0,0\right)$   & 0.01993  \scriptsize{(0.00807)} & 0.00224  \scriptsize{(0.00077)} & 0.00022  \scriptsize{(0.00005)} \\  \cdashline{2-6}
                                                        & \multirow{3}{*}{\rotatebox[origin=c]{90}{$\widehat{\gamma}_{\mathrm{t}}$}} & $\left(0,0.5\right)$ & 0.05750  \scriptsize{(0.00802)} & 0.02227  \scriptsize{(0.00228)} & 0.00809  \scriptsize{(0.00078)} \\ 
                                                        &                                                                                 & $\left(0,0.2\right)$ & 0.03669  \scriptsize{(0.00978)} & 0.01141  \scriptsize{(0.00189)} & 0.00420  \scriptsize{(0.00047)} \\ 
                                                        &                                                                                 & $\left(0,0\right)$   & 0.02641  \scriptsize{(0.00414)} & 0.00257  \scriptsize{(0.00057)} & 0.00024  \scriptsize{(0.00004)} \\  \cdashline{2-6}
                                                        & \multirow{3}{*}{\rotatebox[origin=c]{90}{$\widehat{\gamma}_{\mathrm{r}}$}} & $\left(0,0.5\right)$ & 0.05396  \scriptsize{(0.02910)} & 0.03047  \scriptsize{(0.00118)} & 0.00950  \scriptsize{(0.00030)} \\ 
                                                        &                                                                                 & $\left(0,0.2\right)$ & 0.03064  \scriptsize{(0.00985)} & 0.01895  \scriptsize{(0.00186)} & 0.00540  \scriptsize{(0.00027)} \\ 
                                                        &                                                                                 & $\left(0,0\right)$   & 0.02546  \scriptsize{(0.00295)} & 0.00238  \scriptsize{(0.00053)} & 0.00023  \scriptsize{(0.00001)} \\ 
   \hline
\end{tabular}
\end{table}

%% file: SDarXiv.bbl
\def\polhk#1{\setbox0=\hbox{#1}{\ooalign{\hidewidth
  \lower1.5ex\hbox{`}\hidewidth\crcr\unhbox0}}}
\begin{thebibliography}{}

\bibitem[Aloupis et~al., 2002]{Aloupis_etal2002}
Aloupis, G., Cort\'{e}s, C., G\'{o}mez, F., Soss, M., and Toussaint, G. (2002).
\newblock Lower bounds for computing statistical depth.
\newblock {\em Comput. Statist. Data Anal.}, 40(2):223--229.

\bibitem[Arcones and Gin\'e, 1992]{Arcones_Gine1992}
Arcones, M.~A. and Gin\'e, E. (1992).
\newblock On the bootstrap of {$U$} and {$V$} statistics.
\newblock {\em Ann. Statist.}, 20(2):655--674.

\bibitem[Cascos, 2009]{Cascos2009}
Cascos, I. (2009).
\newblock Depth functions based on a number of observations of a random vector.
\newblock Working Paper 07-29, Universidad Carlos III de Madrid.

\bibitem[Chaudhuri, 1996]{Chaudhuri1996}
Chaudhuri, P. (1996).
\newblock On a geometric notion of quantiles for multivariate data.
\newblock {\em J. Amer. Statist. Assoc.}, 91(434):862--872.

\bibitem[Cuesta-Albertos et~al., 2017]{Cuesta_etal2017}
Cuesta-Albertos, J.~A., Febrero-Bande, M., and Oviedo de~la Fuente, M. (2017).
\newblock The {${\rm DD}^G$}-classifier in the functional setting.
\newblock {\em TEST}, 26(1):119--142.

\bibitem[de~la Pe{\~n}a and Gin{\'e}, 1999]{delaPena_Gine1999}
de~la Pe{\~n}a, V.~H. and Gin{\'e}, E. (1999).
\newblock {\em Decoupling. From dependence to independence}.
\newblock Probability and its Applications (New York). Springer-Verlag, New
  York.

\bibitem[Dudley, 2002]{Dudley2002}
Dudley, R.~M. (2002).
\newblock {\em Real analysis and probability}, volume~74 of {\em Cambridge
  Studies in Advanced Mathematics}.
\newblock Cambridge University Press, Cambridge.
\newblock Revised reprint of the 1989 original.

\bibitem[Dyckerhoff et~al., 2015]{Dyckerhoff_etal2015}
Dyckerhoff, R., Ley, C., and Paindaveine, D. (2015).
\newblock Depth-based runs tests for bivariate central symmetry.
\newblock {\em Ann. Inst. Statist. Math.}, 67(5):917--941.

\bibitem[Eddelbuettel et~al., 2024]{RcppArmadillo}
Eddelbuettel, D., Francois, R., Bates, D., Ni, B., and Sanderson, C. (2024).
\newblock {\em RcppArmadillo: `Rcpp' Integration for the `Armadillo' templated
  linear algebra library}.
\newblock R package version 14.2.2-1.

\bibitem[Ghosh and Chaudhuri, 2005]{Ghosh_Chaudhuri2005}
Ghosh, A.~K. and Chaudhuri, P. (2005).
\newblock On maximum depth and related classifiers.
\newblock {\em Scand. J. Statist.}, 32(2):327--350.

\bibitem[Hubert et~al., 2017]{Hubert_etal2017}
Hubert, M., Rousseeuw, P., and Segaert, P. (2017).
\newblock Multivariate and functional classification using depth and distance.
\newblock {\em Adv. Data Anal. Classif.}, 11(3):445--466.

\bibitem[Killeen and Hettmansperger, 2021]{Killeen_Hettmansperger2021}
Killeen, T.~J. and Hettmansperger, T.~P. (2021).
\newblock A bivariate test for location based on simplicial depth.
\newblock {\em Comm. Statist. Theory Methods}, 50(17):3954--3971.

\bibitem[Koshevoy and Mosler, 1998]{Koshevoy_Mosler1998}
Koshevoy, G. and Mosler, K. (1998).
\newblock Lift zonoids, random convex hulls and the variability of random
  vectors.
\newblock {\em Bernoulli}, 4(3):377--399.

\bibitem[Lee, 1990]{Lee1990}
Lee, A.~J. (1990).
\newblock {\em {$U$}-statistics. {T}heory and practice}, volume 110 of {\em
  Statistics: Textbooks and Monographs}.
\newblock Marcel Dekker, Inc., New York.

\bibitem[Li et~al., 2012]{Li_etal2012}
Li, J., Cuesta-Albertos, J.~A., and Liu, R.~Y. (2012).
\newblock {$DD$}-classifier: nonparametric classification procedure based on
  {$DD$}-plot.
\newblock {\em J. Amer. Statist. Assoc.}, 107(498):737--753.

\bibitem[Liu, 1988]{Liu1988}
Liu, R.~Y. (1988).
\newblock On a notion of simplicial depth.
\newblock {\em Proc. Natl. Acad. Sci. U.S.A.}, 85(6):1732--1734.

\bibitem[Liu, 1990]{Liu1990}
Liu, R.~Y. (1990).
\newblock On a notion of data depth based on random simplices.
\newblock {\em Ann. Statist.}, 18(1):405--414.

\bibitem[Liu et~al., 1999]{Liu_etal1999}
Liu, R.~Y., Parelius, J.~M., and Singh, K. (1999).
\newblock Multivariate analysis by data depth: descriptive statistics, graphics
  and inference.
\newblock {\em Ann. Statist.}, 27(3):783--858.

\bibitem[Malcherczyk et~al., 2021]{Malcherczyk_etal2021}
Malcherczyk, D., Leckey, K., and M\"{u}ller, C.~H. (2021).
\newblock {$K$}-sign depth: from asymptotics to efficient implementation.
\newblock {\em J. Statist. Plann. Inference}, 215:344--355.

\bibitem[Mendro\v{s}, 2025]{Mendros2025}
Mendro\v{s}, E. (2025).
\newblock Sign tests for multivariate data.
\newblock Master's thesis, Charles University.

\bibitem[Mendro\v{s} and Nagy, 2025]{Mendros_Nagy2025}
Mendro\v{s}, E. and Nagy, S. (2025).
\newblock Explicit bivariate simplicial depth.
\newblock {\em J. Multivariate Anal.}, 205:105375.

\bibitem[Mosler and Hoberg, 2006]{Mosler_Hoberg2006}
Mosler, K. and Hoberg, R. (2006).
\newblock Data analysis and classification with the zonoid depth.
\newblock In {\em Data depth: robust multivariate analysis, computational
  geometry and applications}, volume~72 of {\em DIMACS Ser. Discrete Math.
  Theoret. Comput. Sci.}, pages 49--59. Amer. Math. Soc., Providence, RI.

\bibitem[Mozharovskyi et~al., 2015]{Mozharovskyi_etal2015}
Mozharovskyi, P., Mosler, K., and Lange, T. (2015).
\newblock Classifying real-world data with the {$DD\alpha$}-procedure.
\newblock {\em Adv. Data Anal. Classif.}, 9(3):287--314.

\bibitem[M\"{u}ller et~al., 2025]{Muller_etal2025}
M\"{u}ller, C.~F., Nagy, S., and Trippler, S. (2025).
\newblock Bivariate and multivariate sign depth and distribution-free tests.
\newblock {\em Under review}.

\bibitem[Oja and Nyblom, 1989]{Oja_Nyblom1989}
Oja, H. and Nyblom, J. (1989).
\newblock Bivariate sign tests.
\newblock {\em J. Amer. Statist. Assoc.}, 84(405):249--259.

\bibitem[Pokotylo et~al., 2017]{ddalpha}
Pokotylo, O., Mozharovskyi, P., Dyckerhoff, R., and Nagy, S. (2017).
\newblock {\em ddalpha: Depth-based classification and calculation of data
  depth}.
\newblock R package version 1.3.11.

\bibitem[Politis et~al., 1999]{Politis_Romano1999}
Politis, D.~N., Romano, J.~P., and Wolf, M. (1999).
\newblock {\em Subsampling}.
\newblock Springer Series in Statistics. Springer-Verlag, New York.

\bibitem[Rousseeuw and Ruts, 1996]{Rousseeuw_Ruts1996}
Rousseeuw, P.~J. and Ruts, I. (1996).
\newblock {Algorithm {AS} 307: Bivariate location depth.}
\newblock {\em J. R. Stat. Soc., Ser. C}, 45(4):516--526.

\bibitem[Rousseeuw and Struyf, 2004]{Rousseeuw_Struyf2004}
Rousseeuw, P.~J. and Struyf, A. (2004).
\newblock Characterizing angular symmetry and regression symmetry.
\newblock {\em J. Stat. Plan. Inference}, 122(1-2):161--173.

\bibitem[Serfling, 1980]{Serfling1980}
Serfling, R. (1980).
\newblock {\em Approximation theorems of mathematical statistics}.
\newblock John Wiley \& Sons Inc., New York.
\newblock Wiley Series in Probability and Mathematical Statistics.

\bibitem[Tukey, 1975]{Tukey1975}
Tukey, J.~W. (1975).
\newblock Mathematics and the picturing of data.
\newblock In {\em Proceedings of the {I}nternational {C}ongress of
  {M}athematicians ({V}ancouver, {B}. {C}., 1974), {V}ol. 2}, pages 523--531.
  Canad. Math. Congress, Montreal, Que.

\bibitem[Zuo and Serfling, 2000a]{Zuo_Serfling2000}
Zuo, Y. and Serfling, R. (2000a).
\newblock General notions of statistical depth function.
\newblock {\em Ann. Statist.}, 28(2):461--482.

\bibitem[Zuo and Serfling, 2000b]{Zuo_Serfling2000c}
Zuo, Y. and Serfling, R. (2000b).
\newblock On the performance of some robust nonparametric location measures
  relative to a general notion of multivariate symmetry.
\newblock {\em J. Stat. Plan. Inference}, 84(1-2):55--79.

\end{thebibliography}
